\documentclass[conference]{IEEEtran}
\usepackage[T1]{fontenc}
\def\BibTeX{{\rm B\kern-.05em{\sc i\kern-.025em b}\kern-.08em
    T\kern-.1667em\lower.7ex\hbox{E}\kern-.125emX}}
\usepackage{subcaption}
\usepackage{cite}
\usepackage{amsmath,amsthm,amssymb,amsfonts}
\usepackage{lipsum}
\usepackage{algorithm}
\usepackage{algpseudocode}

\DeclareMathOperator*{\argmin}{argmin}
\usepackage{mathtools}
\usepackage{dsfont}
\usepackage{graphicx}
\usepackage{textcomp}
\usepackage{xcolor}
\usepackage{svg}
\usepackage[spaces,hyphens]{url}
\theoremstyle{plain}
\newtheorem{lemma}{Lemma}
\newcommand{\hbarr}{\bar{h}}
\newcommand{\taustar}{\bar{\tau}}

\newcommand{\gbar}{\bar{g}}
\newtheorem{theorem}{Theorem}

\theoremstyle{definition}
\newtheorem{definition}{Definition}

\newtheorem{remark}{Remark}
\newcommand{\cbar}{\bar{C}_{av}}
\newcommand{\abar}{\bar{a}}
\newcommand{\hbarav}{\bar{h}_{av}}
\newcommand{\thetabar}{\bar{\theta}}
\renewcommand{\baselinestretch}{0.93}
\newcommand{\remove}[1]{}

\usepackage{accents}

\newcommand{\bbE}{\mathbb{E}}
\newcommand{\bs}{\mathbf{s}}
\usepackage{bbm}

\begin{document}

\title{ Age Aware Content Fetching and Broadcast in a Sensing-as-a-Service System}

\author{ \IEEEauthorblockN{ Ankita Koley\IEEEauthorrefmark{1}, Anu Krishna\IEEEauthorrefmark{1},  Chandramani Singh\IEEEauthorrefmark{1}, V Mahendran \IEEEauthorrefmark{2}} \IEEEauthorblockA{ \IEEEauthorrefmark{1}Department of Electronic Systems Engineering, Indian Institute of Science, Bangalore, India\\ \IEEEauthorrefmark{2}Department of Computer Science and Engineering, Indian Institute of Technology, Tirupati, India} }
\maketitle
\begin{abstract}
We consider a {\em Sensing-as-a-Service} (S2aaS) system consisting of a sensor, a set of users, and a sensor cloud service provider (SCSP). The sensor updates its content each time it captures a new measurement.  The SCSP occasionally fetches the content from the sensor,  caches the latest fetched version and broadcasts it on being requested by the users. The SCSP incurs content fetching costs while fetching and broadcasting the contents. The SCSP also incurs an age cost if users do not receive the most recent version of the content after requesting. We study a content fetching and broadcast problem, aiming to minimize the time-averaged content fetching and age costs.  The problem can be framed as a Markov decision process but cannot be elegantly solved owing to its multi-dimensional state space and complex dynamics. To address this, we first obtain the optimal policy for the homogeneous case with all the users having the same request probability and age cost. We extend this algorithm for heterogeneous case but the complexity grows exponentially with the number of users. To tackle this, we propose a low complexity Whittle index based algorithm, which performs very close to the optimal. The complexity of the algorithm is linear in number of users and serves as a heuristic for both homogeneous and heterogeneous cases. 
\end{abstract}

\section{Introduction}
 
Time-sensitive information is essential in IoT-enabled systems for real-time monitoring applications, catering to end users such as weather forecasting agencies, agricultural landowners, and industrial plants~\cite{8481588,perera2018designing}. These agencies may require similar measurements  such as temperature and humidity. Also, they may not want to deploy their own sensors and monitoring systems over several geographical regions due to economic and logistics reasons.  To this end, {\em Sensing-as-a-Service} (S2aaS) offers an efficient way for accessing sensory data without owning any physical infrastructure.  In a S2aaS system, an aggregator, called  {\em sensor-cloud service provider} (SCSP), acquires
 measurements from multiple sensors and delivers these to the users on demand.  S2aaS is expected to play a significant role in the upcoming 6G networks for real-time monitoring~\cite{9945983,10387517}.
  
\remove{
In these systems, IoT sensors continuously collect physical attributes in  designated areas and update the contents accordingly. These updates are transmitted to designated users via a \textit{sensor-cloud service provider} (SCSP), which maintains a local cache that stores the latest content updates, replacing outdated ones.  Ensuring content freshness is crucial, as the recent content better reflects the current state of the monitored environment. This timeliness is particularly crucial in applications where control actions depend on the immediacy and accuracy of the information, such as steering autonomous vehicles or remotely controlling actuators in industrial automation. }

SCSP pays sensor owners for getting real-time contents  and also incurs communication costs in fetching and delivering the contents. On the other hand, it charges the users based on their quality of service~(QoS).  Preferred metric of QoS in this context is version age of information~(VAoI) at the usres~\cite{9517796}.  The VAoI measures how many times the content at a sensor has changed since it was last fetched, without needing detailed timing or system knowledge. This makes it a lightweight and practical metric for real-world applications, especially when only meaningful content changes matter, and not just how old the information is. The SCSP aims to maximize its net profit by intelligently balancing the cost of fetching content from sensors with the revenue generated from serving user requests. By leveraging VAoI, the SCSP can assess the number of content changes since the last fetch, allowing it to prioritize updates that carry meaningful information. This enables more strategic decisions about when to fetch and deliver content, helping reduce unnecessary transmissions while ensuring that users receive timely and valuable data.

\remove{
\par {\bf SCSP revenue model}: On one hand the SCSP compensates sensor owners for acquiring real-time measurements. On the other hand, SCSP generates revenue by delivering the data to requested users in accordance with a Quality of Service (QoS) requirement. To this end, SCSP also pays a transmission cost to the network service providers. Furthermore, the SCSP can also cache the  fetched contents in order to serve future requests (especially when fetching new content is deemed costly). 

\par {\bf Content freshness as QoS}: The QoS, in our context, is defined by the freshness of the information namely, version age of information (VAoI)~\cite{9517796}. The VAoI of content captures the difference in the number of updates at the sensor, since the content was fetched by the receiver namely, the SCSP or the end user. It is to be noted that VAoI overcomes a fundamental limitation of AoI metric~\cite{YatesSurvey} which linearly increases with time even when sensor content is not updated. 
}

In S2aaS systems, delivering timely and relevant data to a large number of users is essential to ensure scalability and responsiveness. However, traditional unicast delivery becomes inefficient when multiple users request similar or overlapping information, resulting in redundant transmissions and higher communication costs. Broadcast delivery offers an alternative by transmitting the sensor data to multiple users simultaneously, significantly reducing network load and improving energy efficiency.  Broadcasting sensor data are suitable for local or area-level services, where users within the same region frequently require similar types of data, such as traffic updates, air quality readings, or environmental conditions~\cite{area_service_24}.

\remove{

{\color{blue}
In S2aaS systems, the fetching and delivery of the content also depends upon the value of information~(VoI) of the content. The VoI is defined as an assessment of the utility of an information 
when used in a specific usage context~\cite{voi_19,8606228}, and VoI of the content can vary over time due to several factors. Consider S2aaS system for environmental monitoring, where various sensor owners send real-time air quality measurements to the SCSP. The SCSP fetches a set of air quality readings from this system. At the time of fetching the value of the data may be low if air quality is stable. The user may not need the information immediately. Later, due to environmental change, for example, if a wildfire starts nearby or pollution levels spike unexpectedly, the same previously fetched data becomes highly valuable. Furthermore, the SCSP can be prompted to fetch fresh data.  This demonstrates that in S2aaS systems, the value of sensed content is not static but can vary significantly over time depending on external factors such as environmental events, policy changes, or new user needs~\cite{area_service_24}.}

}
We consider a S2aaS system where the SCSP sources contents from a sensor and  multiple users are associated with it via a broadcast medium. Both, the content updates and the users' requests, are modeled as stochastic processes. Upon receiving requests, the SCSP decides whether to fetch a fresh content from the sensor which incur fetching costs. After fetching the content the SCSP broadcasts the fresh version to every users. If the SCSP does not deliver the latest version of the content, it also incurs an age cost that is an increasing function of the VAoI at the users. We aim to design a content fetching and broadcast policy to minimize the time averaged cost of the SCSP. 

\subsection{Related work}

Mondal et al. investigate cache orchestration in S2aaS deployments over vehicular networks~\cite{9773312}, and Almasaeid studies S2aaS from the perspective of spectrum-sensing efficiency~\cite{ALMASAEID2025103716}. Although these works are relevant to S2aaS, they ignore QoS measures based on content freshness or accuracy, and also do not consider stochastic request processes. In the remainder of this section, we therefore focus on prior work that directly addresses freshness-driven optimization in closely related caching and sensing models.

Delfani et al. \cite{delfaniversion} investigate VAoI minimization 
considers an aggregator (SCSP in our model) and a set of users connected in a ring topology. Only one user can request the aggregator for the content at a time. The aggregator is aware of the updates at the sensor. It can either serve the cached version or can fetch and serve a fresh version. The requesting user can share the received content with one of its neighbours.

Abolhassani et al.\cite{Yehinfocom} study average fetching and age cost minimization subject to cache capacity constraints. Koley et al. \cite{11202652} frame the same problem as in~\cite{Yehinfocom} as a restless multi armed bandits problem and propose a Whittle index based policy.   Our cost structure is similar to~\cite{Yehinfocom}. However, we consider a fixed set of users. Also, when the SCSP fetches a fresh content, the VAoIs of all the users are impacted. So, the overall problem cannot be decoupled into per user problems, and cannot be framed as a restless multi armed bandits problem. 

Krishna et al. \cite{Anu_EAI_valuetools} address version age aware caching and transmission in a multi-user system, but they do not consider a broadcast channel between the cache and the users.

Abolhassani et al.~\cite{AchieveFreshness} consider a broadcast network with a homogeneous set of users, no SCSP, multiple contents and a unit size cache at each user. There, each user caches a fixed content. The set of contents for caching are determined via solving a static optimization problem. The cached contents are served on request irrespective of their version, which is suboptimal.

A more recent work~\cite{Karevvanavar} considers a BS which communicates to the users over a fading wireless channel using NOMA. The BS has the most updated version of the content all the time, so no fetching decisions have to be made. The authors focus on the BS’s transmission strategy subject to the average power consumption constraint.

On the other hand, we introduce a controller (SCSP) between the users and the sensor, which decides when to fetch and broadcast the content, thereby creating a more flexible and general framework. 

\remove{
where users are treated as distributed caches, with the goal of updating the cache while minimizing both age and fetching costs. In this setup, users directly fetch data from the server, and at any given time, the number of requests cannot exceed one. In contrast, our approach imposes no restrictions on the number of requests. Instead, we introduce a controller (SCSP) positioned between the users and the sensor, which decides when to fetch data, and if fetched, whether to transmit it, thereby creating a more flexible and general framework. Additionally, we extend our study to scenarios where the age cost is an increasing function of version age.}
\remove {\par \textcolor{red}{[Non-SCSP model, may we remove this work as that is not relevant to our context?] Abolhassani et al.~\cite{AchieveFreshness} investigate a problem where users are treated as distributed caches, with the goal of updating the cache while minimizing both age and fetching costs. In this setup, users directly fetch data from the server, and at any given time, the number of requests cannot exceed one. In contrast, our approach imposes no restrictions on the number of requests. Instead, we introduce a controller (SCSP) positioned between the users and the sensor, which decides when to fetch data, and if fetched, whether to transmit it, thereby creating a more flexible and general framework. Additionally, we extend our study to scenarios where the age cost is an increasing function of version age.}}

\remove{
\par {\bf Novelty}: In this work, we propose a novel multi-user broadcast-based SCSP system that is capable to accept and serve multiple user request with stochastically updated sensory data.  The objective of the aforementioned service model is to a VAoI-aware content fetching and broadcasting service by minimizing overall average costs involving version age, content fetching, and transmission. 
}

\remove{
\par {\bf contributions}:The aforementioned problem falls under a class of average-cost Markov decision process (MDP). Obtaining an optimal solution to this problem is computationally intractable~\cite{Papadimitriou}. Hence, we propose a heuristic that performs close to the optimal policy.  
}
\remove{  and Given the increasing demand for these applications, fetching the fresh content directly from sensors upon every request can be costly and may worsen backhaul congestion, leading to higher latency. An effective approach to mitigate congestion and latency is caching content closer to the user at the SCSP's local cache rather than directly fetching from the sensors. On the other hand, the SCSP must update the contents in its cache to avoid serving stale or less useful content to the users. The SCSP gets the information about the content updates when it fetches from the sensor. Moreover, the requests from the users are dynamic in nature, and there can be multiple requests for a single content. Depending upon these dynamics, the SCSP follows an efficient approach to deliver the contents to the users when a request arrives. 
 }
\remove{However, in such scenarios, contents are updated at various rates, which renders the older versions of the contents less useful~\cite{AchieveFreshness}. Since the contents are updated dynamically, traditional caching approaches are now almost inefficient. Serving users with the latest version of content is crucial, as contemporary content holds more significant value than older versions. However, constantly providing the most up-to-date content from the back-end sensor can congest the network, compromising the Quality of Experience (QoE). Consequently, there is a pressing need to develop appropriate content delivery strategies that consider content refresh characteristics and age costs.}
\remove{
Age of Information (AoI) is a traditional performance metric to capture the freshness of contents \cite{Yeh}.
However, AoI of content increases with time, even if the contents are updated. Therefore, a new age metric called  Version Age of Information (VAoI)  has been introduced,  which measures how many versions out-of-date the content at the SCSP is, compared to the version at the sensor~\cite{Yates}.}
\remove{
\color{black}
 We assume fixed fetching and transmission cost, and age cost which is an increasing function of version age-of-Information (VAoI) for each content. The SCSP can potentially choose one of the three actions: fetch and transmit, fetch and hold, or do not transmit. The contents are requested according to the Bernoulli process.
When the SCSP broadcast the content, it incurs a cost for fetching and transmission, and all the users update their contents. Consequently, the VAoI of the content of any user is same as that of the content at the SCSP. If the SCSP do not fetch and transmit, an age cost proportional to the number of requests is incurred. If the SCSP fetch and hold the content, an age cost proportional to the number of requests and a fetching cost are incurred.
 We pose this version age aware problem as an average cost Markov decision process~(MDP) problem. Obtaining an optimal solution to this problem is computationally intractable~\cite{Papadimitriou}. We outline our approach here.
 We derive the optimal policy to the version age oblivious variant of this problem. We develop
 an algorithm to compute the optimal policy to the version age oblivious broadcast problem.
 Using the policy from this algorithm, we propose a heuristic approach to solve the version age aware broadcast problem. We also study another fictitious problem to validate the performance of the proposed heuristic solution.
}
\remove{\subsection{Related work}}
\remove{Since traditional metrics such as throughput and delay cannot capture timely status updates of contents, researchers have adopted a new age metric called Age of Information (AoI). 
Numerous works have investigated the optimization of content freshness in energy-constrained IoT networks \cite{PappasSensor,GunduzSensor,BacinogluSensor,BaranSensor,ArafaSensor}. The realization that the objective of timely status updating is not captured by metrics such as throughput and delay has led to the introduction of a new metric called the Age of Information (AoI).
Age of Information has been widely investigated as a metric of freshness for last few decades~\cite{YatesSurvey}. AoI measures the time elapsed since the most recent version from the sensor has been fetched by the SCSP. This age metric increases linearly in time even if the sensor has not received any updated content and the SCSP has the most up-to-date content. 
To overcome this fundamental limitation of AoI metric, a new age metric called  Version Age of Information (VAoI)  has been introduced,  which measures how many versions out-of-date the content at the SCSP is, compared to the version at the sensor~\cite{9517796,Yehinfocom}.}
\remove{While AoI is a meaningful metric for measuring the freshness of content in some systems, there are many real-world scenarios where content does not lose its value simply because time has passed since it was put into the cache. A new freshness metric called Version Age of Information (VAoI), which counts the integer difference between the versions at the database and the local cache, is introduced in \cite{Yehinfocom}.}

\subsection{Our Contributions}
Following is a preview of our contributions.
\begin{enumerate}
\item We formulate a content fetching and  broadcast problem to minimize the average content fetching and age costs in an S2aaS system.

\item We show that the optimal policy is of threshold type. We develop a recursive algorithm that provides the thresholds values of the optimal policy. We show that our proposed algorithm requires $O(N)$ computations in each step for the homogeneous case. 

\item The proposed algorithm requires $O(2^N)$ computations in each step for the heterogeneous case. To tackle this problem, we propose a low complexity Whittle index based policy that is applicable for both homogeneous and heterogeneous cases and requires only $O(N)$ computations. We show that the algorithm performs very close to the optimal policy. We also show that the policy is asymptotically optimal via simulations. 

\end{enumerate}
\remove{

consider the slotted system of wireless network consisting of  $N$ users that are connected to a Base Station (BS) which in turn is connected
to a content server via the core network. Whenever a user needs a content, it request to the BS. BS either delivers fresh content or deliver cached content or it remains idle. Caching decisions are taken by BS at
the slot boundaries. We cast the problem as an MDP whose state space is of $N+1$ dimensions. The users are coupled with parameter $\alpha_i, i \in \{1,2,..,N\}$ which is the probability that the user $i$ gets a content for free. We solve the optimization problem for each user and then provides an algorithm which computes $\alpha_i, i \in \{1,2,..,N\}$ in Section \ref{Section: Multi user}
}
\remove{
\section{System Model and Problem Formulation}
\input{Known age_multiple users}
}
\remove{
\section{Single User Problem }
\label{Section: Single user zero alpha}
}

\remove{   {\color{red}We first consider a single user problem with $\alpha_i=0$. The aggregator takes the decision not to transmit any content or fetch and transmit the fresh content upon receiving a content request from the user.}}
\remove{ Let $q$ be the probability of content being requested at slot $t$. Since there is only one user, the aggregator faces a binary decision: either fetch the
content and transmit if the user requests the content
or does not transmit. Hence, the aggregator and the user have the same version, i.e., $d_1(t)=0 \,\,\forall t.$ Therefore, we consider $s(t)=\tau(t)$, i.e.,  the age of the content at the aggregator at the beginning of time slot $t$. 

A policy $\pi$ is a sequence of mappings $\{u^\pi_t, t = 1,2,\cdots\}$ where $u^\pi_t: \{0,1,2,..\} \to \{0,1\}$. $u^\pi({\tau}(t))=1$, if the aggregator fetches and transmits the cached content if the content is requested at slot $t$ and $u^\pi({\tau}(t))=0$, otherwise.  
\remove{\[
    \begin{cases}
     1, \text{ }\\
      0, \text{otherwise} 
    \end{cases}\] }
     The cost at slot $t$ is:
    \[C^{\pi}(\tau(t))=C_{a}pq\tau(t)(1-u^\pi(\tau(t)))+(C_{f}+C_{tx})qu^\pi(\tau(t))\]
$C_{a}pq\tau(t)$ is the average age cost incurred when a content of version age $p\tau(t)$ is delivered to the user on being requested. $q(C_{f}+C_{tx})$ is the average cost incurred in fetching fresh content and transmitting it on being requested.
We solve the following optimization problem,
\begin{align}
\text{Minimize } & \ \lim_{T \to \infty}\frac{1}{T}\mathbb{E}\left[\sum_{t=0}^T C^{\pi}(\tau(t))\Bigg\vert \tau(0) =\tau\right] \label{eqn:objective}
\end{align}

The above optimization  problem ~\eqref{eqn:objective} can be formulated as
an MDP consisting of a tuple
$<\mathcal{S, A}, P, C>$, where $\mathcal{S}$ is the state space, $\mathcal{A}$ is the set of
actions, $P$ is the state transition probability and $C$
is the cost of MDP.
\paragraph*{State space} State space is $\mathcal{S}$. State $s \in \mathcal{S}$ is the age of content at the aggregator. {\color{black}$\mathcal{S}=\{1, 2, 3, ....\}$.}

\paragraph*{Action space} Action space is $\mathcal{A}$. Action $a \in \mathcal{A}$ is the action taken by the aggregator. $\mathcal{A}=\{0, 1\}$. $a=1$ implies fetch and serve the fresh version of content from the sensor; $a=0$ implies do not serve the content.

\paragraph*{Transition probability} 
Let $r$ be the random variable which denotes the request. $r=1$ w.p. $q$ and $r=0$ w.p $1-q$. $P_{\tau\tau'}(a)=\sum_{r}P_{\tau\tau'}(a,r)P(r)$.

In the absence of an external request, i.e., when $r=0$, state evolution is independent of the action taken. 
When $r=0$, if $\tau(t)=\tau$, then $\tau(t+1)=\tau+1$.
\begin{equation*}
P_{\tau\tau'}(a=0, r=0) = \begin{cases}
               1, & \text{ if } \tau'=\tau+1 \\
               0, & \text{ otherwise.}
                \end{cases}
\end{equation*}
\begin{equation*}
P_{\tau\tau'}(a=1, r=1) = \begin{cases}
                1, & \text{ if } \tau'=1 \\
               0, & \text{ otherwise.}
                \end{cases}
\end{equation*}
Therefore,
\begin{equation*}
P_{\tau\tau'}(a=0) = \begin{cases}
                1, & \text{ if } \tau'=\tau+1 \\
                0, & \text{ otherwise.}
                \end{cases}
\end{equation*}

\begin{equation*}
P_{\tau\tau'}(a=1) = \begin{cases}
                q, & \text{ if } \tau'=1 \\
               1-q, & \text{ if } \tau'=\tau+1 \\
               0, & \text{ otherwise.}
                \end{cases}
\end{equation*}
\remove{
\color{blue}
Suppose if $\tau(t)=\tau$.
If the aggregator chooses an action $a=0$, then $\tau(t+1)=\tau+1$.
Therefore,
\begin{equation*}
P_{\tau\tau'}(a=0) = \begin{cases}
                1, & \text{ if } \tau'=\tau+1 \\
                0, & \text{ otherwise.}
                \end{cases}
\end{equation*}
In the absence of a request from the user, state evolution is independent of the action taken. 
Therefore, if $\tau(t)=\tau$, then $\tau(t+1)=\tau+1$. However, if the aggregator chooses an action $a=1$ upon receiving a request from the user, then $\tau(t+1)=1$. Therefore,
\begin{equation*}
P_{\tau\tau'}(a=1) = \begin{cases}
                q, & \text{ if } \tau'=1 \\
               1-q, & \text{ if } \tau'=\tau+1 \\
               0, & \text{ otherwise.}
                \end{cases}
\end{equation*}

\color{black}}
\paragraph*{Cost} The single stage cost $C(\tau,a)=qC_{a}\tau p(1-a)+q(C_{f}+C_{tx})a$. 

We introduce average cost MDP for Problem~\eqref{eqn:objective}.
The optimal average cost $\theta$ achieved by optimal policy $\pi^{\ast}$ is the same for all initial states, satisfying the Bellman equation
\begin{align}
h(\tau)+\theta=\min_{a \in \{0,1\}}\{C(\tau,a)+\sum_{\tau^{'}}P_{\tau \tau^{'}}(a)h(\tau^{'})\}\label{eqn:bellman}.
\end{align}
$h(\tau)$ is the differential cost function of the MDP problem. Please see Chapter 4 of \cite{BertsekasVol2} for more details.
\begin{lemma}
\label{h_tau}
    $h(\tau)$ is non decreasing in $\tau$.
\end{lemma}
\begin{IEEEproof}
Please see \cite[Appendix~A]{AnuK:2024}.
\end{IEEEproof}
\begin{theorem}
\label{Theorem1}
The optimal policy is a threshold policy such that 
\begin{equation*}
    u^{\pi^{\ast}}(\tau) = \begin{cases}
                1, \text{ if } \tau\geq \tau_{opt}^{\ast} \\
                0, \text{ otherwise.}
                \end{cases} 
\end{equation*}
with $\tau_{opt}^{\ast}=\min_{\tau^{\ast}\in \mathcal{Z_+}}{\frac{C_{a}pq^{2}(\tau^{\ast}-1)\tau^{\ast}+2(C_{f}+C_{tx})q}{2(q(\tau^{\ast}-1)+1)}}$
\end{theorem}
\begin{IEEEproof}
See Appendix~\ref{Appendix:Theorem1}.
\end{IEEEproof}
We now consider a model with $N>1$ users. 
 }
\remove{Contents arrive at the sensor, stay for a random time and expire afterwards. Contents that are alive are updated at the sensor. The sensor communicates the updation to the aggregator.
Aggregator has to decide whether to
request a fresh update from the sensor or serve the external request
with cached content that is stale. 

\paragraph*{Content Dynamics} Content updation
follows a Bernoulli distribution, 
 probability $0 <p< 1$. That is, in each time slot,  with probability $p$, content is updated.

\paragraph*{Content Requests} We assume that at each time
slot, at most, one request from the user will
be served by the aggregator. The service probability of external requests
follows a Bernoulli distribution, 
where the external request from user is
served with probability $0 < q < 1$, and no external request
is served with probability $1-q$. Every time the aggregator receives a fresh content, the new
content is stored, and the previously cached one is evicted. 

\paragraph*{Content caching} 
The aggregator has a {\it cache} where it can store contents. Fresh contents can be cached at slot boundaries while evicting other stale contents to meet the cache capacity constraints. 

\paragraph*{Fetching and Age Costs}
We introduce the key operational and performance costs associated with our caching system. On the operational side, we denote the cost of fetching an item from the sensor to the aggregator by $C_f > 0$. On the performance side, we assume that serving a content from the local cache with an average version age $x$ incurs an age cost of $C_{a}x$ for some $C_a \geq 0$, which grows linearly with the average version age $x$. This ageing cost measures the growing discontent of the user for receiving an older version of the content.
Our objective is to find an optimal policy at the aggregator that minimizes the long run time average cost of serving content. We denote, as $\tau(t)$.} 
\section{System Model}\label{Section:sys_model}

We consider a S2aaS system consisting of a sensor connected to a SCSP serving a population of $N$ users. The SCSP is connected to the users via a broadcast medium~(e.g., a wireless edge)~\cite{AchieveFreshness}. The sensor data, referred to as the content, is updated when the sensor captures a new measurement. The users obtain the content through the SCSP which occasionally fetches and caches the content. We assume a discrete time system in which the users' requests arrive at the slot boundaries. A subset of users request for the content at each time. On receiving requests, the SCSP may or may not fetch the content from the sensor. If the SCSP fetches the content, it  broadcasts the content. We consider a reliable broadcast channel. This assumption is justified by employing reliability-enhancing mechanisms such as automatic repeat request or forward error correction, where packets are transmitted with sufficient redundancy to ensure reliable delivery. Such techniques have been shown to achieve near-perfect reliability in wireless broadcast systems~\cite{reliable_broadcast}.  All the users always possess the last broadcast version of the content. So, if the SCSP does not fetch an updated version from the sensor, it also need not broadcast its cached version. The SCSP takes fetching and broadcast decisions in order to optimize certain time averaged expected costs. We now formally describe the content update and request processes and the costs pertaining to our S2aaS system.

\paragraph*{Content update dynamics}  We assume the content updates in different slots to be independent and identically distributed~( i.i.d.) Bernoulli$(p)$ random variables. So, in each slot,  the content is updated once at the sensor with probability $p$ and is not updated with probability $1-p$.

\paragraph*{Content requests} We assume that all the users have i.i.d. content request processes. For each user, its requests at different times are i.i.d. Bernoulli random variables. At each time,  user $i$ requests with probability $q_i$, $i\in[N]$. Let us define $s_i(t)$ such that, $s_i(t)=1$, if there is a request from user $i$ and $s_i(t)=0$, otherwise.

\paragraph*{Age dynamics}
Age refers to the staleness of the content at the SCSP and the users. Let $\tau(t)$ be the time elapsed since the SCSP last fetched the content until time $t$; $\tau(t) \geq 1$ for all $t$. Let $V(t)$ be the number of content updates in  the interval $\{t-\tau(t)+1,\cdots,t\}$. We call $V(t)$ the VAoI  at the SCSP at time $t$.  Note that $V(t)$ has Binomial ($\tau(t),p$) distribution.

The VAoI at the users refers to the number of content updates since the last version that was broadcast to the users. Since the  SCSP always broadcast the content after fetching, the VAoI of the users and SCSP are same.

\paragraph*{Costs} Following are the three different costs  incurred in a S2aaS system:
\begin{enumerate}
    \item Content fetching  cost: If the SCSP fetches the content from the sensor, it broadcast the content incurring a fetching cost $C_f$. It includes the communication cost and the economic cost of acquiring the content.

    \item Age Cost: When a user that requests the content does not receive the latest version, an age cost depending on the VAoI at the user $V(t)$ is incurred. Let $C_{a_i}(V(t))$ be this age cost for user $i$ where $C_{a_i}(\cdot)$ is a convex, increasing function. So, the total age cost at time $t$ equals $0$ if the SCSP broadcasts the latest version and  $\sum_{i=1}^N s_iC_{a_i}\!\left(V(t)\right)$ otherwise. Note that  $\mathbb{E}\left[C_{a_i}(V(t))\right]=C_{av_i}(\tau(t))$
\end{enumerate}

\paragraph*{Actions} The SCSP does not know the VAoI $V(t)$  unless it fetches the content from the sensor at time $t$. However, it knows $\tau(t)$ and the content update dynamics. 
 Let $S(t)=(\tau(t),s_i(t),i\in[N])$ be the state at time $t$. Let us define $\bs(t)=(s_i(t),i\in[N])$ and hence, $S(t)=\left(\tau(t),\bs(t)\right)$.

\remove{Let $a_f(t)$ and $a_{tx}(t)$ denote the fetching and broadcast decisions, respectively, at time $t$.  Both take  values in $\{0,1\}$; 
\begin{equation*}
a_f(t) =
\begin{cases}
     1 \text{ if  the SCSP fetches the content at $t$,}\\
     0 \text{ if  the SCSP does not fetch the content}\\
\end{cases}
 \end{equation*}
$a_{tx}(t)$ also has similar connotation with respect to the broadcast decision. As discussed earlier, $(a_f(t),a_{tx}(t)) 
\in \{(0,0),(1,0),(1,1)\}$ are the only reasonable action pairs.

\paragraph*{Objective}
The cost incurred at time $t$ is
\[C(t)=\begin{cases}
    m(t)C_a(V(t)+d(t)) \text{ if } a_f(t)=0,\\
    C_{tx}+m(t)C_a(V(t))\text{ if } a_f(t)=0, a_{tx}= 1,\\
    C_f+m(t)C_a(V(t)+d(t)) \text{ if } a_f(t)=1, a_{tx}= 0, \\
    C_f+C_{tx} \text{ if } a_f(t)= a_{tx}(t)=1. 
\end{cases}
\]
We aim to minimize the time averaged expected cost, i.e., to solve
\begin{align}
\min & \lim_{T \to \infty}\frac{1}{T}\mathbb{E}\left[\sum_{t=1}^{T}C(t)\right].\label{eq:original_problem}
\end{align}
We can pose this problem as an average cost MDP with $(\tau(t),d(t),m(t))$ and $(a_f(t),a_{tx}(t))$ as state and action, respectively, at time $t$. It has a three-dimensional state space 
$\mathbb{Z}_{++} \times \mathbb{Z}_+ \times 
[N]$ rendering computation of its optimal policy intractable. We formally describe this MDP in Section~\ref{section: VAoIBroadcast}. Below we introduce a simpler problem wherein whenever the SCSP fetches the content it also transmit the fetched content. We refer to this problem as {\it VAoI oblivious transmission}. Its solution yields a heuristic for the original problem. }
\remove{\begin{table}[]
\label{tab:notation}
 \caption{Notations}
    \centering
    \renewcommand{\arraystretch}{1.3}
    \begin{tabular}{|c|c|}
    \hline
        $C_f$ & fetching cost \\
         \hline
         $C_{a_i}(\cdot)$ & Age cost of user $i$ as a function of VAoI \\
         \hline
        $ \bar{C}_{av_i}(.)$&$\mathbb{E}[C_{a_i}(\cdot)]$\\
        \hline
        $\bar{h}_{av}$ &  relative cost function for homogeneous users MDP \\
         \hline
         $\bar{\theta}$ & optimal costs for homogeneous users MDP\\
         \hline
    \end{tabular}
\end{table}
}
 Let $\abar(t) \in \{0,1\}$ denote the SCSP's action at time $t$, defined as follows.
\begin{equation*}
\abar(t)=
    \begin{cases}
     1 \text{ if the SCSP fetches and transmits at time $t$,}\\
      0 \text{ if the SCSP does not fetch and transmit.}  
    \end{cases}
\end{equation*}
 Let $\bar{C}(\tau,\bs,\abar)$ be the expected single stage cost for given state $(\tau,\bs)$ for action $\bar{a}$. 
\begin{align}\bar{C}(\tau,\bs,\bar{a})=C_f\abar+(1-\abar)\sum_{i=1}^N s_i\bar{C}_{av,i}(\tau)\label{eq:cost_coup_MDP}\end{align}
\remove{The cost incurred at time $t$ is
\[\bar{C}(t)=\begin{cases}
    m(t)C_a(V(t)) \text{ if } \abar(t) =0,\\
    C_f \text{ if } \abar(t) =1. 
\end{cases}
\]
}
We aim to solve 
\begin{align}
\min & \lim_{T \to \infty}\frac{1}{T}\mathbb{E}\left[\sum_{t=1}^{T}\bar{C}(\tau(t),\bs(t),\bar{a}(t))\right].\label{eq:min_cost_problem}
\end{align}
We can pose this problem as an average cost MDP with $(\tau(t),\bs(t))$ and $(\abar(t))$ as state and action, respectively, at time $t$. It has a $N+1$-dimensional state space 
$\mathbb{Z}_{++} \times \{0,1\}^N$, rendering the computation of its optimal policy intractable.
In the following section, we discuss the optimal policy. 
\section{Optimal Policy} 
\label{sec:opt_policy}
We first discuss the optimal policy for the homogeneous case. In this case, the request probability $q_i=q$ and the cost function $C_{av_i}(\cdot)=C_{av}(\cdot)$ for all users $i$. Let $m(t)\in [N]$ be the total number of requests at time $t$ where $[N] \coloneqq \{0, 1,\cdots,N\}$. If the SCSP broadcasts the content then the average age cost at time $t$ becomes $\sum_is_i(t)C_{av_i}(t)=m(t)C_{av}(t)$.  Hence, the  decision regarding fetching the content at $t$ depends on $\tau(t)$ and $m(t)$.
Here we pose the content fetching and broadcast problem  as  an average cost MDP and provide an algorithm to obtain the optimal policy.  $(\tau(t),m(t)) \in \mathbb{Z}_{++} \times 
[N]$ and $\abar(t) \in \{0,1\}$ represent the state and the action, respectively, at time $t$. Recall that, given $(\tau(t),m(t)) = (\tau,m)$, the VAoI at the SCSP and the user $V \sim \text{Binomial}(\tau,p)$. With a slight abuse of notation, we use $\bar{C}(\tau,m,a)$ to denote the expected single stage cost. It is given by 
\begin{equation*} 
     \bar{C}(\tau,m,\bar{a}) = \begin{cases} m\cbar(\tau) \text{ if } \abar=0,\\
     C_f \text{ if } \abar = 1.  
    \end{cases}
\end{equation*}
 Note that $\cbar(.)$ is also a convex, increasing function. Furthermore, given $(\tau(t),m(t)) = (\tau,m)$, the state at time $t+1$ is given by
\begin{equation*}
  (\tau(t+1),m(t+1)) =
    \begin{cases}
      (1,M) \text{ if } \abar(t)=1, \\
      (\tau+1,M) \text{ if } \abar(t)=0.
    \end{cases}
\end{equation*}
where $M \sim \text{Binomial}(N,q)$.

Following~\cite[Chapter~4, Proposition~2.1]{BertsekasVol2}, the optimal average cost  of the MDP is independent of the initial state. Let $\thetabar$ and $\hbarr: \mathbb{Z}_{++} \times 
[N] \to \mathbb{R}$ be the optimal average cost and the relative cost function, respectively, of the  MDP. These satisfy the following Bellman equations~\cite[Chapter~4]{BertsekasVol2}.
\begin{align}
\hbarr(\tau,m)+\thetabar{=}\min\left\{C_f+\sum_{k=0}^{N}B(N,q,k)\hbarr(1,k),\right.\nonumber&\\
\left.m\cbar(\tau) {+}\sum_{k=0}^{N}B(N,q,k)\hbarr(\tau{+}1,k)\right\}&\label{eqn:bellmanAgg}
\end{align}
where 
\[B(N,q,k)\coloneqq\binom{N}{k}q^k(1-q)^{N-k} \text{ for all } k \in [N].\] 
Defining $\hbarav: \mathbb{Z}_{++} \to \mathbb{R}$ as
\begin{equation}
    \hbarav(\tau) \coloneqq \sum_{k=0}^{N}B(N,q,k)\hbarr(\tau,k), \label{eq:h_bar_tau}
\end{equation}
we can rewrite~\eqref{eqn:bellmanAgg} as 
\begin{align}
\hbarr(\tau,m){+}\thetabar{=}\min&\left\{C_f{+}\hbarav(1), m\cbar(\tau) {+}\hbarav(\tau+1)\right\}.\label{eq:bellman_hbar}
\end{align}
Let $\bar{\pi}$ be the optimal policy of the MDP; it is also characterized by the above equation. More precisely, for any  $(\tau,m) \in \mathbb{Z}_{++} \times [N]$,  $\bar{\pi}(\tau,m) = 1$ if the first term in the right hand side is less than or equal to the second term and $\bar{\pi}(\tau,m) = 0$ otherwise.

The following lemma shows monotonicity of $\hbarav(\tau)$. 
\begin{lemma}
\label{lemmaAgg}
$\hbarav(\tau)$ in non-decreasing in $\tau$.
\end{lemma}
\begin{IEEEproof}
For any given $k$, we can show using induction and relative value iteration that $\hbarr(\tau,k)$ is non-decreasing in $\tau$~\cite{bertsekas2011dynamic}. Since $\hbarav(\tau)$ is a convex combination of $\hbarr(\tau,k), k \in [N]$, it is also non-decreasing in $\tau$. 
\end{IEEEproof}

The following theorem characterizes the optimal policy $\bar{\pi}$. The optimal policy is to never fetch if the number of requests is $0$. If there is at least one request then it is optimal to fetch the content if $\tau$ exceeds some threshold value which is a function of the number of requests, $m$. The threshold value is non increasing in $m$. 
\begin{theorem}
\label{TheoremAgg}
1. $\bar{\pi}(\tau,0) =0, \forall \tau \geq 1 $.\\
2. For all $m \in \{1,\cdots,N\}$, 
\[\bar{\pi}(\tau,m) = \begin{cases}
                1, \text{ if } \tau\geq \taustar(m), \\
                0, \text{ otherwise}
                \end{cases} \]
where 
\[\taustar(m) \coloneqq \min\{\tau: \hbarav(\tau+1)\geq C_f+\hbarav(1)-m\cbar(\tau)\}.\]
3. $\taustar(m-1)\geq\taustar(m)$ for all $m \in \{1,\cdots,N\}$.
\end{theorem}
\begin{IEEEproof}
See Appendix~\ref{Appendix:TheoremAgg}.
\end{IEEEproof}

Obtaining the optimal cost and the optimal policy of the MDP using conventional relative value iteration is not viable due to the countable state space. One naive approach is to truncate the state space and to perform relative value iteration. But this would yield only an approximately optimal policy and cost. Below, we develop an algorithm to compute the optimal policy and the optimal cost $\thetabar$ without truncating the state space. Its computational complexity is also less than relative value iteration.    

\subsection {Computation of $\taustar(m)$s}\label{sec:algo}
From Theorem~\ref{TheoremAgg} and~\eqref{eq:bellman_hbar},
\begin{equation}
\label{eqn:htau0}
\hbarr(\tau,0)=\hbarav(\tau+1)-\thetabar \text{ for all } \tau \geq 1.
\end{equation}
Further, it is optimal to fetch and transmit the content in state $(\tau,m)$ if $\tau\geq\taustar(1)$ and $m \geq 1$. So,
\begin{equation}
     \hbarr(\tau,m)+\thetabar=C_f+\hbarav(1)  \text{ for all }  \tau\geq\taustar(1), m \geq 1.\label{eq:bellman_for_tau>taustar}
\end{equation}
From~\eqref{eq:h_bar_tau} we obtain the value of $\hbarav(\tau)$ for $\tau\geq\taustar(1)$. 
\begin{align}
    \hbarav(\tau) =&\sum_{k=0}^{N}B(N,q,k)\hbarr(\tau,k)\nonumber\\
    = &(1-q)^{N}(\hbarav(\tau+1)-\thetabar)\nonumber\\
    &+\sum_{k=1}^{N}B(N,q,k)(C_f+\hbarav(1)-\thetabar)\nonumber\\
    =&(1-q)^{N}\hbarav(\tau{+}1)\nonumber\\
    & +(1-(1-q)^{N})(C_f+\hbarav(1))-\thetabar \label{eqn: tau>taustar1}
\end{align}
where the second equality follows from~\eqref{eqn:htau0} and~\eqref{eq:bellman_for_tau>taustar}. 

The following lemma states that $\hbarav(\tau)$ remains constant for all $\tau \geq \taustar(1)$. We use it to obtain the value of $\taustar(1)$.
\begin{lemma}\label{lemma:cost_func_after_taustar}
   $\hbarav(\tau)=\hbarav(\tau+1)$ for $\tau\geq\taustar(1).$ 
\end{lemma}
\begin{IEEEproof}
    The proof involves using~\eqref{eqn: tau>taustar1} recursively for all $\tau\geq\taustar(1)$. We skip the details for brevity. \remove{See Appendix~\ref{Appendix:lemmaAgg2}.}
\end{IEEEproof}

This lemma suggests that,  given $\taustar(1)$, we can compute the optimal cost and the policy using Bellman's equations~\eqref{eq:bellman_hbar} for states $\{(\tau,m):\tau \leq \taustar(1) \text{ and } m\leq N\}$. Let us define \begin{equation}
    \gbar(\tau)=\hbarav(\tau)-\hbarav(1).  \label{eq:def_g_tau}
\end{equation}
We can rewrite~\eqref{eqn: tau>taustar1} using~\eqref{eq:def_g_tau} as
\begin{equation}
    \gbar(\tau)=C_f-\frac{\thetabar}{(1-(1-q)^{N})}. \label{eq:g_tau_fort=_tau_geq_taustar}
\end{equation}
Equation~\eqref{eq:g_tau_fort=_tau_geq_taustar} gives $\gbar(\tau)$ for all $\tau\geq\taustar(1)$. Using this we can derive $\taustar(1)$ in terms of $\thetabar$. 
From Theorem~\ref{TheoremAgg}, 
\begin{align}
    \taustar(1)&=\min\{\tau:C_f+\hbarav(1)<\cbar(\tau)+\hbarav(\tau+1)\}\nonumber\\
    &=\min\{\tau:C_f<\cbar(\tau)+\gbar(\tau+1)\}\nonumber\\   &=\min\left\{\tau:\cbar(\tau)>\frac{\thetabar}{(1-(1-q)^{N})}\right\}.\label{eq:taustar_1_min_func}
\end{align}
where the second equality follows from the definition of $\gbar(\tau)$ and the third from~\eqref{eq:g_tau_fort=_tau_geq_taustar}.
From~\eqref{eq:taustar_1_min_func},
\begin{equation}
   \taustar(1)= \left\lceil \cbar^{-1}\left(\frac{\thetabar}{ (1-(1-q)^{N})}\right)\right\rceil.\label{eq:taustar_1_in_theta}
\end{equation}

\paragraph*{Example} Let the age cost be quadratic in $V$, say $C_a(V)=c_aV^2$ for all $V \in \mathbb{Z}_+$. In this case, $\cbar(\tau)=  c_a(\tau p(1-p)+(\tau p)^2)$ and
    \[\taustar(1){=}\left\lceil\frac{(1{-}p)}{2p}\left(\sqrt{1+\frac{4\thetabar}{c_a(1-p)^2(1-(1-q)^N)}}-1\right) \right\rceil.\]

From Theorem~\ref{TheoremAgg}, \begin{equation}\taustar(m)=\min\{\tau:C_f<m\cbar(\tau)+\gbar(\tau+1)\} \label{eq:def_taustar_m}
\end{equation}
where we have used the definition of $\gbar(\tau)$ in~\eqref{eq:def_g_tau}. Let us consider a $\tau \in \{\taustar(m),\taustar(m-1)-1\}$. The optimal action in state $(\tau,k)$ is to fetch and transmit the content if $\tau\geq\taustar(m)$ and  $k \geq m$ for any $m \geq 0$. So, 
\begin{equation}
    \hbarr(\tau,k)+\thetabar = 
    \begin{cases} 
    C_f+\hbarav(1) \text{ if } k\geq m, \\
    k\cbar(\tau){+}\hbarav(\tau{+}1) \text{ if } k< m. 
    \end{cases}
    \label{eq:h_tau_k_geq_m}
\end{equation}
Hence, using~\eqref{eq:h_bar_tau},
\begingroup
\allowdisplaybreaks
\begin{align*}
   \hbarav(\tau) =&\sum_{k=0}^{N}B(N,q,k)\hbarr(\tau,k)\nonumber\\
     =&\sum_{k<m}B(N,q,k)(k\cbar(\tau)+\hbarav(\tau+1))\nonumber\\
    &+\sum_{k\geq m}B(N,q,k)(C_f+\hbarav(1))-\thetabar.
\end{align*}
\endgroup
Subtracting $\hbarav(1)$ from both the sides and  using the definition of $\gbar(\tau)$ in~\eqref{eq:def_g_tau} we get 
\begin{align}
    \gbar(\tau)=&\sum_{k<m}B(N,q,k)(k\cbar(\tau)+\gbar(\tau+1))\nonumber\\
    &+\sum_{k\geq m}B(N,q,k)(C_f)-\thetabar\nonumber\\
    =&\sum_{k<m}B(N,q,k)(k\cbar(\tau)+\gbar(\tau+1)-C_f)\nonumber\\
    &+C_f-\thetabar. \label{eqn:taustarm<tau<taustarm1}
\end{align}
The above equation provides a relation between $\gbar(\tau)$ and $\gbar(\tau+1)$ for $\tau\in \{\taustar(m),\taustar(m-1)-1\}$. Using this recursively for $m \geq 1$ we can express $\thetabar$ in terms of $\gbar(2)$. There can be two cases as follows.
\begin{enumerate}
    \item $\taustar(N)>1$: In this case, it is optimal not to transmit for $\tau=1$ irrespective of the number of requests, $m$. Hence, from~\eqref{eqn:taustarm<tau<taustarm1},
    \begin{align}
        \gbar(1)&=\sum_{k=0}^NB(N,q,k)(k\cbar(1)+\gbar(2))-\thetabar \nonumber\\
        &=Nq\cbar(1)+\gbar(2)-\thetabar.\label{eq:g_2_g_1_and_theta}
    \end{align}
    By the definition of $\gbar(\tau)$ in~\eqref{eq:def_g_tau}, $\gbar(1)=0$. Hence, from~\eqref{eq:g_2_g_1_and_theta}, 
    \begin{equation}
\thetabar=Nq\cbar(1)+\gbar(2).\label{eq:g_2_and_theta_first_case}
    \end{equation}
    \item  $\taustar(N)=1$: In this case, define $m = \min\{k \geq 1: \taustar(k)=1\}$.  Clearly, for $\tau=1$ it is optimal to fetch and transmit the content if and only if the number of requests is at least $m$. Hence, from~\eqref{eqn:taustarm<tau<taustarm1},
     \begin{align*} \gbar(1)=&\sum_{k<m}B(N,q,k)(k\cbar(1)+\gbar(2))\\
    &+\sum_{k\geq m}B(N,q,k)C_f-\thetabar \end{align*}
    and therefore, 
    \begin{align}
    \thetabar=&\sum_{k<m}B(N,q,k)(k\cbar(1)+\gbar(2))\nonumber\\&+\sum_{k\geq m}B(N,q,k)C_f. \label{eq:g_2_and_theta_second_case}
    \end{align}
\end{enumerate}
 In either case $\thetabar$ can be expressed in terms of $\gbar(2)$ using~\eqref{eq:g_2_and_theta_first_case} or~\eqref{eq:g_2_and_theta_second_case}. 
 As discussed earlier, $\gbar(\tau)$ can be expressed in terms of $\gbar(\tau+1)$ for all $\tau \in \{2,\cdots,\taustar(1)-1\}$. Also, $\gbar(\taustar(1))$ is a function of $\thetabar$ as in~\eqref{eq:g_tau_fort=_tau_geq_taustar}. We  thus have a fixed point equation in $\thetabar$, say $\thetabar = f(\thetabar)$, which can be used to obtain $\thetabar$ and also the thresholds $\taustar(1),\cdots\taustar(N)$. However, we cannot explicitly solve this fixed point equation. Therefore, we propose an iterative algorithm to solve it~(see~Algorithm~\ref{alg:fixed_point}).
 
 \begin{algorithm}[h]
\caption{Algorithm to compute $\taustar(1),\dots,\taustar(N)$}\label{alg:fixed_point}
\begin{algorithmic}[1]
\Require $ \theta^0, \epsilon, \alpha$
\Repeat 
\State $\theta \gets \theta^l$
\State $\taustar(1) \gets \left\lceil \cbar^{-1}\left(\frac{\theta}{ (1-(1-q)^{N})}\right)\right\rceil $ \label{taustar_in_terms_theta}
\State $\gbar(\taustar(1)) \gets C_f-\frac{\theta}{1-(1-q)^N}$ \label{g_taustar_in_terms_theta}
\State $m \gets 1$
\State $\tau\gets\taustar(1)$
\While{$\tau > 1$}
\While {$\gbar(\tau)\leq C_f-(m+1)\cbar(\tau-1) \text{ and }  m<N$ } \label{while_connd_to_inc_m}
    \State $m \gets m+1$
    \State $\taustar(m) \gets \tau$\label{update_taustar}
\EndWhile
\If {$\tau>2$}  
\State \lefteqn{\gbar(\tau-1) \gets C_f-\theta}\[ +\sum_{k\leq m}B(N,q,k)(k\cbar(\tau-1)-C_f+\gbar(\tau))\] \label{update_g_tau_minus_1}
\Else 
   \State $\taustar(m+1)=\dots=\taustar(N)=1$  
\EndIf
\State $\tau \gets \tau-1$ 
\EndWhile
\If {$\taustar(N)>1$}
    \State $f(\theta) \gets N\cbar(1)q+\gbar(2)$ \label{cond:tau_N>1} 
\Else
   \State $\,f(\theta) \gets \sum_{k\leq m}B(N,q,k)(k\cbar(1)+\gbar(2))$\[+\sum_{k> m}B(N,q,k)C_f\] \label{cond:tau_N=1}
  
\EndIf

\State $\theta^{l+1}\gets \alpha f(\theta)+(1-\alpha)\theta$ 
\State $l\gets l+1$
\Until {$|\theta^{l}-\theta^{l-1}|\leq \epsilon$} 
\end{algorithmic}
\end{algorithm}
 The computational complexity of Algorithm~\ref{alg:fixed_point} is linear in \( N \).
 For ease of understanding, we briefly describe the steps in Algorithm~\ref{alg:fixed_point} below.
\paragraph{Description of Algorithm~\ref{alg:fixed_point}}  
\begin{enumerate}
    \item Start with $l=0$ and an arbitrary $\theta^l$.
    \item \label{algo:taustar} Compute $\taustar(1)$ from~\eqref{eq:taustar_1_in_theta}. 
    \item \label{algo:g_taustar} Compute $\gbar(\taustar(1))$ from~\eqref{eq:g_tau_fort=_tau_geq_taustar}. 
    \item \label{algo:g_tau} Compute $\gbar(\tau)$ for $\tau \in \{2,\cdots,\taustar(1)-1\}$ from~\eqref{eqn:taustarm<tau<taustarm1}. For this we initialize $m=1$ and $\tau=\taustar(1)$. After obtaining $\taustar(m)$ for a $m \geq 1$ there can be two scenarios.

\begin{enumerate}
     \item \label{algo:cond_sat} $\gbar(\taustar(m))\leq C_f-(m+1)\cbar(\taustar(m)-1)$: This implies that if $\tau = \taustar(m)-1$ and the number of requests is $m+1$ then it is optimal not to fetch and transmit~(see the definition of $\taustar(m)$ in~\eqref{eq:def_taustar_m}). Hence   $\taustar(m+1)=\taustar(m)$.
     Similarly, $\taustar(k+1)=\taustar(m)$
     for all $k$ for which $\gbar(\taustar(m))\leq C_f-(k+1)\cbar(\taustar(m)-1)$. On hitting a $k$ such that $\gbar(\taustar(m)) > C_f-(k+1)\cbar(\taustar(m)-1)$
follow~\ref{algo:cond_nonsat}. 
     
     \item \label{algo:cond_nonsat} $\gbar(\taustar(m)) > C_f-(m+1)\cbar(\taustar(m)-1)$:  This implies that if $\tau = \taustar(m)-1$ and the number of requests is  $m+1$ then it is optimal to fetch and transmit and consequently, $\taustar(m+1) \leq \taustar(m) -1$ (see the definition of $\taustar(m)$ in~\eqref{eq:def_taustar_m}). In this case, compute $\gbar(\taustar(m)-1)$ using \eqref{eqn:taustarm<tau<taustarm1} and reduce $\tau$ by $1$. Continue to compute $\gbar(\tau-1)$ for $\tau \leq \taustar(m)-1$ as long as $\gbar(\tau) > C_f-(m+1)\cbar(\tau-1)$.
    On hitting a $\tau$ such that $\gbar(\tau) \leq C_f-(m+1)\cbar(\tau-1)$
follow~\ref{algo:cond_sat}. 
     \end{enumerate}
    
     \item \label{algo:compute_ftheta} Compute $f(\theta^l)$ using~\eqref{eq:g_2_and_theta_first_case} if $\taustar(N)>1$ and using~\eqref{eq:g_2_and_theta_second_case} otherwise.  
 \item \label{algo:update_theta} Update $\theta^l$ and call it $\theta^{l+1}$.  
 \item Repeat Steps~\ref{algo:taustar}-\ref{algo:update_theta} until $\theta^l$s converge. 
\end{enumerate}
 
The following theorem shows the existence and uniqueness of the fixed point of $f(\theta)$.  
\begin{theorem}\label{theorem:fixedpoint}
There exists a unique solution to $f(\theta)=\theta$. 
\end{theorem}
\begin{IEEEproof}
 See Appendix~\ref{Appendix:fixedpoint}. 
\end{IEEEproof}
\begin{remark}
The fixed point iteration $\theta^{l+1} \gets f(\theta^l)$ need not converge. To address this, we consider a damped fixed point iteration with the damping parameter $\alpha \in (0,1)$~(Line~24 in Algorithm~\ref{alg:fixed_point})~\cite{4237146}. It  converges for sufficiently small $\alpha$ which must be chosen based on the variability of $f(\theta)$. An elegant way of choosing $\alpha$ has eluded us so far.
\end{remark}

For illustration, we plot $f(\theta)$ vs $\theta$ considering linear age cost  $\cbar(\tau)=c_a p \tau$ in Figure~\ref{fig:ftheta-theta}. We have set $c_a=10$, $C_f=250$, $p=0.2$, $q=0.5$, and $N=100$. It can that $f(\theta)$ decreases with $\theta$ and $\thetabar=174.5$ is the unique fixed point. Algorithm~\ref{alg:fixed_point} also yields the same fixed point. 
\begin{figure}[h]
    \centering
 
    \includegraphics[scale=0.3]{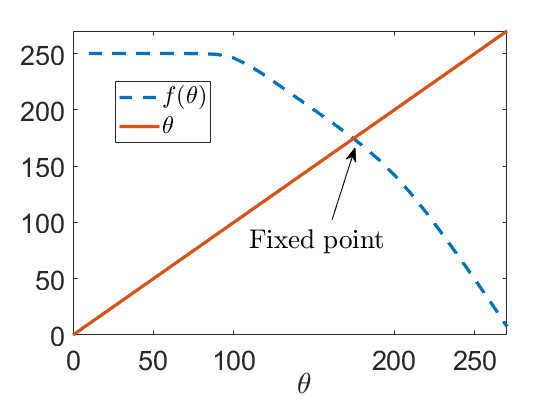}
  
 \caption{$f(\theta)$ is a non-increasing function of $\theta$ and there exists a fixed point.} 
    \label{fig:ftheta-theta}
\end{figure}
\subsection{Heterogeneous users}

In Section~\ref{sec:algo}, we have considered that the users have identical parameters, i.e., same request rate and age cost.  We relax this assumption and let $q_i, \bar{C}_{av,i}(.), i\in\{1,2,\dots,N\}$ be the request probability and cost function of the user $i$. 
Let us denote
 \[
\mathcal{M} = \{(x_1, x_2, \dots, x_N) : x_i \in \{0,1\}\}.
\]
Each vector \( x \in \mathcal{M} \) represents a request configuration. 
The cost associated with configuration \( x \) is given by
\[
C(x) = \sum_{i : x_i = 1} \bar{C}_{\mathrm{av},i}(\tau).
\]
We arrange the costs of all the  configurations in \( \mathcal{M} \) in ascending order 
and index them by \( m = 0, 1, \dots, 2^N - 1 \).
Here $m$ is analogous to the number of requests in Algorithm~\ref{alg:fixed_point}. Then we can apply Algorithm~\ref{alg:fixed_point} to compute the optimal policy for heterogeneous users by varying $m$ from $1$ to $2^N-1$ in the while loop (see line~\ref{while_connd_to_inc_m} of Algorithm~\ref{alg:fixed_point}).  The optimal policy for the heterogeneous case is indeed a threshold policy – for any given set of requesting users, the optimal action is to fetch and broadcast if the time since last fetch exceeds a threshold and not to fetch otherwise. So, the optimal policy is characterized by $2^{N-1}$ thresholds on “time since last fetch”. The computational complexity of Algorithm~\ref{alg:fixed_point} is linear in \( N \) for the homogeneous case. 
However, for heterogeneous users, the complexity grows exponentially as \( 2^N \). 
Therefore, we aim to develop a low-complexity heuristic that performs close to the optimal policy. 
In the following section, we propose a Whittle index-based policy that can be applied to both homogeneous and heterogeneous user scenarios. Furthermore, the complexity of Whittle index policy is also linear in $N$.

\section{Whittle index-based policy}
The Whittle index-based policy acts as a low complexity heuristic to restless multi-armed bandit (RMAB) problems and offers close to optimal performance in most cases. However, the problem~\eqref{eq:min_cost_problem} does not belong to the class of RMAB~\cite{whittle1988restless}. In RMAB, the arms are coupled through a common constraint on the number of arms. While one arm is pulled, the cost incurred depends on that particular arm only. In contrast, in~\eqref{eq:min_cost_problem}, the users are coupled through a common action, the fetching cost $C_f$ is common while the age cost corresponds to each user. To reformulate~\eqref{eq:min_cost_problem} into a user action-specific problem, we need to divide the fetching cost among them and then consider each user's actions separately. By leveraging the basic principle behind the Whittle index policy, we develop a heuristic for~\eqref{eq:min_cost_problem}. This requires solving the single-user problem for each user and combining their solutions.

We reformulate the original problem~\eqref{eq:min_cost_problem} by considering user-specific actions under the constraint that all users take identical actions.  
Let $a_i(t)$ be the action of user $i$ at time $t$. Note that $a_i(t)=a_j(t)$ for all $i\neq j$, equivalently, $a_i(t)=\frac{1}{N}\sum_{i=1}^Na_j(t)\,   \forall i.$ Let $0<\alpha_i<1$ be such that $\sum_{i=1}^N\alpha_i=1.$
We can write the equivalent MDP of~\eqref{eq:min_cost_problem} as follows. 
\begingroup
\allowdisplaybreaks
\begin{align}
    \min\lim_{T\to \infty}\frac{1}{T}&\bbE\Big[\sum_{t=1}^T\sum_{i=1}^N\left(\alpha_iC_f a_i(t)\nonumber\right.\\
    &\left.\quad+ (1-a_i(t))s_i(t)\bar{C}_{av,i}(\tau)\right)\Big] \label{eq:eqv_mdp}\\
    \text{subject to }& a_i(t)=\frac{1}{N}\sum_{i=1}^Na_j(t)\, \forall t,\, \forall i.\label{eq:N_constrt}
\end{align}
\endgroup
In the following, we relax the problem by relaxing the hard constraint~\eqref{eq:N_constrt} to a time average constraint.  
\begin{align}
  &\min\lim_{T\to \infty}\frac{1}{T}\bbE\left[\sum_{t=1}^T\sum_{i=1}^Nc_i(\tau(t),s_i(t),a_i(t))\right] \label{eq:eqv_mdp_rel}\\
    \text{s. t. }& \lim_{T\to\infty}\frac{1}{T}\bbE\left[\sum_{t=1}^Ta_i(t)\right]=\lim_{T\to\infty}\frac{1}{T}\bbE\left[\sum_{t=1}^T\frac{1}{N}\sum_{j=1}^Na_j(t)\right]\,  \forall i.\label{eq:N_constrt_rel}
\end{align}
where \begin{equation}
c_i(\tau(t),s_i(t),a_i(t))=\alpha_iC_f a_i(t){+} (1{-}a_i(t))s_i(t)\bar{C}_{av,i}(\tau).\label{eq:cost_shrt_form}
\end{equation}

We write the Lagrangian of the above relaxed problem with Lagrange multiplier $\beta_i$ for $i=1,2,\dots,N$. 
\begingroup
\allowdisplaybreaks
\begin{align}
    &\min\sum_{i=1}^N\lim_{T\to \infty}\frac{1}{T}\bbE\left[\sum_{t=1}^T\Big(c_i(\tau(t),s_i(t),a_i(t))\right.\nonumber\\
    &\quad \left.+\beta_i(a_i(t)-\frac{1}{N}\sum_{j=1}^Na_j(t)\Big)\right] \label{eq:eqv_mdp_lag}\\
    &=\min\sum_{i=1}^N\lim_{T\to \infty}\frac{1}{T}\bbE\left[\sum_{t=1}^T\Big(c_i(\tau(t),s_i(t),a_i(t))\right.\nonumber\\
&\quad \left.+a_i(t)(\beta_i-\frac{1}{N}\sum_{j=1}^N\beta_j)\Big)\right]\nonumber\\
&=\sum_{i=1}^N\min\lim_{T\to \infty}\frac{1}{T}\bbE\Big[\sum_{t=1}^T\Big(c_i(\tau(t),s_i(t),a_i(t))\nonumber+a_i(t)\lambda_i\Big)\Big]\\&\quad \coloneqq \sum_{i=1}^NV^i(\lambda_i)=D(\lambda).
\end{align}
\endgroup
where $\lambda_i=\beta_i-\frac{1}{N}\sum_{j=1}^N\beta_j$ and \begin{equation}
V^i(\lambda_i)=\min\lim_{T\to \infty}\frac{1}{T}\bbE\Big[\sum_{t=1}^Tc_i(\tau(t),s_i(t),a_i(t))+a_i(t)\lambda_i\Big].\label{eq:sngl_usr}
\end{equation}
The dual problem corresponding to~\eqref{eq:eqv_mdp_rel} subject to~\eqref{eq:N_constrt_rel} is \[\max_{\lambda}D(\lambda)\]
For a fixed value of $\lambda$, $D(\lambda)$ decouples the multi user problem~\eqref{eq:eqv_mdp} in $N$ different MDPs.
Maximizing $D(\lambda)$ requires obtaining $V^i(\lambda_i)$ for each individual user and maximizing $\sum_{i=1}V^i(\lambda_i)$ over all $\lambda$. However, the solution to the dual problem is not always feasible to the original problem due to the hard constraints~\eqref{eq:N_constrt}. The Whittle index-based policy requires solving per-user MDPs and combining the solutions in such a way that the policy is feasible for the multi user problem~\eqref{eq:min_cost_problem} subject to~\eqref{eq:N_constrt}. In the following, we discuss single user problem and obtain the optimal policy and Whittle indices corresponding to each problem. 
\subsection{Single user problem} From~\eqref{eq:sngl_usr} we aim to minimize \begin{align}
&\min\lim_{T\to \infty}\frac{1}{T}\bbE\Big[\sum_{t=1}^Tc_i(\tau(t),s_i(t),a_i(t))+a_i(t)\lambda_i\Big]\nonumber\\
&=\min\lim_{T\to \infty}\frac{1}{T}\bbE\Big[\sum_{t=1}^T\Big((\alpha_iC_f+\lambda_i) a_i(t)\nonumber\\
&\quad \quad+ (1-a_i(t))s_i(t)\bar{C}_{av,i}(\tau)\Big)\Big].
\end{align}
The above equality is obtained by substituting $c_i(\tau(t),s_i(t),a_i(t))$ from~\eqref{eq:cost_shrt_form}.  
The optimal policy for the single user problem is as follows. 
\begin{lemma}\label{lemma:opt_pol_sngle_user}
 If $\lambda_i+\alpha_i C_f\leq 0$, then \[\pi^{\ast}(\tau,s)=1.\]
 If $\lambda_i+\alpha_i C_f> 0$, then \[\pi^{\ast}(\tau,s_i)=\begin{cases}
     1, \text{ if }\tau\geq f_i(\alpha_i C_f+\lambda_i)\text{ and }s_i  =1,\\
     0, \text{ otherwise}
 \end{cases}\]
 where $f_i(\alpha_i C_f+\lambda_i)={\arg\min}_\tau\frac{q_i\sum_{x=1}^{\tau-1}C_{av,i}(x)+\alpha_i C_f+\lambda_i}{\tau+\frac{1-q_i}{q_i}}$.
\end{lemma}
\begin{IEEEproof}
    See Appendix~\ref{app:proof_opt_pol_sngl_user}.
\end{IEEEproof}
\begin{definition}[Passive sets]
    The passive set for a user $i$ for a given $\lambda_i$, $\mathcal{P}^i(\lambda_i)$, is the set of states in which the optimal action does not require the content to be fetched~\cite{zhao2022multi}. More precisely, 
$\mathcal{P}^i(\lambda_i):=\left\{(\tau,s_i): \pi^{\ast}_n(\tau,s)=0\right\}.$
\end{definition}
\begin{definition}[Whittle index]\label{def:WI_index}
    The Whittle index associated with state $(\tau,s_i)$ of user $i$, $W^i(\tau,s_i)$, is the minimum $\lambda$ for which $(\tau,s_i)$ is in the passive set. In other words, 
\[W^i(\tau,s_i) \coloneqq \min\{\lambda:(\tau,s_i) \in \mathcal{P}^i(\lambda)\}.\]
\end{definition}
The following lemma provides these Whittle indices. 
\begin{lemma}\label{lemma:WI_indices_mult_user} \begin{enumerate}
    \item The Whittle indices for the state $(\tau,s_i)$ are as follows. 
   \[W^i(\tau,s_i)=\begin{cases}
       g_i(\tau)-\alpha_i C_f, \text{ if }s_i=1,\\
       -\alpha_i C_f \text{ if }s_i=0
   \end{cases}\]
   where $g_i(\tau)=f_i^{-1}(\tau)$ and $f_i(\cdot)$ is as defined in Lemma~\ref{lemma:opt_pol_sngle_user}. 
   \item $\sum_{i=1}^NW^i(\tau,s_i)$ is independent of $\alpha$. 
\end{enumerate}
\end{lemma}
\begin{IEEEproof}
See Appendix~\ref{app:lemma_WI}. 
\end{IEEEproof}
\subsection{Whittle index policy for the multi content problem}
Let $\pi_{WI}$ be the Whittle index-based policy. 
\[ \pi_{WI}(\tau(t),\bs(t))=\begin{cases}1, \text{ if }\sum_{i=1}^NW^i(\tau(t),s_i(t))>0\\
\quad\text{ and }\sum_{i=1}^Ns_i>0,\\
0, \text{ otherwise.} 
\end{cases}
\]

\begin{remark}[Scalability]
    As we mentioned in Lemma~\ref{lemma:WI_indices_mult_user},  this policy is independent of $\alpha$. This enhances its scalability since the same index function can be applied uniformly across all users. Since the computational complexity of Whittle index policy is linear in $N$, it is easily implementable in large-scale systems.
\end{remark}

\begin{remark}[Optimality]
The Whittle index-based policy is optimal when $q_i=1$ $\forall$ $i$. Let us consider the homogeneous case and $q<1$. As we increase $N$ and $C_f$ also increases proportionally with $N$.  Let $C_f=NC_1$, then the content is fetched when $\sum_{i=1}^Ng(\tau)-C_f>0$ from Lemma~\ref{lemma:WI_indices_mult_user}. Let $M\sim Binomial(N,q)$ be the number of requests, then $Mg(\tau)-NC_1>0\Rightarrow \frac{M}{N}g(\tau)-C_1>0$. Then $\lim_{N\to \infty}\frac{M}{N}g(\tau)-C_1>0\Rightarrow qg(\tau)-C_1>0\,w.p.\, 1$. The Whittle index policy fetches the content periodically, irrespective of the number of requests.  We obtain the optimal policy from Algorithm~\ref{alg:fixed_point} and obtain the fetching intervals from the simulations and plot their variance (see Figure~\ref{fig:plot_var}). This shows that the optimal policy also becomes periodic.  For this, we consider $p=0.7,q=0.12,C_f=N$ and $C_a=10$. 
 \begin{figure}[h]
\begin{subfigure}[t]
{0.5\linewidth}
  \centering
\includegraphics[width=\linewidth]{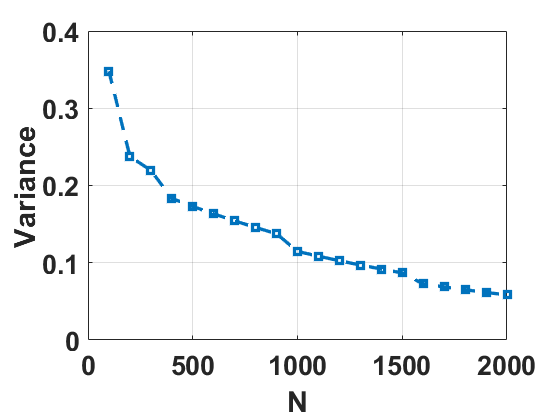}
\caption{Variance of fetching intervals under the optimal policy}
\label{fig:plot_var}
\end{subfigure}
\begin{subfigure}[t]
{0.49\linewidth}
  \centering
\includegraphics[width=\linewidth]{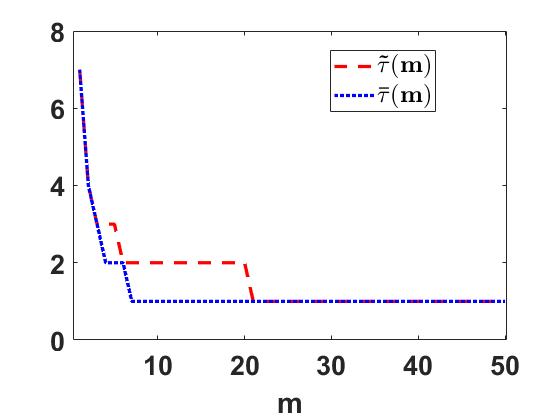}
\caption{Fetching thresholds of optimal and Whittle index policy}
\label{fig:tautil_vs_taum}
\end{subfigure}
\caption{ Asymptotic optimality of WI policy for homogeneous users
}
 \end{figure}
 
Let $\tilde{\tau}(m)$ and $\taustar(m)$ denote the fetching time thresholds under the Whittle index and optimal policies, respectively, for a given number of requests $m$. We fix $N = 50$, $C_f = 50$ while keeping the other parameters same as above. We plot $\tilde{\tau}(m)$ and $\taustar(m)$ in Figure~\ref{fig:tautil_vs_taum}. We observe that $\tilde{\tau}(m)$ and $\taustar(m)$ exhibit similar structures and closely match each other, eventually becoming identical beyond a certain value of $m$. 
 This explains why the WI policy gives close to optimal performance (see Figures~\ref{fig:percent_error} and~\ref{fig:subopt_vs_N} ) in large user settings. However, its theoretical optimality analysis has eluded us so far. 
\end{remark}

{\color{blue}

}
\remove{
\section{VAoI Aware Transmission}\label{section: VAoIBroadcast}
Now we address the original problem, Problem~\eqref{eq:original_problem}, where the SCSP observes the VAoI after fetching the content and uses it to decide whether to transmission the content or not. 
As mentioned in Section~\ref{Section:sys_model}, this problem can also be framed as  an average cost MDP. $s(t) \coloneqq (\tau(t),d(t),m(t)) \in \mathbb{Z}_+ \times \mathbb{Z}_+ \times 
[N]$ and $(a_f(t),a_{tx}(t)) \in \{0,1\} \times \{0,1\}$ represent this MDP's state and action, respectively, at time $t$. However, unlike standard MDPs, we take the two subactions $a_f(t)$ and $a_{tx}(t)$ sequentially and can potentially have two transitions at $t$, first after subaction $a_f(t)$ and the second after subaction $a_{tx}(t)$. Let $s'(t)$ denote the MDP's intermediate state after 
subaction $a_f(t)$. The MDP moves to state $s(t+1)$  after subsction $a_{tx}(t)$.

We now describe the state transitions and the expected single stage costs. Let the state at time $t$ be $s(t) = (\tau,d,m)$. If $a_f(t) = 0$ then $s'(t) = s(t)$. In this case, the VAoI at the SCSP $V \sim \text{Binomial}(\tau,p)$ is unknown to the SCSP, and the expected
single stage cost is given by
\begin{equation}
    C(\tau,d,m) = \begin{cases}
        m C_{av}(\tau,d) \text{ if }a_{tx}(t)=0,\\
        C_{tx}+ m C_{av}(\tau,0)  \text{ if }a_{tx}(t)=1.
    \end{cases}
\end{equation}
where $C_{av}(\tau,d) \coloneqq \mathbb{E}[C_a(d+V)]$.  On the other hand, if $a_f(t) = 1$, the SCSP
learns $V$ and transitions to $s'(t) = (0,d + V,m)$. Given $s'(t) = (0,d',m')$, the single stage cost is given by
\begin{equation}
    C'(d',m',a_{tx}) = \begin{cases}
    C_f+ m'C_a(d') \text{ if } a_{tx}=0, \\
      C_f \text{ if } a_{tx}=1.
    \end{cases}
\end{equation} 
Further, given $s'(t) = (\tau',d',m')$, the next state $s(t+1)$, after taking subaction $a_{tx}(t)$, is given as follows.
\begin{equation*}
    s(t+1) = \begin{cases}
        (\tau'(t)+1,d'(t),M) \text{ if } a_{tx}(t) = 0, \\
        (\tau'(t)+1,0,M) \text{ if } a_{tx}(t) = 1,
    \end{cases}
\end{equation*}
where $M \sim \text{Binomial}(N,q)$. 

\remove{

We now describe the single stage expected cost and the state transitions more formally. Recall that,
given $(d(t),\tau(t),m(t)) = (d,\tau,m)$, the VAoIs at the SCSP and at the user are $V$ and $d+V$, respectively, where $V \sim \text{Binomial}(\tau,p)$.  We use $C'(d,\tau,m)$ to denote the expected single stage cost when $a_f(t) = 0$. It is given by 
\begin{equation}
    C'(d,\tau,m) = m C_{av}(d,\tau)
\end{equation}
where $C_{av}(\tau,m) \coloneqq \mathbb{E}[C_a(d+V)]$. If 

When $a_f(t) = 0$, 

\begin{equation*} 
     C(d,\tau,m,a_f,a_{tx}) = \begin{cases} \cbar'(d,\tau) \text{ if } a_f =0,\\
     C_f \text{ if } a = 1  
    \end{cases}
\end{equation*}

Recall that, given $(\tau(t),m(t)) = (\tau,m)$, the VAoI at the SCSP and the user $V(t) \sim \text{Binomial}(\tau,p)$. With a slight abuse of notation we use $\bar{C}(\tau,m,a)$ to denote the expected single stage cost. It is given by 
\begin{equation*} 
     \bar{C}(\tau,m,a) = \begin{cases} \cbar(\tau,m) \text{ if } a=0,\\
     C_f \text{ if } a = 1  
    \end{cases}
\end{equation*}
where $\cbar(\tau,m) \coloneqq \mathbb{E}[C_a(V)]$ and $C_f \coloneqq C_f + C_{tx}$. Note that $\cbar$ is also a convex, increasing function.

Note that, the VAoI at time $t$, $V(t)$ is not directly observable by the SCSP. 
The expected cost incurred at time $t$ if the SCSP chooses not to transmit $\mathbb{E}[C_a(V(t)+d(t))]=C_{av}(\tau(t),d(t))$, where $C_{av}$ is convex and increasing function of $\tau(t)$ and $d(t)$. However, if the SCSP chooses to fetch, then it can observe the VAoI and in that case $V(t)=v$ is known and the expected cost for not transmitting will be $C_a(v+d(t))$ where $v\in\{1,2,\dots,\tau(t)\}$. 
Hence we can reformulate the MDP problem~\eqref{eq:original_problem} considering $s(t)=(\tau(t),d(t),m(t))$ as state as follows. 

\paragraph*{Evolution of states}
Let $s(t)=(\tau,d,m)$.
The state at $(t+1)$ is $s(t+1)=(\tau(t+1),d(t+1),m(t+1))$
\begin{align*}\tau(t+1)&=\begin{cases}
    \tau(t)+1 ,&\text{if }u(t)=0,\\
    1, &\text{if }u(t)\in\{1,2\}, 
\end{cases}\\d(t+1)&=\begin{cases}
    d ,&\text{if }u(t)=0,\\
    0, &\text{if }u(t)=1,\\
    d+v, &\text{if }u(t)=2.
\end{cases}
\end{align*}
 \[\text{where }v\sim Binomial(\tau,p) \text{ and } m(t+1)\sim Binomial(N,q).\]
  Hence the expected cost at time $t$ is
\[C(\tau,d,m)=\begin{cases}
    mC_{av}(\tau,d), &\text{ if }u(t)=0,\\
    C_f+C_{tx}, &\text{ if }u(t)=1,\\
    C_f+mC_a(v+d), &\text{ if }u(t)=2.
\end{cases}\]
}

Let $\theta^{\ast}$ and $h: \mathbb{Z}_{++} \times \mathbb{Z}_+ \times 
[N] \to \mathbb{R}$ be the optimal average cost and the differential cost function, respectively, of the  MDP. These satisfy the following Bellman equations~\cite[Chapter~4]{BertsekasVol2}.
We skip the detailed arguments for brevity.
\begin{align}
    h(\tau,d,m){+}\theta^{\ast}\
    {=}&\min\Big\{mC_{av}(\tau,d)+h_{av}(\tau+1,d),\nonumber\\
    &C_{tx}+mC_{av}(\tau,0)+h_{av}(\tau+1,0), C_f +\nonumber\\
    \sum_i&B(\tau,p,i)\min\left\{m C_a(d{+}i){+}h_{av}(1,d{+}i),\right.\nonumber\\
    &\left. C_{tx}+h_{av}(1,0)\right\}\Big\}.\label{eq:bellman_original}
\end{align}
where $h_{av}(\tau,d)\coloneqq\sum_k B(N,q,k)h(\tau,d,k)$. As mentioned earlier, the three dimensional state space and complex state transitions and cost structure of the MDP make derivation of the optimal policy hard. Below, we propose a heuristic based on the optimal solution of the VAoI oblivious transmission problem.
\begin{enumerate}
    \item Compute  $\hbarav(\tau)$ for the VAoI oblivious transmission problem~\eqref{eq:bellman_hbar} using Algorithm~\ref{alg:fixed_point}. 
    \item \label {def:htilde_assign} Set $h_{av}(\tau,d)=\hbarav([\frac{d}{p}]+\tau)$.
    \item Obtain a fetching and transmission policy $\pi = (\pi_f,\pi_{tx})$ from~\eqref{eq:bellman_original} using  $h_{av}(\cdot)$ obtained in Step 2.  Suppose, $\pi_f(\tau,d,m)=0$ then 
     \begin{align*}
    \lefteqn{\pi_{tx}(\tau,d,m)} \ \ \ &\\
    &= \begin{cases}
     &0 \text{ if }  m C_a(\tau,d)+h_{av}(\tau+1,d) < C_{tx}\\
    &\quad \quad +mC_{av}(\tau,0)+h_{av}(\tau+1,0),  \\   
    & 1 \text{ otherwise.}
     \end{cases}
         \end{align*}
  If $\pi_f(\tau,d,m)=1$ then   
    \begin{align*}
    \lefteqn{\pi_{tx}(0,d',m')} \ \ \ &\\
    &= \begin{cases}
     0 \text{ if }  m' C_a(d'){+}h_{av}(1,d') < C_{tx}+h_{av}(1,0),  \\   
     1 \text{ otherwise.}
     \end{cases}
         \end{align*}
         Note that $m'=m$ and $d'$ is the observed VAoI after fetching. 
 \end{enumerate}
    Since the heuristic employs a {\it policy improvement} step over the policy obtained from the Algorithm~\ref{alg:fixed_point}, the average cost of~\eqref{eq:original_problem} under the heuristic is lower than the optimal cost of~\eqref{eq:min_cost_problem}~\cite{bertsekas2011dynamic}. 
    Next, we introduce a fictitious problem, called {\it genie-aided transmission}, whose optimal cast is a lower bound on the optimal cost of the original problem. We use it to access the performance of the proposed heuristic. 
    
\subsection{Genie-aided Transmission}
Let us consider a fictitious content fetching and transmission problem in which the SCSP regularly learns about updates of the sensor's contents for free. So, effectively,  $V(t) = 0$ at all $t$. Moreover, the SCSP only needs to decide at each $t$ whether to transmission or not; it fetches the content whenever it decides to transmission. Formally, let $\tilde{a}(t) \in \{0,1\}$ and $\tilde{C}(t)$ denote the SCSP's action and the cost incurred at time $t$, respectively. $\tilde{a}(t)$ has the same connotation as $\abar(t)$ in Section~\ref{subsection :VAoI oblivious} and  
$\tilde{C}(t)$ is as follows.
\[\tilde{C}(t)=\begin{cases}
    m(t)C_a(d(t)) \text{ if } \tilde{a}(t) =0,\\
    C_f \text{ if } \tilde{a}(t) =1. 
\end{cases}
\]
We again aim to minimize the time averaged expected cost.

We can pose the genie-aided transmission problem 
also  as  an average cost MDP. $(d(t),m(t)) \in \mathbb{Z}_+ \times 
[N]$ and $\tilde{a}(t) \in \{0,1\}$ represent the state and the action, respectively, at time $t$. Given $(d(t),m(t)) = (d,m)$, the state at time $t+1$ is given by
\begin{equation*}
  (d(t+1),m(t+1)) =
    \begin{cases}
      (X,M) \text{ if } \tilde{a}(t)=1, \\
      (d+X,M) \text{ if } \tilde{a}(t)=0,
    \end{cases}
\end{equation*}
where $X \sim \text{Bernoulli}(p)$ and $M \sim \text{Binomial}(N,q)$. We again slightly abuse the notation and use $\tilde{C}(d,m,a)$ to denote the single stage cost. It is given by 
\begin{equation*} 
     \tilde{C}(d,m,a) = \begin{cases} mC_a(d) \text{ if } a=0,\\
     C_f \text{ if } a = 1 . 
    \end{cases}
\end{equation*}

Let $\tilde{h}: \mathbb{Z}_+ \times 
[N] \to \mathbb{R}$, $\tilde{\theta}$ and $\tilde{\pi}$  be the differential cost function, the optimal average cost and the optimal policy, respectively, of the  MDP.  
 These satisfy the following Bellman's equations~\cite[Chapter~4]{BertsekasVol2}.
\begin{align}
\tilde{h}(d,m)+\tilde{\theta}{=}&\min\left\{C_f+\sum_{k=0}^{N}B(N,q,k)\tilde{h}_{av}(0,k),\right.\nonumber\\
m\tilde{C}(d) & \left.{+}\sum_{k=0}^{N}B(N,q,k)\tilde{h}_{av}(d,k)\right\}\label{eqn:bellmanAgg2}
\end{align}
 where $\tilde{h}_{av}(d,k)\coloneqq p\tilde{h}(d+1,k) + (1-p)\tilde{h}(d,k)$. 
 Notice the similarity between~\eqref{eqn:bellmanAgg} and~\eqref{eqn:bellmanAgg2}. In view of it, we can provide a similar algorithm as Algorithm~\ref{alg:fixed_point} to obtain  $\tilde{\theta}$ and $\tilde{\pi}$. We omit the details to save space.  

\begin{remark}
\label{remark2} (a) The optimal cost of the genie-aided transmission problem gives a lower bound on  the optimal cost of the original problem. So, this can be used to assess the performance of the heuristic for the original problem. In particular, the difference between the cost of the heuristic  and $\tilde{\theta}$ gives a bound on the suboptimality gap of the heuristic.  \\ 
(b) The genie-aided transmission problem also approximates the original problem when the fetching cost is small. Recall that the  VAoI oblivious transmission problem also approximates the original problem when the transmission cost is small or when the content
updates are predictable. So, we can use the solutions to both these problems to arrive at a better heuristic for the original problem for general parameter values. This is a potential future work.
\end{remark}
}

\remove{\subsection{Optimal Policy for functions of VAoI}
  We consider the same model as Section~\ref{Section:sys_model} and develop an algorithm to derive optimal policy considering functions of VAoI. 
Let $H$ be a continuous increasing function of $V_c$. Let the age cost per request be $C_a\hbarr(V(t))$.   Given that there are total $m(t)$ requests at time $t$, then the total age cost incurred by the SCSP is $m(t)C_aV(t)$. 
  Furthermore,  \[\mathbb{E}[m(t)C_a\hbarr(V(t))]=C_aF_p(\tau(t))m(t)\]
  where $F_p(\tau(t))$ is a continuous, convex and increasing function of $\tau(t)$. Hence, $F_p(\tau(t))$ is also invertible. 
In this case, Bellman's equation will be as follows.
  \begin{align}
\hbarr(\tau,m)+\thetabar=\min&\left\{C_f+\sum_{k=0}^{N}B(N,q,k)\hbarr(1,k),\right.\nonumber\\
&\left.C_aF_p(\tau) m{+}\sum_{k=0}^{N}B(N,q,k)\hbarr(\tau{+}1,k)\right\}.\nonumber
\end{align}
Following similar approaches as in Section~\ref{sec:opt_policy} we can establish Lemmas~\ref{lemmaAgg},~\ref{lemma:cost_func_after_taustar} and Theorems~\ref{TheoremAgg},~\ref{TheoremAgg2}. 
Hence, using the following algorithm (similar to Algorithm~\ref{alg:fixed_point}) we can compute $\taustar(m)$ and $\thetabar$. 
 \begin{algorithm}[h]
\caption{Algorithm to compute $\taustar(1),\dots,\taustar(N)$}\label{alg:fixed_point_func_of_VAoI}
\begin{algorithmic}[1]
\Require $ \thetabar^0, \epsilon, \alpha$
\Ensure $|\thetabar^{l+1}-\thetabar^{l}|\geq \epsilon$
\State $\thetabar^l\gets \thetabar$
\State $\taustar(1) \gets \left \lceil{\cbar^{-1}\left(\frac{\thetabar}{C_ap(1-(1-q)^N)}\right)}\right \rceil $ 
\State $\gbar(\taustar(1)) \gets C_f-\frac{\thetabar}{1-(1-q)^N}$ 
\State $m \gets 1$
\State $\tau\gets\taustar(1)$
\While{$\tau > 1$}

\While {$\gbar(\tau)\leq C_f-C_a(m+1)F_p((\tau-1)) \, \text{and}\, m<N$ } 
    \State $m \gets m+1$
    \State $\taustar(m) \gets \tau$
\EndWhile
\If {$\tau>2$}  
\State \lefteqn{\gbar(\tau-1) \gets C_f-\thetabar}\[ +\sum_{k\leq m}B(N,q,k)(kC_aF_p((\tau-1))-C_f+\gbar(\tau))\] 
\Else 
   \State $\taustar(m+1)=\dots=\taustar(N)=1$  
\EndIf
\State $\tau \gets \tau-1$ 
\EndWhile
\If {$\taustar(N)>1$}
    \State $f(\thetabar) \gets NC_aF_p(1)q+\gbar(2)$ 
\Else
   \State $\,f(\thetabar) \gets \sum_{k\leq m}B(N,q,k)(kC_aF_p(1)+\gbar(2))$\[+\sum_{k> m}B(N,q,k)C_f\]
  
\EndIf

\State $\thetabar^{l+1}\gets \alpha f(\thetabar)+(1-\alpha)\thetabar$ 
\end{algorithmic}
\end{algorithm}}

\remove{
\subsection{Computation of $\alpha$}
\section {Transmission at two power level}
Consider the scenario where there are $N_1$ user within a distance $r$ with respect to the SCSP and there are $N_2$ user at a distance more than $r$ with respect to the SCSP. The SCSP needs to transmission to the nearby user at lower transmission power. However, to reach the faraway user, it needs to transmission at higher transmission power.
Whenever the SCSP transmission a content at $C_{tx_1}$, all $N_1$ user gets the content while $N_2$ do not get the content. However, whenever the SCSP transmission a content at $C_{tx_2}$, all $N_1+N_2$ user gets the content. 

Let $d(t)$ be the difference in version age of content of faraway $N_2$ user in slot $t$. Each user can request content from the SCSP in a slot with probability $q$.
Let $m_1(t)$ and $m_2(t)$ be the number of requests fron $N_1$ user and $N_2$ user, respectively, in slot $t$, such that $0\leq m_1(t)\leq N_1, 0\leq m_2(t)\leq N_2,\forall t \geq 0$. $\tau(t)$ is the age of content at the SCSP in slot $t$.
Let $\textbf{s}(t) \in [N]\times\mathbb{Z}_+\times \{0,1,...,N_1\}\times \{0,1,...,N_2\}$ be the state at slot $t$. We define $\textbf{s}(t)=(\tau(t),d(t),m_1(t),m_2(t))$.
 The policy $\pi$ is a sequence $\{u^\pi_t, t = 1,2,\cdots\}$ where $u^\pi_t: [N]\times\mathbb{Z}_+\times \{0,1,...,N_1\}\times \{0,1,...,N_2\} \to \{0,1\}$. The SCSP takes the following actions:
\begin{align*}
  &u^\pi(\textbf{s}(t)){=}
    \begin{cases}
     2, \text{ fetch and transmission the content at $C_{tx_1}$ }\\
     1, \text{ fetch and transmission the content at $C_{tx_2}$ }\\
      0, \text{ idle} 
    \end{cases}
\end{align*}

 Given, $\textbf{s}(t)$, the state at time $t+1$, $\textbf{s}(t+1)$ evolves as following:
\begin{align*}
  \textbf{s}(t+1)=
    \begin{cases}
      (1,0,m'_1(t),m'_2(t)), \text{ if $u^\pi(\textbf{s}(t))=2$} \\
      (1,d(t)+x(t),m'_1(t),m'_2(t)), \text{ if $u^\pi(\textbf{s}(t))=1$} \\
      (\tau(t)+1,d(t),m'_1(t),m'_2(t)), \text{ if $u^\pi(\textbf{s}(t))=0$} \\
    \end{cases}
\end{align*}
where $x(t)$ is a random variable of binomial distribution with parameters $\tau(t)$ and $p$, and $m'_i(t), i \in\{1,2\}$ is a random variable of binomial distribution with parameters $N_i$ and $q$
The total cost $C^{\pi}(\textbf{s}(t))$, incurred at the end of slot $t$ under the policy $\pi$.
\begin{align}
\label{eqn:CpistAgg2}
   & C^\pi(\textbf{s}(t))=(C_f+C_{tx_2})\mathds{1}\{u^\pi(\textbf{s}(t))=2\}\nonumber\\
   &+(C_{a}m_2(t)(p\tau(t)+d(t))+C_f+C_{tx_1})\mathds{1}\{u^\pi(\textbf{s}(t))=1\}\nonumber\\
    &+(\cbar(\tau(t))(m_1(t)+m_2(t))+C_{a}d(t)m_2(t))\mathds{1}\{u^\pi(\textbf{s}(t))=0\}
\end{align}

We solve the following optimisation problem, 
\begin{align}
\text{Minimize } & \ \lim_{T \to \infty}\frac{1}{T}\mathbb{E}\left[\sum_{t=0}^T C^{\pi}(\textbf{s}(t))\Bigg\vert \textbf{s}(0) =(\tau,d,m_1,m_2)\right] \label{eqn:objectiveAgg2}
\end{align}
We introduce average cost MDP for Problem~\eqref{eqn:objectiveAgg2}.
The optimal average cost $\thetabar$ achieved by optimal policy $\bar{\pi}$ is the same for all initial states, satisfying the Bellman equation
\begin{align}
&\hbarr(\tau,d,m_1,m_2)+\thetabar=\min\{(C_f+C_{tx_2})\mathds{1}\{a=2\}+\hbarav(1,0),\nonumber\\
&(C_{a}m_2(t)(p\tau(t)+d(t))+C_f+C_{tx_1})\mathds{1}\{a=1\}\nonumber\\
&+\sum_{x=0}^{\tau}\binom{\tau}{x}p^x(1-p)^{\tau-x}\hbarav(1,d+x),\nonumber\\ 
&(\cbar(\tau(t))(m_1(t)+m_2(t))+C_{a}d(t)m_2(t))\mathds{1}\{a=0\}\nonumber\\
&+\hbarav(\tau+1,d)\}\label{eqn:bellmanAgg}.
\end{align}
where 
$\hbarr(\tau,d,m_1,m_2)$ is the differential cost function of the MDP problem and $\hbarav(\tau,d):=\sum_{i=0}^{N_1}\binom{N_1}{i}q^i(1-q)^{N_1-i}\sum_{j=0}^{N_2}\binom{N_2}{j}q^i(1-q)^{N_2-j}\hbarr(\tau,d,i,j)$.

\begin{lemma}
$\hbarav(\tau,d)$ is non-decreasing in $\tau$ and $d$.
\end{lemma}
\begin{theorem}
\label{TheoremAgg2}
\begin{enumerate}
    \item $\exists d^{\ast}(\tau,m_1,m_2)$ such that $\forall d>d^{\ast}(\tau,m_1,m_2)$, $u^{\bar{\pi}}(\tau,d,m_1,m_2)=2.$
    \item $\exists \taustar_1(d,m_1,m_2)$ and $\taustar_0(d,m_1,m_2)$ such that $\forall \tau>\taustar_1(d,m_1,m_2)$, $u^{\bar{\pi}}(\tau,d,m_1,m_2)=2$ and $\forall \tau \leq \taustar_0(d,m_1,m_2)$, $u^{\bar{\pi}}(\tau,d,m_1,m_2)=0.$ 
\end{enumerate}
    
\end{theorem}
}
\remove{
\begin{lemma}
    $\gbar(\tau_1^{(0)}(1)-1)>\gbar(\tau_2^{(0)}(1)-1)$ if $\thetabar_1^{(0)}<\thetabar_2^{(0)}$.
\end{lemma}
\begin{IEEEproof}
    If $\thetabar_1^{(0)}<\thetabar_2^{(0)}$, clearly, $\tau_1^{(0)}(1)<\tau_2^{(0)}(1)$ and $\gbar(\tau_1^{(0)}(1))>\gbar(\tau_2^{(0)}(1))$.
\begin{enumerate}
    \item $\tau_1^{(0)}(1)>\tau_1^{(0)}(2)$ and $\tau_2^{(0)}(1)>\tau_2^{(0)}(2)$: In this case, $\gbar(\tau_1^{(0)}(1)-1)$ and $\gbar(\tau_2^{(0)}(1)-1)$ are as follows.
    \begin{align}
    \label{equation: Case1g1}
    &\gbar(\tau_1^{(0)}(1)-1)=C_f-\thetabar_1^{(0)}\nonumber\\
    &+\sum_{k\leq1}B(N,q,k)k\cbar(\tau_1^{(0)}(1)-1)\nonumber\\
    &-\sum_{k\leq1}B(N,q,k)\frac{\thetabar_1^{(0)}}{1-(1-q)^N}\nonumber\\
    \end{align}

    \begin{align}
    \label{equation: Case1g2}
    &\gbar(\tau_2^{(0)}(1)-1)=C_f-\thetabar_2^{(0)}\nonumber\\
    &+\sum_{k\leq1}B(N,q,k)k\cbar(\tau_2^{(0)}(1)-1)\nonumber\\
    &-\sum_{k\leq1}B(N,q,k)\frac{\thetabar_2^{(0)}}{1-(1-q)^N}\nonumber\\
    \end{align}
    From \eqref{equation: Case1g1} and \eqref{equation: Case1g2}, 
    \begin{align}
    \label{equation: Case1g1g2}
     &\gbar(\tau_1^{(0)}(1)-1)- \gbar(\tau_2^{(0)}(1)-1)= \thetabar_2^{(0)}-\thetabar_1^{(0)}  \nonumber\\
     & +(1-q)^{N}\frac{\thetabar_2^{(0)}-\thetabar_1^{(0)}}{1-(1-q)^N}\nonumber\\
     &=\frac{\thetabar_2^{(0)}-\thetabar_1^{(0)}}{1-(1-q)^N}>0\nonumber\\
    \end{align}
 \item $\tau_1^{(0)}(1)=\tau_1^{(0)}(2)$ and $\tau_2^{(0)}(1)>\tau_2^{(0)}(2)$: In this case, $\gbar(\tau_1^{(0)}(1)-1)$ and $\gbar(\tau_2^{(0)}(1)-1)$ are as follows.

 \begin{align}
    \label{equation: Case2g1}
    &\gbar(\tau_1^{(0)}(1)-1)=C_f-\thetabar_1^{(0)}\nonumber\\
    &+\sum_{k\leq2}B(N,q,k)k\cbar(\tau_1^{(0)}(1)-1)\nonumber\\
    &-\sum_{k\leq2}B(N,q,k)\frac{\thetabar_1^{(0)}}{1-(1-q)^N}\nonumber\\
    \end{align}

    \begin{align}
    \label{equation: Case2g2}
    &\gbar(\tau_2^{(0)}(1)-1)=C_f-\thetabar_2^{(0)}\nonumber\\
    &+\sum_{k\leq1}B(N,q,k)k\cbar(\tau_2^{(0)}(1)-1)\nonumber\\
    &-\sum_{k\leq1}B(N,q,k)\frac{\thetabar_2^{(0)}}{1-(1-q)^N}\nonumber\\
    \end{align}

    Since $\tau_1^{(0)}(1)=\tau_1^{(0)}(2)$, $\gbar(\tau_1^{(0)}(1))\leq C_f-2\cbar(\tau_1^{(0)}(1)-1)$.  
    From \eqref{equation: Case2g1} and \eqref{equation: Case2g2}, 
    \begin{align}
    \label{equation: Case2g1g2}
     &\gbar(\tau_1^{(0)}(1)-1)- \gbar(\tau_2^{(0)}(1)-1)\nonumber\\
     &=\frac{\thetabar_2^{(0)}-\thetabar_1^{(0)}}{1-(1-q)^N}\nonumber\\
     &+\binom{N}{2}q^2(1-q)^{N-2}(2\cbar(\tau_1^{(0)}(1)-1)-C_f-C_f)\nonumber\\
    &+\binom{N}{2}q^2(1-q)^{N-2}(C_f-\frac{\thetabar_1^{(0)}}{1-(1-q)^N})\nonumber\\
     \end{align}
     \color{red} $\binom{N}{2}q^2(1-q)^{N-2}(2\cbar(\tau_1^{(0)}(1)-1)-C_f-C_f+\gbar(\tau_1^{(0)}(1)))\leq 0$.
     \color{black}
 \item $\tau_1^{(0)}(1)=\tau_1^{(0)}(k_1)$ and $\tau_2^{(0)}(1)=\tau_2^{(0)}(k_2)$: Suppose $k_1=k_2$. In this case, $\gbar(\tau_1^{(0)}(1)-1)$ and $\gbar(\tau_2^{(0)}(1)-1)$ are as follows.   

 \begin{align}
    \label{equation: Case3ag1}
    &\gbar(\tau_1^{(0)}(1)-1)=C_f-\thetabar_1^{(0)}\nonumber\\
    &+\sum_{k\leq k_1}B(N,q,k)k\cbar(\tau_1^{(0)}(1)-1)\nonumber\\
    &-\sum_{k\leq k_1}B(N,q,k)\frac{\thetabar_1^{(0)}}{1-(1-q)^N}\nonumber\\
    \end{align}

    \begin{align}
    \label{equation: Case3ag2}
    &\gbar(\tau_2^{(0)}(1)-1)=C_f-\thetabar_2^{(0)}\nonumber\\
    &+\sum_{k\leq k_1}B(N,q,k)k\cbar(\tau_2^{(0)}(1)-1)\nonumber\\
    &-\sum_{k\leq k_1}B(N,q,k)\frac{\thetabar_2^{(0)}}{1-(1-q)^N}\nonumber\\
    \end{align}
    
\end{enumerate}

\end{IEEEproof}
}
\remove{
\begin{lemma}
    $\gbar(2)$ is decreasing in $\thetabar^{(0)}$.
\end{lemma}
\begin{IEEEproof}
    $\gbar(\tau)$ is non-increasing in $\tau$. Therefore, if $\gbar(\taustar)(1)$ is decreasing in $\thetabar^{(0)}$, $\gbar(2)$ is decreasing in $\thetabar^{(0)}$. From the algorithm, $\gbar(\taustar(1))=C_f-\frac{\thetabar^{(0)}}{1-(1-q)^N}$, which is decreasing in $\thetabar^{(0)}$. 
\end{IEEEproof}
\begin{theorem}
    Fixed point equation $\thetabar^{(1)}= t\thetabar^{(1)}+(1-t)\thetabar^{(0)}$ has unique solution.
\end{theorem}
\begin{IEEEproof}
    From algorithm, if $\taustar(N) > 1$, $\thetabar^{(1)} \gets NC_apq+\gbar(2)$.   
If $\taustar(N) = 1$, $\thetabar^{(1)} \gets \sum_{k\leq M}B(N,q,k)(kC_ap+\gbar(2))+\sum_{k> M}B(N,q,k)C_f$. In either case, from the above lemma, since $\gbar(2)$ is decreasing in $\thetabar^{(0)}$, $\thetabar^{(1)}$ is decreasing in $\thetabar^{(0)}$.   Therefore, $\thetabar^{(1)}= t\thetabar^{(1)}+(1-t)\thetabar^{(0)}$ has unique solution.
\end{IEEEproof}
}
\remove{Action $\hat{\textbf{a}}(t)=(\hat{a}_1(t),\hat{a}_2(t),...,\hat{a}_N(t))$ under the heuristic policy is given below.
\begin{align*}
    \hat{a}_i(t)=
    \begin{cases}
      2,  \text{ if } (u_f^{\hat{\pi}}(s(t)),u_i^{\hat{\pi}}(s(t)))=(1,1)\\
      1, \text{ if } (u_f^{\hat{\pi}}(s(t)),u_i^{\hat{\pi}}(s(t)))=(1,0)\\
      0, \text{ if } (u_f^{\hat{\pi}}(s(t)),u_i^{\hat{\pi}}(s(t)))=(0,0) \text{ or } (0,1)\\
    \end{cases}
\end{align*}
}
\color{black}\remove{
\appendices
\section{Proof of Lemma 1} \label{Appendix:lemma1}
We prove the lemma by value iteration.
\begin{align*}
    Q_{t+1}(d,\tau,0)=& qC_a(p\tau+d)+(1-\alpha)h_t(d,\tau+1)\\
    &+\alpha\sum_{i=0}^{\tau}\binom{\tau}{i} p^i(1-p))^{\tau-i}h_t(d+i,\tau)
\end{align*}
\begin{align*}
    Q_{t+1}d,\tau,1)=&qC_a(p\tau)+qC_f+q\alpha h_t(0,1)+q(1-\alpha)h_t(0,\tau+1)\\
    &+(1-q)(1-\alpha)h_t(d,\tau+1)\\
    &+(1-q)\alpha\sum_{i=0}^{\tau}\binom{\tau}{i} p^i(1-p))^{\tau-i}h_t(d+i,\tau)
\end{align*}
\begin{align*}
    Q_{t+1}(d,\tau,2)=&qC_f+qC_f+q h_t(0,1)+(1-q)(1-\alpha)h_t(d,\tau+1)\\
    &+(1-q)\alpha\sum_{i=0}^{\tau}\binom{\tau}{i} p^i(1-p))^{\tau-i}h_t(d+i,\tau)
\end{align*}
\begin{align*}
    h_{t}(d,\tau,)+\thetabar=\min\{Q_t(d,\tau,0), Q_t(d,\tau,1), Q_t(d,\tau,2)\}
\end{align*}
Let $h_0(d,\tau)$ be $0$, $\forall d, \tau$.
Therefore,
\begin{align*}
    Q_1(d,\tau,0)=qC_a(p\tau+d)
\end{align*}
\begin{align*}
    Q_1(d,\tau,1)=qC_a p\tau+qC_f
\end{align*}
\begin{align*}
    Q_1(d,\tau,2)=qC_f+qC_f
\end{align*}
Clearly, $Q_1(d,\tau,a)$ is non decreasing in $d$, for $a \in \{0,1,2\}$. Therefore, $h_1(d,\tau)$ is non decreasing in $d$.
Suppose $h_t(d,\tau)$ is non decreasing in $d$, $\forall t \leq T$. We will now show that $h_{T+1}(d,\tau)$ is non-decreasing in $d$.
\begin{align*}
    Q_{T+1}(d,\tau,0)=& qC_a(p\tau+d)+(1-\alpha)h_T(d,\tau+1)\\
    &+\alpha\sum_{i=0}^{\tau}\binom{\tau}{i} p^i(1-p))^{\tau-i}h_T(d+i,\tau)
\end{align*}
Since each of the terms in the summation is non-decreasing in $d$, $Q_{T+1}(d,\tau,0)$ is non-decreasing in $d$. Same argument holds true for $Q_{T+1}(d,\tau,1)$ and $Q_{T+1}(d,\tau,2)$. Therefore, $h_{T+1}(d,\tau)$ is non decreasing in $d$.

\section{Proof of Lemma 2} \label{Appendix:lemma2}
Similar arguments as lemma 1.
\section{Proof of Lemma 3} \label{Appendix:lemma3}
Given $u^{\ast}(d,\tau)=0$.
This implies $Q(d,\tau,0)\leq Q(d,\tau,1)$ and $Q(d,\tau,0)\leq Q(d,\tau,2)$.
\begin{itemize}
\item Given $Q(d,\tau,0)\leq Q(d,\tau,1)$: 
\end{itemize}
\begin{align*}
   &Q(d,\tau,0)\leq Q(d,\tau,1)\\
 \implies &qC_a(p\tau+d)+(1-\alpha)\hbarr(d,\tau+1)\\
 &+\alpha\sum_{i=0}^{\tau}\binom{\tau}{i} p^i(1-p))^{\tau-i}\hbarr(d+i,1)\\
 &\leq qC_a(p\tau)+qC_f+q\alpha \hbarr(0,1)+q(1-\alpha)\hbarr(0,\tau+1)\\
    &+(1-q)(1-\alpha)\hbarr(d,\tau+1)\\
    &+(1-q)\alpha\sum_{i=0}^{\tau}\binom{\tau}{i} p^i(1-p))^{\tau-i}\hbarr(d+i,1)\\
    \implies &qC_ad+q(1-\alpha)\hbarr(d,\tau+1)\\
 &+q\alpha\sum_{i=0}^{\tau}\binom{\tau}{i} p^i(1-p))^{\tau-i}\hbarr(d+i,1)\\
 &\leq qC_f+q\alpha \hbarr(0,1)+q(1-\alpha)\hbarr(0,\tau+1)\\
 \stackrel{(a)}\implies &qC_a(d-\delta)+q(1-\alpha)\hbarr(d-\delta,\tau+1)\\
 &+q\alpha\sum_{i=0}^{\tau}\binom{\tau}{i} p^i(1-p))^{\tau-i}\hbarr(d-\delta+i,1) \\
 &\leq qC_f+q\alpha \hbarr(0,1)+q(1-\alpha)\hbarr(0,\tau+1), \forall \delta  \in [0, d] \\ 
\end{align*}
Inequality (a) holds since $\hbarr(d)$ is non-decreasing in $d$. Therefore, if $Q(d,\tau,0)\leq Q(d,\tau,1)$, then $Q(d-\delta,\tau,0)\leq Q(d-\delta,\tau,1), \forall \delta \in  [0, d]$.

\begin{itemize}
\item Given $Q(d,\tau,0)\leq Q(d,\tau,2)$: 
\end{itemize}

\begin{align*}
   &Q(d,\tau,0)\leq Q(d,\tau,2)\\
 &\implies qC_a(p\tau+d)+(1-\alpha)\hbarr(d,\tau+1)\\
 &+\alpha\sum_{i=0}^{\tau}\binom{\tau}{i} p^i(1-p))^{\tau-i}\hbarr(d+i,1)\\
 &\leq qC_f+qC_f+q\hbarr(0,1)+(1-q)(1-\alpha)\hbarr(d,\tau+1)\\
    &+(1-q)\alpha\sum_{i=0}^{\tau}\binom{\tau}{i} p^i(1-p))^{\tau-i}\hbarr(d+i,1)\\
    &\implies qC_ad+q(1-\alpha)\hbarr(d,\tau+1)\\
 &+q\alpha\sum_{i=0}^{\tau}\binom{\tau}{i} p^i(1-p))^{\tau-i}\hbarr(d+i,1)\\
 &\leq qC_f-qC_a(p\tau)+q\alpha \hbarr(0,1)\\
 &\stackrel{(b)}\implies qC_a(d-\delta)+q(1-\alpha)\hbarr(d-\delta,\tau+1)\\
 &+q\alpha\sum_{i=0}^{\tau}\binom{\tau}{i} p^i(1-p))^{\tau-i}\hbarr(d-\delta+i,1) \\
 &\leq qC_f-qC_a(p\tau)+q\alpha \hbarr(0,1), \forall \delta  \in [0, d] \\ 
\end{align*}
Inequality (b) holds since $\hbarr(d)$ is non-decreasing in $d$. Therefore, if $Q(d,\tau,0)\leq Q(d,\tau,2)$, then $Q(d-\delta,\tau,0)\leq Q(d-\delta,\tau,2), \forall \delta \in  [0, d]$.

Therefore, if $u^{\ast}(d,\tau)=0$, then $u^{\ast}(d-\delta,\tau)=0, \forall \delta \in  [0, d].$
\section{Proof of Lemma 5} \label{Appendix:lemma5}
Given $u^{\ast}(d,\tau)=2$.
This implies $Q(d,\tau,2)\leq Q(d,\tau,1)$ and $Q(d,\tau,2)\leq Q(d,\tau,0)$.
\begin{itemize}
    \item Given $Q(d,\tau,2)\leq Q(d,\tau,1)$: 
\end{itemize}
\begin{align*}
    &Q(d,\tau,2)\leq Q(d,\tau,1)\\
    \implies & qC_f+qC_f+q\hbarr(0,1)+(1-q)(1-\alpha)\hbarr(d,\tau+1)\\
    &+(1-q)\alpha\sum_{i=0}^{\tau}\binom{\tau}{i} p^i(1-p))^{\tau-i}\hbarr(d+i,1)\\
    &\leq qC_a(p\tau)+qC_f+q\alpha \hbarr(0,1)+q(1-\alpha)\hbarr(0,\tau+1)\\
    &+(1-q)(1-\alpha)\hbarr(d,\tau+1)\\
    &+(1-q)\alpha\sum_{i=0}^{\tau}\binom{\tau}{i} p^i(1-p))^{\tau-i}\hbarr(d+i,1)\\
    \implies & qC_f+q(1-\alpha)\hbarr(0,1)\leq qC_a(p\tau)+q(1-\alpha)\hbarr(0,\tau+1)\\
    &\stackrel{(c)}\leq qC_a(p(\tau+\delta))+q(1-\alpha)\hbarr(0,\tau+\delta+1), \forall \delta \geq 0.
\end{align*}
Inequality (c) holds since $\hbarr(d, \tau)$ is non-decreasing in $\tau$. Therefore, if $Q(d,\tau,2)\leq Q(d,\tau,1)$, then $Q(d,\tau+\delta,2)\leq Q(d,\tau+\delta,1), \forall \delta \geq 0$.
\begin{itemize}
    \item Given $Q(d,\tau,2)\leq Q(d,\tau,0)$: 
\end{itemize}
\begin{align*}
     &Q(d,\tau,2)\leq Q(d,\tau,0)\\
      \implies & qC_f+qC_f+q\hbarr(0,1)+(1-q)(1-\alpha)\hbarr(d,\tau+1)\\
    &+(1-q)\alpha\sum_{i=0}^{\tau}\binom{\tau}{i} p^i(1-p))^{\tau-i}\hbarr(d+i,1)\\
    &\leq qC_a(p\tau+d)+(1-\alpha)\hbarr(d,\tau+1)\\
 &+\alpha\sum_{i=0}^{\tau}\binom{\tau}{i} p^i(1-p))^{\tau-i}\hbarr(d+i,1)\\
 \implies & qC_f+qC_f-qC_ad+q\hbarr(0,1)\\
    &\leq q\cbar(\tau)+q(1-\alpha)\hbarr(d,\tau+1)\\
 &+q\alpha\sum_{i=0}^{\tau}\binom{\tau}{i} p^i(1-p))^{\tau-i}\hbarr(d+i,1)\\
 &\stackrel{(d)}\leq q\cbar(\tau+\delta)+q(1-\alpha)\hbarr(d,\tau+\delta+1)\\
 &+q\alpha\sum_{i=0}^{\tau+\delta}\binom{\tau+\delta}{i} p^i(1-p))^{\tau+\delta-i}\hbarr(d+i,1)\\
\end{align*}
Inequality (d) holds since $\hbarr(d, \tau)$ is non-decreasing in $\tau$. Therefore, if $Q(d,\tau,2)\leq Q(d,\tau,0)$, then $Q(d,\tau+\delta,2)\leq Q(d,\tau+\delta,0), \forall \delta \geq 0$.
Therefore, if $u^{\ast}(d,\tau)=2$, then $u^{\ast}(d,\tau+\delta)=2$, $\delta \geq 0$.
\section{Proof of Lemma 6} \label{Appendix:lemma6}

WLOG, $u^{\ast}(d,\tau)=argmin\{Q(d, \tau, 1), Q(d, \tau, 2)\}$.
\begin{align*}
    Q(d,\tau,1)- Q(d,\tau,2)
    &= qC_a(p\tau)+q(1-\alpha)\hbarr(0,\tau+1)\\
     &-(qC_f+q(1-\alpha)\hbarr(0,1))\\
\end{align*}
Clearly, $Q(d,\tau,1)- Q(d,\tau,2) \geq 0$ for some $\tau \geq \taustar$.}

\section{Numerical Results}
\label{sec:num-results}
In this section, we validate our analytical results via simulations. 
\subsection {Optimal Policy }
In this section, we illustrate the optimal policy for VAoI oblivious broadcast discussed in section \ref{sec:opt_policy}. We consider linear age cost as discussed in examples in Section~\ref{sec:opt_policy}, assuming $c_a$ as the coefficient of the cost. 

\begin{figure}[h]
\begin{subfigure}[t]
{0.5\linewidth}
  \centering
\includegraphics[width=\linewidth]{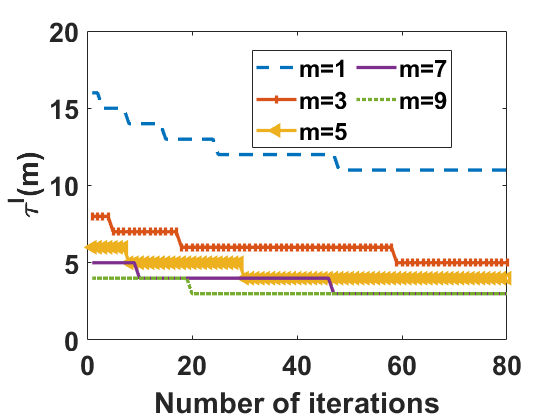}
\caption{${\tau}^{l}(m)$ vs $l$.} 
\label{fig:taustarm_m}
\end{subfigure}
\begin{subfigure}[t]
{0.49\linewidth}
  \centering
\includegraphics[width=\linewidth]{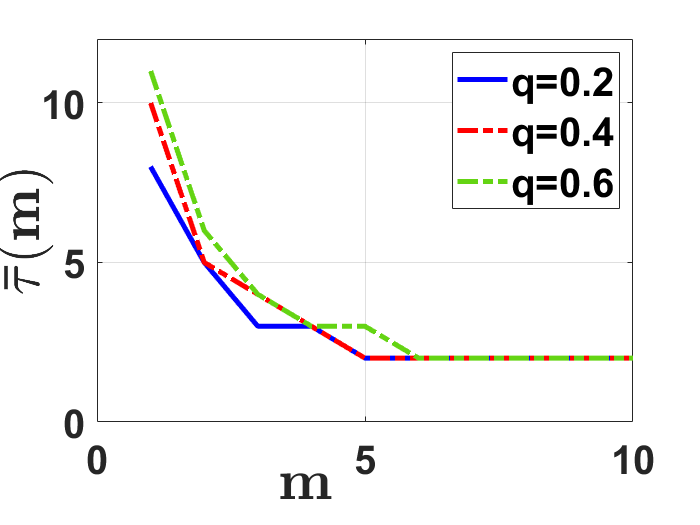}
\caption{$\taustar(m)$ vs. $m$ for various $q$}
\label{fig:taustarm_vs_q}
\end{subfigure}
\caption{ Structure of the optimal policy
}
 \end{figure}

We use the following values for numerical evaluation: $c_a=10$,  $C_f=100$, $q=0.1$ and $p=0.3$, and $N=10$.  Note that, for $m=0$, the optimal policy is to remain idle from Theorem~\ref{TheoremAgg}. Hence, we plot $\taustar(m)$ for $m\in\{1,2,..,10\}$ obtained from Algorithm~\ref{alg:fixed_point}.  In Figure \ref{fig:taustarm_m}, we observe that the convergence of $\tau^{(l)}(m)\to\taustar(m), m\in\{1,2,..,10\}$ happens in a very few iterations. This further establishes that the Algorithm~\ref{alg:fixed_point} performs well without requiring value iteration or truncation of state space.  We also observe from Figure \ref{fig:taustarm_m} that $\taustar(m)$ is non-increasing in $m$ as we establish in Theorem~\ref{TheoremAgg}.  
    
We further illustrate that $\taustar(m)$ is non-increasing in $m$ for various $q$ in Figure \ref{fig:taustarm_vs_q}.{color{blue} As the request rate increases, the age cost also grows, forcing more frequent fetching.} So, the fetching thresholds decrease. For this analysis, we use the same parameters as above and we vary  $q$ from $0.2$ to $0.8$, keeping $p=0.6$. We observe that for any $m$, $\taustar(m)$ is non-decreasing in $q$.  Note that, $\taustar(1)$ depends on both $\theta$  and $q$ (see~\eqref{eq:taustar_1_in_theta}). The effect of increasing $q$ is dominated by the increase in $\theta$. The same follows for $\taustar(m)$ for $m\geq 2$.

\remove{
\begin{figure}[H]
    \centering
    
    \includegraphics[scale=0.4]{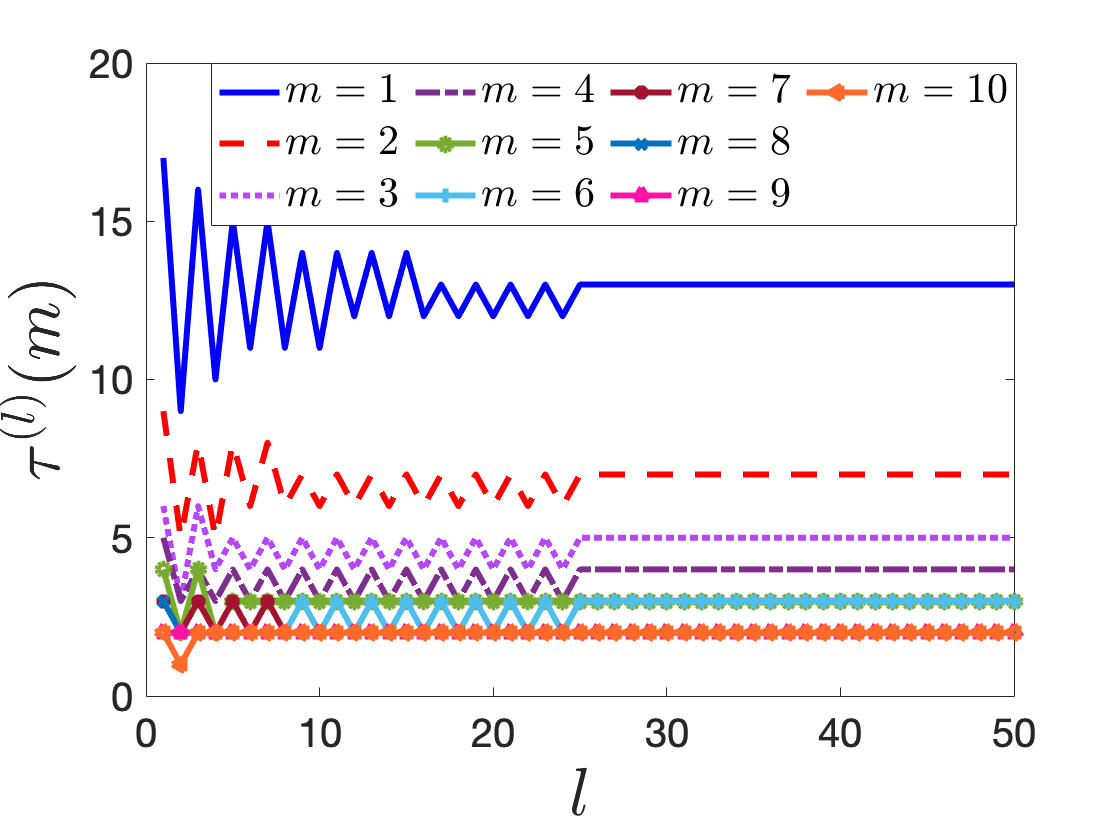}
     
 \caption{$\tau^{(t)}(m)$ vs $t$. We observe that $\tau^{(t)}(m)\to\taustar(m)$. $\taustar(m)$ is reported as $\taustar(1)=13$, $\taustar(2)=7$,$\taustar(3)=5$,$\taustar(4)=4$,$\taustar(m)=3, m \in \{5,6,7,8\}$ and $\taustar(9)=\taustar(10)=2$.}
    \label{fig:Taustarm}
\end{figure}

\begin{figure}[H]
    \centering
    
    \includegraphics[scale=0.4]{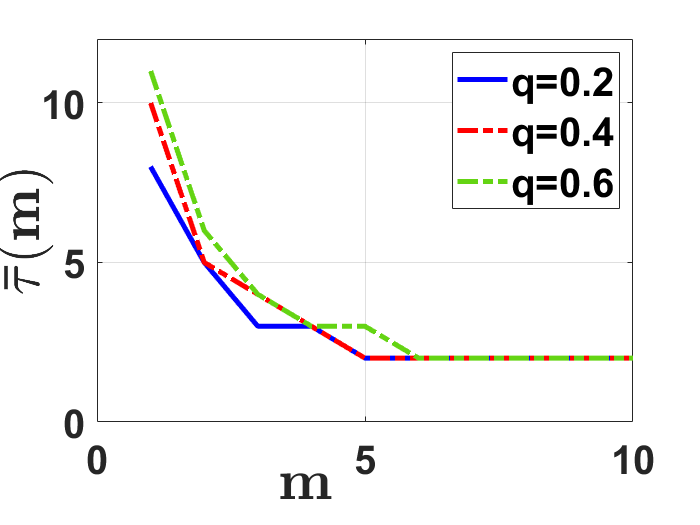}
     
 \caption{$\taustar(m)$ vs $m$ for various $q$.}
    \label{fig:Taustarm_q}
\end{figure}
}
\remove{\subsection {Illustration of Algorithm \ref{alg:fixed_point}}
Figure \ref{fig:ftheta-theta} illustrates how $f(\theta^{(0)})$ and $\theta^{(1)}$ varies with $\theta^{(0)}$. We use the following values: $Ca=10$, $Ctx=200$,$\alpha=0.1$, $p=0.7$ and $q=0.2$. In \ref{fig:ftheta-theta}, we observe that $f(\theta^{(0)})$ is non-increasing in $\theta^{(0)}$, confirming the property was proved in \ref{theorem: theoremA3}. We also observe that, for $\alpha=0.1$, $\theta^{(1)}$ increases with $\theta^{(0)}$.}
\subsection{Properties of the optimal policy }
\begin{figure}[h]
\begin{subfigure}[t]
{0.5\linewidth}
  \centering
\includegraphics[width=\linewidth]{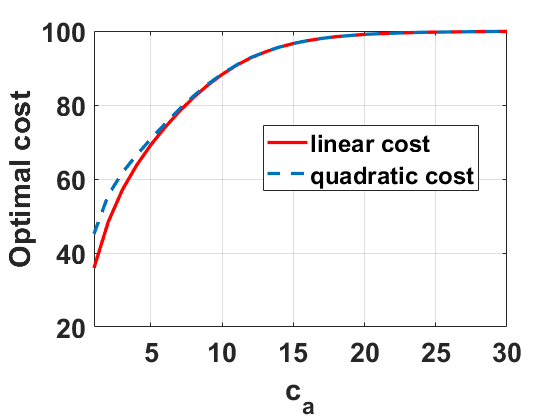}
\caption{Optimal cost vs. $c_a$}
\label{fig:theta_vs_ca}
\end{subfigure}
\begin{subfigure}[t]
{0.49\linewidth}
  \centering
\includegraphics[width=\linewidth]{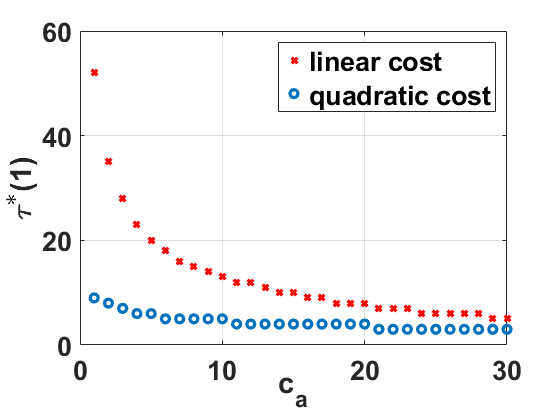}
\caption{$\taustar(1)$ vs. $c_a$}
\label{fig:taustar_vs_ca}
\end{subfigure}
\caption{ Variation in the optimal cost due to age cost
}
 \end{figure}

\begin{figure}[h]
\begin{subfigure}[t]
{0.5\linewidth}
  \centering
\includegraphics[width=\linewidth]{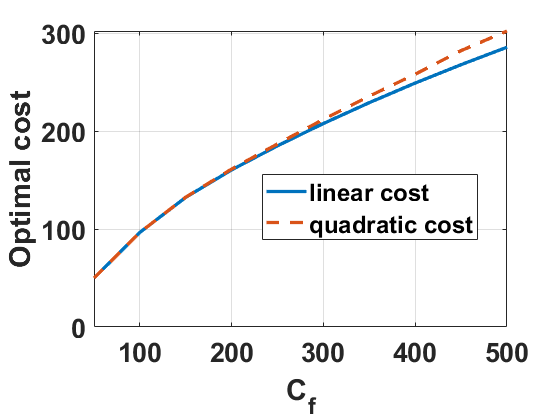}
\caption{Optimal cost vs. $C_f$}
\label{fig:theta_vs_cftx}
\end{subfigure}
\begin{subfigure}[t]
{0.49\linewidth}
  \centering
\includegraphics[width=\linewidth]{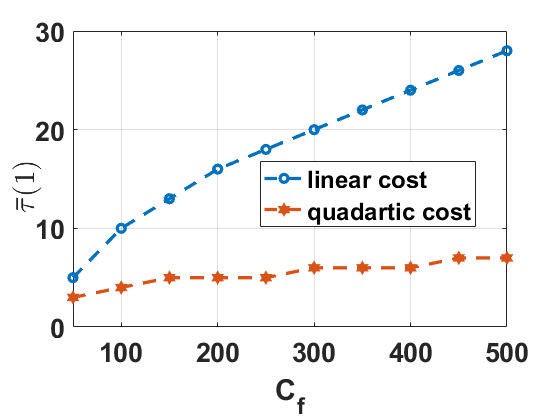}
\caption{$\taustar(1)$ vs. $C_f$}
\label{fig:taustar_vs_ctx}
\end{subfigure}
\caption{ Variation in the optimal cost due to $C_f$
}
 \end{figure}

In this subsection, we investigate how the optimal costs vary with  $c_a$, $C_f$ and $p$. We consider $N=100$, $q=0.12$ and $\alpha=0.1$. We consider two different cost structures, i.e., linear and quadratic (see examples in~\ref{sec:opt_policy}). 
\paragraph{Effect of variation in the age cost coefficient $c_a$}
We fix $p=0.7$ and $C_f=100$ and vary $c_a$ from $1$ to $30$.
 We observe in Figure~\ref{fig:theta_vs_ca} that the optimal cost increases with $c_a$. For lower values of $c_a$, the optimal cost increases rapidly and then saturates as $c_a$ exceeds $20$. This is because, beyond $20$, the age costs at $\taustar(1)$ for linear and quadratic cases become very close to $C_f$. As $c_a$ increases, $\taustar(1)$ decreases (see Figure~\ref{fig:taustar_vs_ca} ), and this further implies that the optimal policy is more inclined towards fetching and transmitting the content. Finally, the optimal cost reaches $C_f=100$ as the maximum value of the optimal cost is $C_f$, and in this case, the optimal policy always fetches and transmits.   
\paragraph{Effect of variation in the transmission cost $C_f$}
We fix $c_a=15$, $p=0.7$ and vary $C_f$ from $50$ to $500$.
  As $C_f$ increases, the optimal cost increases (see Figure~\ref{fig:theta_vs_cftx}) and $\taustar(1)$ (see Figure~\ref{fig:taustar_vs_ctx}) increases as the policy becomes less inclined towards fetching and transmitting. Fix a $C_f$, we observe that the age cost in the quadratic case is higher than in the linear case. As a result, the optimal policy in the quadratic case tends to favor fetching and transmitting content more aggressively compared to the linear case. This gap becomes more pronounced as $C_f$ increases. Consequently, the optimal cost in the quadratic case exceeds that of the linear case, and the gap becomes larger as $C_f$ increases. 
 \subsection{Performance of the Whittle index based policy}
 In this section, we consider $C_{av}(\tau)=C_a\tau$ and compare the average cost of the Whittle index based policy (WI) with the optimal policy. 
 \subsubsection{Homogeneous users} 
\begin{figure}[h]
\begin{subfigure}[t]
{0.5\linewidth}
  \centering
\includegraphics[width=\linewidth]{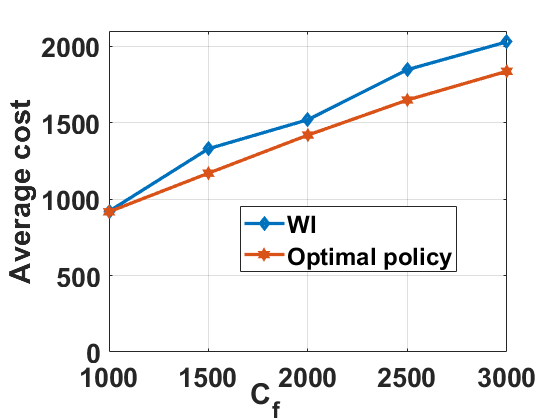}
\caption{Average cost vs. $C_f$}
\label{fig:WI_avg}
\end{subfigure}
\begin{subfigure}[t]
{0.49\linewidth}
  \centering
\includegraphics[width=\linewidth]{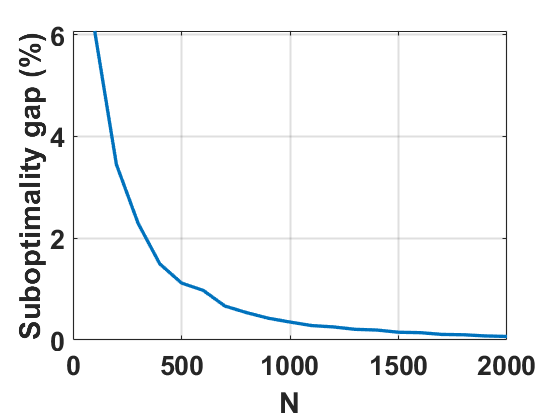}
\caption{Sub optimality gap vs $N$}
\label{fig:percent_error}
\end{subfigure}
\caption{ Performance of WI policy for homogeneous users
}
 \end{figure}

We consider $N=1000$, $      q=0.12,p=0.7,C_a=10$ and vary $C_f$ from $500$ to $3000$. The average costs of both policies increase with $C_f$ (see Figure~\ref{fig:WI_avg}). The WI policy performs close to the optimal policy, and the optimality gap is at most~$10$ percent (see Figure~\ref{fig:WI_avg}). 

 We also show that the WI policy is asymptotically optimal for homogeneous users. For this we vary $N$ from $100$ to $2000$ and $C_f=N$. We show that the percentage difference reduces to zero as we increase $N$ from $100$ to $2000$ (see Figure~\ref{fig:percent_error}). 
 \subsubsection{Heterogeneous users} 
\begin{figure}[h]
\begin{subfigure}[t]
{0.5\linewidth}
  \centering
\includegraphics[width=\linewidth]{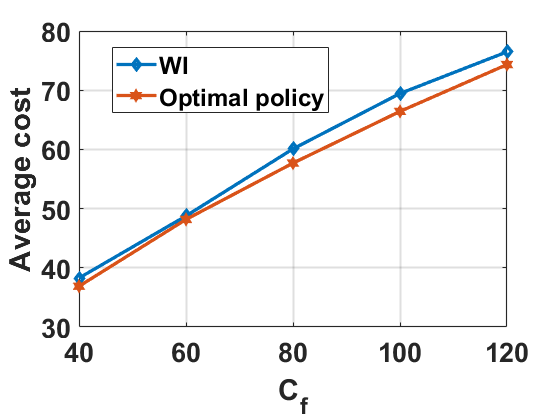}
\caption{Average cost vs. $C_f$}
\label{fig:avg_cst_vs_ctx}
\end{subfigure}
\begin{subfigure}[t]
{0.49\linewidth}
  \centering
\includegraphics[width=\linewidth]{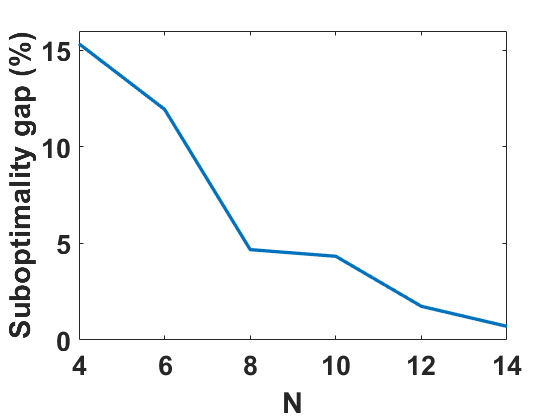}
\caption{Sub optimality gap vs. $N$}
\label{fig:subopt_vs_N}
\end{subfigure}
\caption{ Performance of WI policy for heterogeneous users
}
 \end{figure}

Since the the complexity of the optimal policy is exponential $(2^N)$, we restrict the number of users to $10$. We consider three classes of users as class A, class B, and class C. Users of each class have the same request rate and cost. We consider $q_A=0.12$, $q_B=0.3$ and $q_C=0.9$ and $C_{a_A}=15,C_{a_B}=13$ and $C_{a_C}=10$.     In Figure~\ref{fig:avg_cst_vs_ctx}, we fix $p=0.7$ and vary $C_f$ from $40$ to $120$. Average costs of both policies increase with $C_f$. The WI policy performs very close to the optimal policy despite the small number of users (see Figure~\ref{fig:avg_cst_vs_ctx}).

 We fix $C_f = 100$ and vary $N$ from $4$ to $14$, considering two classes $A$ and $B$, each with an equal number of users. The average costs of the WI and optimal policies are plotted in Figure~\ref{fig:subopt_vs_N}.
 As $N$ increases, the sub-optimality gap reduces. 
 \remove{
 \subsection{Performance of VAoI Aware Policy}
 In this section, we compare the average cost of VAoI oblivious broadcast and VAoI aware broadcast under Algorithm~\ref{alg:fixed_point} and the proposed heuristic, respectively. For this,
 we consider linear cost and $c_a=10,\,p=0.3,\,q=0.12,\, C_f=80,\text{ and }N=100$. 
 We vary $C_{tx}$ from $0$ to $270$ and plot the average cost under VAoI oblivious optimal policy in which $C_f=C_f+C_{tx}$ and VAoI aware heuristic. Observe that, if there are no transmission costs, the heuristic will always transmit and hence for $C_{tx}=0$, the average costs under these policies are identical (see Figure~\ref{fig:comp_vary_ctx}). But as $C_{tx}$ increases, the heuristic becomes less inclined towards transmitting after fetching and observing the VAoI while the VAoI oblivious optimal policy always transmits whenever it fetches. Hence, the gap between these costs increases (see Figure~\ref{fig:comp_vary_ctx}).

  In Figure~\ref{fig:comp_vary_p}, we also compare the performance of the heuristic with the Genie-aided update aware policy in which the optimal cost of the latter is a lower bound of the optimal cost of the VAoI aware broadcast~\eqref{eq:original_problem}. We consider $c_a=10,\,q=0.12,\, C_f=80,\,C_{tx}=120\text{ and }N=100$ and vary $p$ from $0.5$ to $1$. The  VAoI for $p=1$ at any time is the same as the time elapsed since the last fetch. Hence, in all three policies, the SCSP is aware of the VAoI without fetching it.  As we increase $p$ from $0.5$ to $1$, the VAoI becomes closer to the time elapsed since the last fetch, and hence the gap between the average costs of Genie-aided update aware policy and VAoI aware broadcast policy reduces. This implies that for higher values of $p$, the proposed heuristic performs really well. 
 \begin{figure}[h]
\begin{subfigure}[t]
{0.5\linewidth}
  \centering
\includegraphics[width=\linewidth]{comp_heuristic_new.png}
\caption{Average cost vs. $C_{tx}$}
\label{fig:comp_vary_ctx}
\end{subfigure}
\begin{subfigure}[t]
{0.49\linewidth}
  \centering
\includegraphics[width=\linewidth]{c_tx_60_comp_new.png}
\caption{Average cost vs. $p$}
\label{fig:comp_vary_p}
\end{subfigure}
\caption{ Performance of the proposed heuristic
}
 \end{figure}
 \remove{\paragraph{Effect of variation in the update rate $p$}
 {\color{blue} $C_a=10, C_tx=20, p=0.05$ to$ 0.6$}
Figure \ref{subfig:AvgerageCost_P} illustrates the variation of optimal average cost and optimal policy when the number of requests is $1$ with respect to $p$. We use $C_a=100$ and $C_f=100$. As $p$ increases, the optimal average cost increases. This is because, when $p$ is low, the probability of content updation is low and hence the age cost is low. Therefore, the SCSP does not have the incentive to fetch and broadcast the fresh content. We observe the same trend in \ref{subfig:AvgerageCost_P}.
\begin{figure}[h]
\begin{subfigure}[t]
{0.5\linewidth}
  \centering
\includegraphics[width=\linewidth]{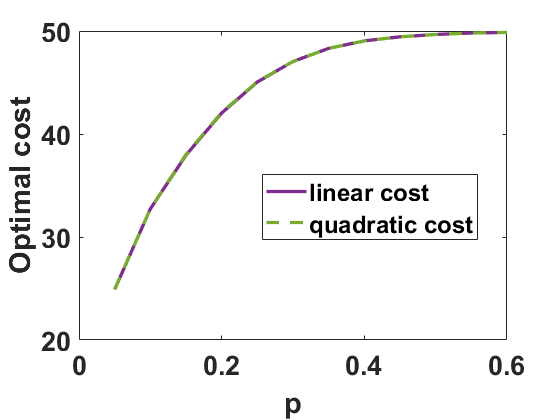}
\caption{}
\label{}
\end{subfigure}
\begin{subfigure}[t]
{0.49\linewidth}
  \centering
\includegraphics[width=\linewidth]{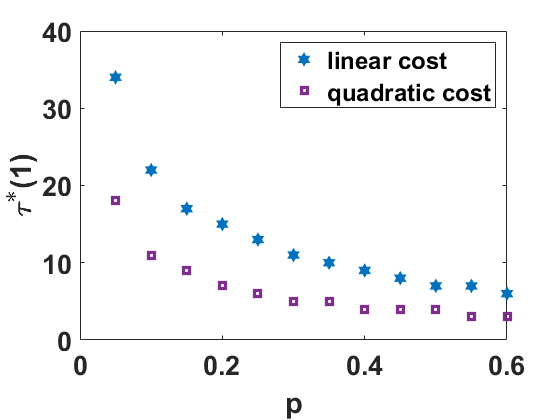}
\caption{}
\label{}
\end{subfigure}
\caption{ 
}
 \end{figure}

 } 
\color{black}
\remove{
\begin{figure}[h]
    \centering
    
    \includegraphics[scale=0.4]{ftheta-theta.eps}
     
 \caption{$f(\theta)$ and $\theta$ vs $\theta$.} 
    \label{fig:ftheta-theta}
\end{figure}

\begin{figure*}[h]   
\begin{subfigure}[t]{.3\textwidth}
    \centering
    \includegraphics[width=\linewidth, height=4.2cm]{thetataustar1Ca.eps}
     \caption{$\theta \text{ and } \taustar(1) \text{ vs } C_a$}\label{subfig:AvgerageCost_Ca}
\end{subfigure}
\hfill
\begin{subfigure}[t]{.3\textwidth}
    \centering
    \includegraphics[width=\linewidth, height=4.2cm]{thetataustar1Ctx.eps}
     \caption{$\theta \text{ and } \taustar(1) \text{ vs } C_f$}\label{subfig:AvgerageCost_Ctx}
\end{subfigure}
\hfill
\begin{subfigure}[t]{.3\textwidth}
    \centering
    \includegraphics[width=\linewidth, height=4.2cm]{thetataustar1P.eps}
     \caption{$\theta \text{ and } \taustar(1) \text{ vs } p$}\label{subfig:AvgerageCost_P}
\end{subfigure}
\remove{
\begin{subfigure}[t]{.35\textwidth}
    \centering
    \includegraphics[width=\linewidth, height=6.2cm]{thetaP.eps}
     \caption{$p$}\label{subfig:AvgerageCost_P}
\end{subfigure}
}
\caption{Variation of the optimal average cost $\theta$ with respect to different system parameters. Each plot shows how $\theta$ evolves as the respective parameter changes, indicating policy transitions and cost sensitivity.}
    \label{fig:Average Cost}
\end{figure*}

\begin{figure}[H]
    \centering
    
    \includegraphics[scale=0.4]{PerformanceEvaluation_Ca1.eps}
     
 \caption{Performance Evaluation for various values of $C_{a,1}$. We use $q_1=0.1$ ; $C_{a,2}=30$ and  $q_2=0.2$; $C_{a,3}=40$ and  $q_3=0.3$. The fetching cost $C_f=100$, transmission cost is $C_{tx,i}=100, i\in\{1,2,3\}$ and content update probability is $p=0.7$. }
    \label{fig:PerformanceCa1}
\end{figure}
\begin{figure}[H]
    \centering
    
    \includegraphics[scale=0.4]{PerformanceEvaluation_Ctx1.eps}
     
 \caption{Performance Evaluation for various values of $C_{tx,1}$. We use $C_{a,1}=10$ and $q_1=0.1$; $C_{a,2}=30$ and $q_2=0.2$; $C_{a,3}=40$ and $q_3=0.3$. The fetching cost is $C_f=100$ and the content update probability is $p=0.7$. }
    \label{fig:PerformanceCtx1}
\end{figure}
}
\remove{
\subsection{Performance evaluation of Optimal Policy of Multi user system}

We analyze the performance of the proposed policy in this section. We consider a candidate policy obtained in the following fashion. We fix the parameters $c_a$, $C_f$, $C_f$ , $p$ and $q$.
We follow Algorithm 1 to obtain $\alpha_i^1$. We use these $\alpha_i^1$ along with the aforementioned parameters to compute the optimal cost of user $i$. 

We then show that the total optimal cost of all users under our proposed policy to be less than the 
total optimal cost under the candidate policy.}
}
\section{Conclusion}
We considered a content fetching and broadcast problem to minimize the average fetching and age costs in an S2aaS system. We first investigated a homogeneous problem and obtained its optimal policy. We extended this algorithm for the heterogeneous case owing to exponential computational complexity. To tackle this, we developed a low complexity algorithm applicable to both heterogeneous and homogeneous cases, achieving near optimal performance. 

In future, we would like to extend our S2aaS system model to have multiple sensors. We would also like to consider wireless channel failures and transmission power control.

\remove{Looking ahead, our future endeavors will focus on a more generalized setting, where the content request probability is not necessarily close to zero. }
\bibliographystyle{IEEEtran}
\bibliography{main}

@inproceedings{ross,
  title={Stochastic Processes},
  author={Ross, Sheldon M.},
  booktitle={Wiley Series},
  pages={405--418},
  year={2016},
  organization={Wiley}
}

@INPROCEEDINGS{Karevvanavar,
  author={Karevvanavar, Gangadhar and Pable, Hrishikesh and Patil, Om and Bhat, Rajshekhar V and Pappas, Nikolaos},
  booktitle={2024 22nd International Symposium on Modeling and Optimization in Mobile, Ad Hoc, and Wireless Networks (WiOpt)}, 
  title={Version Age of Information Minimization Over Fading Broadcast Channels}, 
  year={2024},
  volume={},
  number={},
  pages={170-176},
  doi={}}

@article{Yates,
  author       = {Roy D. Yates},
  title        = {The Age of Gossip in Networks},
  journal      = {CoRR},
  volume       = {abs/2102.02893},
  year         = {2021},
  url          = {https://arxiv.org/abs/2102.02893},
  eprinttype    = {arXiv},
  eprint       = {2102.02893},
  bibsource    = {dblp computer science bibliography, https://dblp.org}
}

@INPROCEEDINGS{PappasSensor,
  author={Stamatakis, George and Pappas, Nikolaos and Traganitis, Apostolos},
  booktitle={2019 IEEE Globecom Workshops (GC Wkshps)}, 
  title={Control of Status Updates for Energy Harvesting Devices That Monitor Processes with Alarms}, 
  year={2019},
  volume={},
  number={},
  pages={1-6},
  doi={10.1109/GCWkshps45667.2019.9024463}}

@INPROCEEDINGS{Yehinfocom,
  author={Abolhassani, Bahman and Tadrous, John and Eryilmaz, Atilla and Yeh, Edmund},
  booktitle={IEEE INFOCOM '21 - Proceedings of the IEEE Conference on Computer Communications}, 
  title={Fresh Caching for Dynamic Content}, 
  year={2021},
  volume={},
  number={},
  pages={1--10},
  doi={10.1109/INFOCOM42981.2021.9488731}}

@ARTICLE{GunduzSensor,
  author={Gunduz, Deniz and Stamatiou, Kostas and Michelusi, Nicolo and Zorzi, Michele},
  journal={IEEE Communications Magazine}, 
  title={Designing intelligent energy harvesting communication systems}, 
  year={2014},
  volume={52},
  number={1},
  pages={210-216},
  doi={10.1109/MCOM.2014.6710085}}

@article{BacinogluSensor,
  title={Achieving the Age-Energy Tradeoff with a Finite-Battery Energy Harvesting Source},
  author={Baran Tan Bacinoglu and Yin Sun and Elif Uysal-Biyikoglu and Volkan Mutlu},
  journal={2018 IEEE International Symposium on Information Theory (ISIT)},
  year={2018},
  pages={876-880},
  url={https://api.semanticscholar.org/CorpusID:3623747}
}

@INPROCEEDINGS{ArafaSensor,
  author={Arafa, Ahmed and Yang, Jing and Ulukus, Sennur},
  booktitle={2018 IEEE International Conference on Communications (ICC)}, 
  title={Age-Minimal Online Policies for Energy Harvesting Sensors with Random Battery Recharges}, 
  year={2018},
  volume={},
  number={},
  pages={1-6},
  doi={10.1109/ICC.2018.8422086}}

@ARTICLE{BaranSensor,
  author={Bacinoglu, Baran Tan and Sun, Yin and Uysal, Elif and Mutlu, Volkan},
  journal={Journal of Communications and Networks}, 
  title={Optimal status updating with a finite-battery energy harvesting source}, 
  year={2019},
  volume={21},
  number={3},
  pages={280-294},
  doi={10.1109/JCN.2019.000033}}

@article{perera2018designing,
  title={Designing the sensing as a service ecosystem for the internet of things},
  author={Perera, Charith and Barhamgi, Mahmoud and De, Suparna and Baarslag, Tim and Vecchio, Massimo and Choo, Kim-Kwang Raymond},
  journal={IEEE Internet of Things Magazine},
  volume={1},
  number={2},
  pages={18--23},
  year={2018},
  publisher={IEEE}
}

@book{BertsekasVol2, author = {Bertsekas, Dimitri P.}, title = {Dynamic Programming and Optimal Control, Vol. II}, year = {2007}, isbn = {1886529302}, publisher = {Athena Scientific}, edition = {3rd} }

@ARTICLE{Yeh,
  author={Abolhassani, Bahman and Tadrous, John and Eryilmaz, Atilla and Yeh, Edmund},
  journal={IEEE/ACM Transactions on Networking}, 
  title={Fresh Caching of Dynamic Content Over the Wireless Edge}, 
  year={2022},
  volume={30},
  number={5},
  pages={2315--2327},
  doi={10.1109/TNET.2022.3170245}}

@inproceedings{delfaniversion,
  author       = {Erfan Delfani and
                  Nikolaos Pappas},
  title        = {Version Age-Optimal Cached Status Updates in a Gossiping Network with
                  Energy Harvesting Sensor},
  booktitle    = {WiOpt '23: Proceedings of the International Symposium on Modeling and Optimization in Mobile,
                  Ad Hoc, and Wireless Networks},
  pages        = {143--150},
  year         = {2023},
  doi          = {10.23919/WIOPT58741.2023.10349895},
}

@INPROCEEDINGS{AchieveFreshness,
  author={Abolhassani, Bahman and Tadrous, John and Eryilmaz, Atilla},
  booktitle={WiOpt '20: Proceedings of the  18th International Symposium on Modeling and Optimization in Mobile, Ad Hoc, and Wireless Networks}, 
  title={Achieving Freshness in Single/Multi-User Caching of Dynamic Content over the Wireless Edge}, 
  year={2020},
  volume={},
  number={},
  pages={1--8},
  doi={}}

@INPROCEEDINGS{Papadimitriou,
  author={Papadimitriou, C.H. and Tsitsiklis, J.N.},
  booktitle={IEEE SCT '94: Proceedings of IEEE 9th Annual Conference on Structure in Complexity Theory}, 
  title={The complexity of optimal queueing network control}, 
  year={1994},
  volume={},
  number={},
  pages={318--322},
  doi={10.1109/SCT.1994.315792}}

@MISC{AnuK:2024,
author = {Krishna, Anu and Koley, Ankita and Singh, Chandramani},
title = {Version Age Optimal Content Update and
Transmission in an Edge Caching System},
howpublished={\url{https://tinyurl.com/2dfpb43s}}

}

@ARTICLE{YatesSurvey,
  author={Yates, Roy D. and Sun, Yin and Brown, D. Richard and Kaul, Sanjit K. and Modiano, Eytan and Ulukus, Sennur},
  journal={IEEE Journal on Selected Areas in Communications}, 
  title={Age of Information: An Introduction and Survey}, 
  year={2021},
  volume={39},
  number={5},
  pages={1183-1210},
  doi={10.1109/JSAC.2021.3065072}}

@INPROCEEDINGS{9517796,

  author={Yates, Roy D.},

  booktitle={IEEE ISIT '21: Proceedings of the IEEE International Symposium on Information Theory}, 

  title={The Age of Gossip in Networks}, 

  year={2021},

  volume={},

  number={},

  pages={2984--2989},

  doi={10.1109/ISIT45174.2021.9517796}}

@article{bertsekas2011dynamic,
  title={Dynamic programming and optimal control 3rd edition, volume ii},
  author={Bertsekas, Dimitri P and others},
  journal={Belmont, MA: Athena Scientific},
  volume={1},
  year={2007}
}

@ARTICLE{8481588,

  author={Chakraborty, Aishwariya and Mondal, Ayan and Roy, Arijit and Misra, Sudip},

  journal={IEEE Transactions on Services Computing}, 

  title={Dynamic Trust Enforcing Pricing Scheme for Sensors-as-a-Service in Sensor-Cloud Infrastructure}, 

  year={2021},

  volume={14},

  number={5},

  pages={1345--1356},

  doi={10.1109/TSC.2018.2873763}}

@ARTICLE{4237146,

  author={Kumar, Anurag and Altman, Eitan and Miorandi, Daniele and Goyal, Munish},

  journal={IEEE/ACM Transactions on Networking}, 

  title={New Insights From a Fixed-Point Analysis of Single Cell IEEE 802.11 WLANs}, 

  year={2007},

  volume={15},

  number={3},

  pages={588-601},

  doi={10.1109/TNET.2007.893091}}

@ARTICLE{9945983,

  author={Dong, Fuwang and Liu, Fan and Cui, Yuanhao and Wang, Wei and Han, Kaifeng and Wang, Zhiqin},

  journal={IEEE Transactions on Wireless Communications}, 

  title={Sensing as a Service in 6G Perceptive Networks: A Unified Framework for ISAC Resource Allocation}, 

  year={2023},

  volume={22},

  number={5},

  pages={3522-3536},

  doi={10.1109/TWC.2022.3219463}}

@ARTICLE{10387517,

  author={Dong, Fuwang and Liu, Fan and Cui, Yuanhao and Lu, Shihang and Li, Yunxin},

  journal={IEEE Network}, 

  title={Sensing as a Service in 6G Perceptive Mobile Networks: Architecture, Advances, and the Road Ahead}, 

  year={2024},

  volume={38},

  number={2},

  pages={87-96},

  doi={10.1109/MNET.2024.3352092}}

@INPROCEEDINGS{area_service_24,

  author={Corici, Marius and Eichhorn, Fabian and Buhr, Hauke and Troudt, Eric and Magedanz, Thomas},

  booktitle={2024 IEEE Conference on Standards for Communications and Networking (CSCN)}, 

  title={Area Services: A Novel Service Class for B5G \& 6G Mobile Networks}, 

  year={2024},

  volume={},

  number={},

  pages={224-229},

  doi={10.1109/CSCN63874.2024.10849759}}

@inproceedings{voi_19,
author = {Ayan, Onur and Vilgelm, Mikhail and Kl\"{u}gel, Markus and Hirche, Sandra and Kellerer, Wolfgang},
title = {Age-of-information vs. value-of-information scheduling for cellular networked control systems},
year = {2019},
isbn = {9781450362856},
publisher = {Association for Computing Machinery},
address = {New York, NY, USA},
url = {https://doi.org/10.1145/3302509.3311050},
doi = {10.1145/3302509.3311050},
booktitle = {Proceedings of the 10th ACM/IEEE International Conference on Cyber-Physical Systems},
pages = {109–117},
numpages = {9},
location = {Montreal, Quebec, Canada},
series = {ICCPS '19}
}

@ARTICLE{8606228,

  author={Bharti, Sourabh and Pattanaik, K. K. and Bellavista, Paolo},

  journal={IEEE Transactions on Emerging Topics in Computing}, 

  title={Value of Information Based Sensor Ranking for Efficient Sensor Service Allocation in Service Oriented Wireless Sensor Networks}, 

  year={2021},

  volume={9},

  number={2},

  pages={823-838},

  doi={10.1109/TETC.2019.2891716}}

@article{reliable_broadcast,
author = {Zribi, Amin and Pyndiah, Ramesh Mahendra and Saoudi, Samir and Lagrange, Xavier},
title = {Erasure coding for reliable adaptive retransmission in wireless broadcast/multicast systems},
year = {2017},
issue_date = {August    2017},
publisher = {Kluwer Academic Publishers},
address = {USA},
volume = {65},
number = {4},
issn = {1018-4864},
url = {https://doi.org/10.1007/s11235-016-0253-4},
doi = {10.1007/s11235-016-0253-4},
journal = {Telecommun. Syst.},
month = aug,
pages = {591–604},
numpages = {14}
}

@ARTICLE{11202652,

  author={Koley, Ankita and Singh, Chandramani},

  journal={IEEE Transactions on Networking}, 

  title={Fresh Caching of Dynamic Contents Using Restless Multi-Armed Bandits Over Wireless Access}, 

  year={2025},

  volume={},

  number={},

  pages={1-13},

  doi={10.1109/TON.2025.3616225}}

@article{whittle1988restless,
  title={Restless bandits: Activity allocation in a changing world},
  author={Whittle, Peter},
  journal={Journal of applied probability},
  volume={25},
  number={A},
  pages={287--298},
  year={1988},
  publisher={Cambridge University Press}
}

@book{zhao2022multi,
  title={Multi-armed bandits: Theory and applications to online learning in networks},
  author={Zhao, Qing},
  year={2022},
  publisher={Springer Nature}
}

@article{ALMASAEID2025103716,
title = {Reliable and cost-efficient session provisioning in CRNs using spectrum sensing as a service},
journal = {Ad Hoc Networks},
volume = {168},
pages = {103716},
year = {2025},
issn = {1570-8705},
doi = {https://doi.org/10.1016/j.adhoc.2024.103716},
url = {https://www.sciencedirect.com/science/article/pii/S1570870524003275},
author = {Hisham M. Almasaeid},
}

@ARTICLE{9773312,
  author={Mondal, Anudipa and Misra, Sudip and Das, Goutam},
  journal={IEEE Transactions on Green Communications and Networking}, 
  title={CALM: QoS-Aware Vehicular Sensor-as-a-Service Provisioning in Cache-Enabled Multi-Sensor Cloud}, 
  year={2022},
  volume={6},
  number={3},
  pages={1564-1573},
  doi={10.1109/TGCN.2022.3174734}}

@InProceedings{Anu_EAI_valuetools,
author="Krishna, Anu
and Koley, Ankita
and Singh, Chandramani",
editor="Gribaudo, Marco
and Iacono, Mauro
and Sedigh Sarvestani, Sahra",
title="Version Age Optimal Content Update and Transmission in an Edge Caching System",
booktitle="Performance Evaluation Methodologies and Tools",
year="2026",
publisher="Springer Nature Switzerland",
address="Cham",
pages="323--344",
isbn="978-3-032-06818-7"
}
\appendices
\remove{
We use relative value iteration algorithm to prove this lemma. Let us define state action function at $k^{th}$  iteration $Q_k(\tau,a), a\in\{0,1\}$ as follows.
\begin{align*}
    Q_{k+1}(\tau,0)=qC_{a}\tau p+h_{k}(\tau+1)
\end{align*}
\begin{align*}
    Q_{k+1}(\tau,1)=q(C_{f}+C_f)+qh_k(1)+(1-q)h_k(\tau+1)
\end{align*}
\begin{align*}
    h_{k+1}(\tau)=&\min\{Q_{k+1}(\tau,0),Q_{k+1}(\tau,1)\nonumber\\
    &-\min\{Q_{k+1}(j,0),Q_{k+1}(j,1)\}
\end{align*}
From Proposition 4.3.2 \cite{BertsekasVol2}, $\lim_{k \to \infty}h_k(\tau)\to \hbarr(\tau), \forall\tau$. Therefore, it is sufficient to show that $h_{k}(\tau)$ is non decreasing in $\tau, \forall k \in \{0,1,2,...\}$.

Let $h_0(\tau)=0,  \forall \tau$. Therefore, $Q_{1}(\tau,0)=qC_{a}\tau p$ and $Q_{1}(\tau,1)=q(C_{f}+C_f)$. Clearly, $Q_1(\tau,a)$ is non decreasing in $\tau$. Suppose $Q_t(\tau,a)$ is non decreasing in $\tau, \forall k\leq K$.
It is easy to see that $Q_{K+1}(\tau,0)$ and $Q_{K+1}(\tau,1)$ are non decreasing in $\tau$. Therefore $h_{K+1}(\tau)$ is non decreasing in $\tau$.
}
\remove{
\section{Proof Theorem 1}
\label{Appendix:Theorem1}

Recall, $\hbarr(\tau)+\theta=\min\{q(C_{f}+C_f)+q\hbarr(1)+(1-q)\hbarr(\tau+1),qC_{a}\tau p+\hbarr(\tau+1)\}$. Since, $\hbarr(\tau)$ is non decreasing in $\tau$, there exists a threshold $\taustar$ such that $u^{\pi}(\tau)=1$ for all $\tau \geq \taustar$. Let $\tau_{opt}^{\ast}$ be the optimal threshold.
For $\tau \geq \taustar$,
\begin{align*}
 & \hbarr(\tau)=q(C_{f}+C_f)+q\hbarr(1)-\theta+(1-q)\hbarr(\tau+1) \nonumber \\
    \implies &\hbarr(\tau)=\frac{q(C_{f}+C_f)+q\hbarr(1)-\theta}{q} \label{eqn:hs1}
\end{align*}
For $\tau < \taustar$, $\hbarr(\tau)=qC_{a}\tau p-\theta+\hbarr(\tau+1).$

From the above equation, we see that 
\[\hbarr(1)=C_{a}pq-\theta+\hbarr(2) \label{eqn:h1}
\]

Without loss of generality, $\hbarr(1)=0 {\implies} \hbarr(2)=\theta-C_apq $.
\begin{align}
\label{eqn:h_2}
     \hbarr(2)=2C_{a}pq-\theta+\hbarr(3).
\end{align}
We substitute $\hbarr(2)$ in~\eqref{eqn:h_2} to yield
\begin{align*}
     \hbarr(1)=C_{a}pq(1+2)-2\theta+\hbarr(3).
\end{align*}
Repeating this procedure $\taustar-2$ times yields the following 
\begin{align}
\label{eqn: hs0}
     &\hbarr(1)=C_{a}pq(1+2+..+\taustar-1)-(\taustar-1)\theta+\hbarr(\taustar)\nonumber\\
     &=\frac{C_{a}pq(\taustar-1)\taustar}{2}-(\taustar-1)\theta+\hbarr(\taustar)  
\end{align}
Substituting $ \hbarr(\taustar)=\frac{q(C_{f}+C_f)+q\hbarr(1)-\theta}{q}$ and $\hbarr(1)=0$ in~\eqref{eqn: hs0}, we obtain $\theta$ as given below:
\begin{align}
\label{eqn: theta_s}
 \theta=\frac{C_{a}pq^{2}(\taustar-1)\taustar+2q(C_{f}+C_f)}{2(q(\taustar-1)+1)}
\end{align}
Optimal threshold value $\taustar_{opt}$ is the threshold that minimise $\theta$ in~\eqref{eqn: theta_s}.
\begin{align*}
    \taustar_{opt}=\argmin_{\taustar\in \mathbb Z_{+}}\theta
\end{align*}
}
\remove{
\section{Proof of Lemma 2} 
\label{Appendix: Lemma 2}
We employ relative value iteration algorithm in a similar fashion that we used to prove Lemma 1. Details of the proof can be found in \cite{AnuK:2024}.
}
\remove{
\begin{align*}
    Q_{k+1}(d,\tau,0)=& qC_a(p\tau+d)+(1-\alpha)h_k(d,\tau+1)\\
    &+\alpha\sum_{i=0}^{\tau}\binom{\tau}{i} p^i(1-p)^{\tau-i}h_k(d+i,\tau)
\end{align*}
\begin{align*}
    &Q_{k+1}d,\tau,1)=
    q(1-\alpha)C_ap\tau+qC_f+q\alpha h_k(0,1)\\
    &+q(1-\alpha)h_k(0,\tau+1)
    +(1-q)(1-\alpha)h_k(d,\tau+1)\\
    &+(1-q)\alpha\sum_{i=0}^{\tau}\binom{\tau}{i} p^i(1-p)^{\tau-i}h_k(d+i,\tau)
\end{align*}
\begin{align*}
   & Q_{k+1}(d,\tau,2)=qC_f+qC_f+q h_k(0,1)\\
    &+(1-q)(1-\alpha)h_k(d,\tau+1)\\
    &+(1-q)\alpha\sum_{i=0}^{\tau}\binom{\tau}{i} p^i(1-p)^{\tau-i}h_k(d+i,\tau)
\end{align*}
Pick any $(d',\tau')\in \mathcal{S}.$
\begin{align*}
    h_{k}(d,\tau)&=\min\{Q_k(d,\tau,0), Q_k(d,\tau,1), Q_k(d,\tau,2)\}\\
    &-\min\{Q_k(d',\tau',0), Q_k(d',\tau',1), Q_k(d',\tau',2)\}
\end{align*}
Let $h_0(d,\tau)$ be $0$, $\forall d, \tau$.
Therefore,
\begin{align*}
    Q_1(d,\tau,0)=qC_a(p\tau+d)
\end{align*}
\begin{align*}
    Q_1(d,\tau,1)=q(1-\alpha)C_a p\tau+qC_f
\end{align*}
\begin{align*}
    Q_1(d,\tau,2)=qC_f+qC_f
\end{align*}
Clearly, $Q_1(d,\tau,a)$ is non decreasing in $d$, for $a \in \{0,1,2\}$. Therefore, $h_1(d,\tau)$ is non decreasing in $d$.
Suppose $h_k(d,\tau)$ is non decreasing in $d$, $\forall k \leq K$. We will now show that $h_{K+1}(d,\tau)$ is non-decreasing in $d$.
\begin{align*}
    Q_{K+1}(d,\tau,0)=& qC_a(p\tau+d)+(1-\alpha)h_K(d,\tau+1)\\
    &+\alpha\sum_{i=0}^{\tau}\binom{\tau}{i} p^i(1-p)^{\tau-i}h_K(d+i,\tau)
\end{align*}
Since each term in the summation is non-decreasing in $d$, $Q_{K+1}(d,\tau,0)$ is non-decreasing in $d$. Same argument holds true for $Q_{K+1}(d,\tau,1)$ and $Q_{K+1}(d,\tau,2)$. Therefore, $h_{K+1}(d,\tau)$ is non decreasing in $d$. Hence, $\hbarr(d,\tau)$ is non-decreasing in $d$.

Using similar argument, it is easy to see that $h_{k}(d,\tau)$ is non decreasing in $\tau$, $\forall k \leq K$, then $h_{K+1}(d,\tau)$ is non decreasing in $\tau$. Hence $\hbarr(d,\tau)$ is non decreasing in $\tau$.
}
\remove{
\section{Proof of Lemma 3} \label{Appendix:lemma3}
\paragraph{Proof of Lemma 3.1}
Given $u^{\ast}(d,\tau)=0$.
This implies $Q(d,\tau,0)\leq Q(d,\tau,1)$ and $Q(d,\tau,0)\leq Q(d,\tau,2)$.
From~\eqref{eq:Q_action_0} and~\eqref{eq:Q_action_1}, 
\begin{align*}
   &Q(d,\tau,0)\leq Q(d,\tau,1)\\
 \implies &qC_ad+q(1-\alpha)\hbarr(d,\tau+1)\\
 &+q\alpha\sum_{i=0}^{\tau}\binom{\tau}{i} p^i(1-p)^{\tau-i}\hbarr(d+i,1) \\
 \stackrel{(b)}\leq& qC_f+q\alpha \hbarr(0,1)-q\alpha C_ap\tau+q(1-\alpha)\hbarr(0,\tau+1) 
\end{align*}
 R.H.S in inequality (b) is independent of $d$, and L.H.S is non-decreasing in $d$. Therefore, if $Q(d,\tau,0)\leq Q(d,\tau,1)$, then $Q(d-\delta,\tau,0)\leq Q(d-\delta,\tau,1), \forall \delta \in  [0, d]$.
Similarly, from~\eqref{eq:Q_action_0} and~\eqref{eq:Q_action2},
\begin{align*}
   &Q(d,\tau,0)\leq Q(d,\tau,2)\\
    &\implies qC_ad+q(1-\alpha)\hbarr(d,\tau+1)\\
 &+q\alpha\sum_{i=0}^{\tau}\binom{\tau}{i} p^i(1-p))^{\tau-i}\hbarr(d+i,1)\\
 &\leq qC_f-qC_ap\tau+q\alpha \hbarr(0,1)\\
 &\implies qC_a(d-\delta)+q(1-\alpha)\hbarr(d-\delta,\tau+1)\\
 &+q\alpha\sum_{i=0}^{\tau}\binom{\tau}{i} p^i(1-p)^{\tau-i}\hbarr(d-\delta+i,1) \\
 &\leq qC_f-qC_ap\tau+q\alpha \hbarr(0,1), \forall \delta  \in [0, d] 
\end{align*}
\remove{Inequality (c) holds true since $\hbarr(d,\tau)$ is non decreasing in $d$. Therefore, if $Q(d,\tau,0)\leq Q(d,\tau,2)$, then $Q(d-\delta,\tau,0)\leq Q(d-\delta,\tau,2), \forall \delta \in  [0, d]$.}
Therefore, if $u^{\ast}(d,\tau)=0$, then $u^{\ast}(d-\delta,\tau)=0, \forall \delta \in  [0, d].$
\paragraph{Proof of Lemma 3.2} 
Given $u^{\ast}(d,\tau)=2$.
This implies $Q(d,\tau,2)\leq Q(d,\tau,1)$ and $Q(d,\tau,2)\leq Q(d,\tau,0)$.
Given $Q(d,\tau,2)\leq Q(d,\tau,1)$, it can be shown from~\eqref{eq:Q_action2} and~\eqref{eq:Q_action_1},
\begin{align*}
    Q(d,\tau,2)&\leq Q(d,\tau,1)\\
   \remove{ \implies & qC_f+qC_f+q\hbarr(0,1)+(1-q)(1-\alpha)\hbarr(d,\tau+1)\\
    &+(1-q)\alpha\sum_{i=0}^{\tau}\binom{\tau}{i} p^i(1-p))^{\tau-i}\hbarr(d+i,1)\\
    &\leq q(1-\alpha)C_a(p\tau)+qC_f+q\alpha \hbarr(0,1)+q(1-\alpha)\hbarr(0,\tau+1)\\
    &+(1-q)(1-\alpha)\hbarr(d,\tau+1)\\
    &+(1-q)\alpha\sum_{i=0}^{\tau}\binom{\tau}{i} p^i(1-p))^{\tau-i}\hbarr(d+i,1)}
    \implies  q(C_f+(1-\alpha)\hbarr(0,1))
    &\stackrel{(d)}\leq q(1-\alpha)C_ap\tau\\
    &+q(1-\alpha)\hbarr(0,\tau+1)
\remove{&\stackrel{(d)}\leq q(1-\alpha)C_a(p(\tau+\delta))+q(1-\alpha)\hbarr(0,\tau+\delta+1), \forall \delta \geq 0.}
\end{align*}
Inequality (d) holds true since L.H.S. of the above inequality is independent of $\tau$ and $\hbarr(d, \tau)$ is non-decreasing in $\tau$. Therefore, if $Q(d,\tau,2)\leq Q(d,\tau,1)$, then $Q(d,\tau+\delta,2)\leq Q(d,\tau+\delta,1), \forall \delta \geq 0$.

Given $Q(d,\tau,2)\leq Q(d,\tau,0)$, from~\eqref{eq:Q_action2} and~\eqref{eq:Q_action_0},
\begin{align*}
     &Q(d,\tau,2)\leq Q(d,\tau,0)\\
     \remove{
      \implies & qC_f+qC_f+q\hbarr(0,1)+(1-q)(1-\alpha)\hbarr(d,\tau+1)\\
    &+(1-q)\alpha\sum_{i=0}^{\tau}\binom{\tau}{i} p^i(1-p))^{\tau-i}\hbarr(d+i,1)\\
    &\leq qC_a(p\tau+d)+(1-\alpha)\hbarr(d,\tau+1)\\
 &+\alpha\sum_{i=0}^{\tau}\binom{\tau}{i} p^i(1-p)^{\tau-i}\hbarr(d+i,1)\\}
 \implies & qC_f+qC_f-qC_ad+q\hbarr(0,1)\\
    &\leq qC_ap\tau+q(1-\alpha)\hbarr(d,\tau+1)\\
 &+q\alpha\sum_{i=0}^{\tau}\binom{\tau}{i} p^i(1-p)^{\tau-i}\hbarr(d+i,1)
\remove{&\stackrel{(e)}\leq qC_ap(\tau+\delta)+q(1-\alpha)\hbarr(d,\tau+\delta+1)\\
 &+q\alpha\sum_{i=0}^{\tau+\delta}\binom{\tau+\delta}{i} p^i(1-p))^{\tau+\delta-i}\hbarr(d+i,1)}
\end{align*}
Inequality (e) holds true since $\hbarr(d, \tau)$ is non-decreasing in $\tau$. Therefore, if $Q(d,\tau,2)\leq Q(d,\tau,0)$, then $Q(d,\tau+\delta,2)\leq Q(d,\tau+\delta,0), \forall \delta \geq 0$.
Therefore, if $u^{\ast}(d,\tau)=2$, then $u^{\ast}(d,\tau+\delta)=2$, $\delta \geq 0$.
}
\remove{
\section{Proof of Lemma 4} \label{Appendix:lemma6}
Without loss of generality, $u^{\ast}(d,\tau)=\arg\min\{Q(d, \tau, 1), Q(d, \tau, 2)\}$.
From~\eqref{eq:Q_action2} and~\eqref{eq:Q_action_1},
\begin{align}\label{eq:Q1Q2}
   & Q(d,\tau,1)- Q(d,\tau,2)
    = qC_ap\tau(1-\alpha)\nonumber\\
     &+q(1-\alpha)\hbarr(0,\tau+1)-q(C_f+(1-\alpha))\hbarr(0,1)
\end{align}
Note that $Q(d,\tau,1)- Q(d,\tau,2)$ is independent of $d$.
Therefore, from~\eqref{eq:Q1Q2}, $Q(d,\tau,1)- Q(d,\tau,2) \geq 0$ for some $\tau \geq \tau_{opt}^{\ast}$.

}

\section{Proof of Theorem 1} \label{Appendix:TheoremAgg}
We will first prove parts~$2$ and~$3$. 
\paragraph*{Proof of Part $2)$} Recall from~\eqref{eqn:bellmanAgg} and~\eqref{eq:h_bar_tau} that
\[\hbarr(\tau,m)+\theta=\min\{\cbar(\tau) m+\hbarav(\tau+1),C_f+\hbarav(1)\}.\]
 Since $\hbarav(\tau)$ is non-decreasing in $\tau$ from Lemma~\ref{lemmaAgg}, $\cbar(\tau) m+\hbarav(\tau+1)$ is strictly increasing in $\tau$. Furthermore, $C_f+\hbarav(1)$ is a constant. Hence, for any $m \geq 1$, there exists a $\taustar(m)$ such that $\cbar(\tau) m+\hbarav(\tau+1)\geq C_f+\hbarav(1)$
if and only if $\tau \geq \taustar(m)$. 
Hence 
\begin{align*}
     \taustar(m)=\min\{\tau:\hbarav(\tau+1)\geq C_f+\hbarav(1)-\cbar(\tau) m\}.
\end{align*}
\paragraph*{Proof of Part~$3)$}
We prove that $\taustar(m)$ is non-increasing in $m$ by contradiction.
Let us assume that $\taustar(m)<\taustar(m+1)$. This implies that there exists a $\tau$ such that $\taustar(m)\leq\tau<\taustar(m+1)$.
Since $\tau\geq\taustar(m)$, $\hbarav(\tau+1)+\cbar(\tau) m\geq C_f+\hbarav(1)$.
Similarly $\tau<\taustar(m+1)$ implies
$\hbarav(\tau+1)+\cbar(\tau) (m+1)< C_f+\hbarav(1)$.
These two inequalities contradict each other and thus also contradict the definition of $\taustar(m)$. Hence $\taustar(m)\geq \taustar(m+1)$. So, $\taustar(m)$ is non-decreasing in $m$.
\paragraph*{Proof of Part $1)$}
We first prove the following lemma. 
\begin{lemma}\label{lemmaAgg2} For any $(\tau,m)$,
  $\hbarr(\tau,m)\leq \hbarr(1,m)+C_f$.   
\end{lemma}
\begin{IEEEproof}
From \eqref{eqn:bellmanAgg}, there are three possible cases. 
\paragraph*{Case (a)} The optimal action in state $(\tau,m)$ is $0$. Then \begin{equation}\hbarr(\tau,m)+\theta=m\cbar(\tau)+\hbarav(\tau+1).\label{eq:first_case}\end{equation}
Since $\tau\geq 1$, $\hbarr(1,m)+\theta=mC_ap+\hbarav(2)$ from part~$3$ of Theorem~\ref{TheoremAgg} (proved above). Therefore,
\begin{align*}
    \hbarr(\tau,m)-\hbarr(1,m)&=m\cbar(\tau-1)+\hbarav(\tau+1)-\hbarav(2)\\
    & < m\cbar(\tau)+\hbarav(\tau+1)-\hbarav(2)\\
    &\leq C_f+\hbarav(1)-\hbarav(2)\leq C_f. 
\end{align*}
The second last inequality follows from~\eqref{eq:bellman_hbar} and~\eqref{eq:first_case}.
\paragraph*{Case (b)} The optimal action in both the states, $(\tau,m)$ and $(1,m)$, is $1$. In this case, $\hbarr(\tau,m)+\theta=C_f+\hbarav(1)$ and $\hbarr(1,m)+\theta=C_f+\hbarav(1)$ from~\eqref{eqn:bellmanAgg}.
In this case, clearly, $\hbarr(\tau,m)-\hbarr(1,m)=0$. Therefore, $\hbarr(\tau,m)-\hbarr(1,m)\leq C_f$.
\paragraph*{Case (c)} The optimal actions in states $(\tau,m)$ and $(1,m)$ are $1$ and $0$, respectively. $\hbarr(\tau,m)+\theta=C_f+\hbarav(1)$ and $\hbarr(1,m)+\theta=m\cbar(1)+\hbarav(2)$ from~\eqref{eqn:bellmanAgg}. 
In this case,
\begin{align*}
    \hbarr(\tau,m)-\hbarr(1,m)&=C_f+\hbarav(1)-m\cbar(1)-\hbarav(2)\\
    &\leq C_f.    
\end{align*}
\end{IEEEproof}
Now, from~\eqref{eqn:bellmanAgg},
\begin{align}
\hbarr(\tau,0)+\theta=\min&\left\{C_f+\sum_{k=0}^{N}B(N,q,k)\hbarr(1,k),\right.\nonumber\\
&\left. \sum_{k=0}^{N}B(N,q,k)\hbarr(\tau{+}1,k)\right\}.\label{eq:bellman_for_m_0}
\end{align}
But, from Lemma~\ref{lemmaAgg2}, 
\[\sum_{k=0}^{N}B(N,q,k)(\hbarr(\tau{+}1,k)-\hbarr(1,k))-C_f\leq 0.\]
Hence,
$\hbarr(\tau,0)+\theta
=\sum_{k=0}^{N}B(N,q,k)\hbarr(\tau{+}1,k) $
and $\bar{\pi}(\tau,0)=0$.

\remove{\section{Proof of Corollary}
\label{Appendix:Corollary}.
Since $\taustar(1)=\taustar(2)$
\begin{align*}
    &\gbar(\taustar(1)) \leq C_f-2C_ap(\taustar(1)-1)\nonumber\\
    & \implies C_f-C_ap\taustar(1)\leq C_f-2C_ap(\taustar(1)-1)\nonumber\\
    & \implies \taustar(1) \leq 2
\end{align*}
Since from Theorem \ref{TheoremAgg}, $\taustar(2)\ge\taustar(3)$, $\taustar(1)=1 \implies \taustar(j)=1, \forall j \in\{2,3,...,N\}$
Since $\taustar(2)>\taustar(3)$, $\taustar(1)=\taustar(2)=2$. From Theorem \ref{TheoremAgg}, $\taustar(j)=1, \forall j \in\{3,...,N\}$

If $\taustar(1)=\taustar(k)$, from Theorem \ref{TheoremAgg}, $\taustar(1)=\taustar(2)=...=\taustar(k)$.
\begin{align*}
    &\gbar(\taustar(1)) \leq C_f-kC_ap(\taustar(1)-1)\nonumber\\
    & \implies C_f-C_ap\taustar(1)\leq C_f-kC_ap(\taustar(1)-1)\nonumber\\
    & \implies \taustar(1) \leq \frac{k}{k-1}
\end{align*}
Since $k>2, \taustar(1)=1$.
Therefore, from Theorem \ref{TheoremAgg}, $\taustar(j)=1, \forall j \in\{1,...,N\}$

}

 \remove{

\section {Proof of Lemma 2}\label{Appendix:lemmaAgg2}
From \ref{TheoremAgg}, $\bar{\pi}(\tau,m)=1,$ $\forall \tau \geq \taustar(1)$. Therefore, $\forall \tau \geq \taustar(1)$,
$\hbarr(\tau,m)=C_f+\hbarav(1)-\theta$. Therefore, from~\eqref{eqn: tau>taustar1}
\begingroup
\allowdisplaybreaks
\begin{align}\label{equation:hbartau_large}
    \hbarav(\tau)
    =&(1-q)^N\hbarav(\tau+1)+\sum_{k=1}^{N}B(N,q,k)(C_f+\hbarav(1)-\theta)\nonumber\\
    =&(1-q)^N\hbarav(\tau+1)\nonumber\\
    &+(1-(1-q)^N)(C_f+\hbarav(1)-\theta)\nonumber\\
    \overset{(a)}=&(1-q)^{KN}\hbarav(\tau+K)\nonumber\\
    &+\sum_{k=0}^{K}(1-q)^{Nk}((1-(1-q)^N)(C_f+\hbarav(1)-\theta))\nonumber\\
    =&(1-q)^{KN}\hbarav(\tau+K)\nonumber\\&+(1-(1-q)^{N(K+1)})((C_f+\hbarav(1)-\theta))\nonumber\\
\end{align}
\endgroup

Equality (a) is obtained by repeating the recursion for $K$ for times. 

As $K\to \infty$, $(1-q)^{KN}\hbarav(\tau+K)\to 0$ as $\hbarr(\tau)<\infty$ due to single recurrent class of optimally controlled Markov chain.
}
\section{Proof of Theorem~\ref{theorem:fixedpoint}}
\label{Appendix:fixedpoint}
We show that $f(\theta)$ is non-increasing in $\theta$.  For this, we first show that $\gbar(2)$  as obtained from Algorithm~\ref{alg:fixed_point} is non-increasing in $\theta$.
 Let $\gbar_1(2)$ and $\gbar_2(2)$ denote the values of $\gbar(2)$ corresponding to $\theta = \theta_1$ and $\theta = \theta_2$, respectively. We will show the following.
 \begin{lemma}
     If $\theta_1<\theta_2$, then $\gbar_1(2)\geq \gbar_2(2)$. 
 \end{lemma}
 \begin{IEEEproof}
     Observe from Line~\ref{taustar_in_terms_theta} of  Algorithm~\ref{alg:fixed_point} that   
$\bar{\tau}_1(1) \leq \bar{\tau}_2(1)$.
Also,  from Line~\ref{g_taustar_in_terms_theta} of Algorithm~\ref{alg:fixed_point},  $\gbar(\taustar(1))$ is inversely related to $\theta$  and hence,  $
\gbar_1(\bar{\tau}_1(1)) \geq \gbar_2(\bar{\tau}_2(1))$. We will show that  
$
\gbar_1(\tau) \geq \gbar_2(\tau)$ for all $ \tau = 2,\cdots,\bar{\tau}_1(1)$ by induction.  

Base case: We show that $\gbar_1(\bar{\tau}_1(1)) \geq \gbar_2(\bar{\tau}_1(1))$.
Since $\hbarav(\tau)$ is non-decreasing in $\tau$ from Lemma~\ref{lemmaAgg},  
and 
$\gbar(\tau) = \hbarav(\tau) - \hbarav(1)$, it is easy to see that  $\gbar(\tau)$ is non-decreasing in $\tau$.  Since $\bar{\tau}_1(1) \leq \bar{\tau}_2(1)$, we have
$\gbar_2(\bar{\tau}_1(1)) \leq \gbar_2(\bar{\tau}_2(1))$. We have already shown that $\gbar_1(\bar{\tau}_1(1)) \geq \gbar_2(\bar{\tau}_2(1))$, and hence
$\gbar_2(\bar{\tau}_1(1)) \leq \gbar_1(\bar{\tau}_1(1))$. 
  
Induction step: Let us fix a $\tau$ such that $2<\tau\leq\taustar(1)$ and assume that $\gbar_1(k) \geq \gbar_2(k)$ for all $\tau\leq k\leq \taustar(1)$. We will show that  
$\gbar_1(\tau - 1) \geq \gbar_2(\tau - 1)$.
From Line~\ref{update_g_tau_minus_1} of Algorithm~\ref{alg:fixed_point},
\begingroup
\allowdisplaybreaks
\begin{align}
\gbar_1(\tau - 1) &= C_f {-} \theta_1+\nonumber\\\sum_{k=1}^{m_1}& B(N,q,k) \Big(k \cbar (\tau{-}1){-}C_f + \gbar_1(\tau)\Big), \label{eq:g_1_tau}\\
\gbar_2(\tau - 1) &= C_f {-} \theta_2 +\nonumber\\&\sum_{k=1}^{m_2} B(N,q,k) \Big(k \cbar (\tau{-}1) {-}C_f + \gbar_2(\tau)\Big) \label{eq:g_2_tau}
\end{align}
\endgroup
where $m_1$ and $m_2$ are values of $m$ in instances $1$ and $2$ respectively, when $\bar{\tau}_1(m_1)=\bar{\tau}_2(m_2)=\tau$. 
From Line~\ref{update_g_tau_minus_1} of Algorithm 1,  
$\gbar_1(\tau) \leq C_f - k \cbar(\tau-1)$ for  all $k \leq m_1$ and $\gbar_2(\tau) \leq C_f - k \cbar(\tau-1)$ for all $k \leq m_2$. So, the terms under summations in~\eqref{eq:g_1_tau} and~\eqref{eq:g_2_tau}  are non-positive. Furthermore,  $\gbar_1(\tau) \geq \gbar_2(\tau)$ from the induction hypothesis.
This implies that
\[
k \cbar(\tau-1) -C_f + \gbar_1(\tau) \geq k \cbar (\tau-1) -C_f + \gbar_2(\tau).
\]
Since $ \theta_1 < \theta_2 $, to show $ \gbar_1(\tau - 1) \geq \gbar_2(\tau - 1) $ it suffices to show that $m_1 \leq m_2$. This can be argued as follows.
From the induction hypothesis, we have $\gbar_1(k)\geq \gbar_2(k)$ for all $k\leq \taustar(1)$. Hence, if  the condition in Line~\ref{while_connd_to_inc_m} in Algorithm~\ref{alg:fixed_point} is satisfied for $\gbar_1(k)$ , then it must also be satisfied for $\gbar_2(k)$ but the converse is not true. This is true for all $k = \tau, \cdots,  \taustar(1)$. Hence $m_1\leq m_2$.

We have thus shown for all $\tau = 2,\cdots,\taustar(1)$ that if $\gbar_1(k) \geq \gbar_2(k)$ for all $k = \tau,\cdots, \taustar(1)$ then  $\gbar_1(\tau - 1) \geq \gbar_2(\tau - 1)$.
This further implies that $\gbar_1(2)\geq \gbar_1(2)$. Thus we conclude that if $\theta_1\leq \theta_2$ then $\gbar_1(2)\geq \gbar_1(2)$.  
\end{IEEEproof}
In the following we show that $f(\theta)$ is non decreasing in $\gbar(2)$. For this we need the following lemma.
\begin{lemma}
    If $\taustar_1(N)>1$ then $\taustar_2(N)>1$.
\end{lemma}
\begin{IEEEproof}
    Suppose the above statement if false. Then $\taustar_1(N)>1$ and $\taustar_2(N)=1$ is a possibility. In this case $\gbar_2(2)>C_tx-N\cbar(1)$ from the definition of $\taustar(2)$ in Theorem~\ref{TheoremAgg} or from the condition in Line~\ref{while_connd_to_inc_m} in Algorithm~\ref{alg:fixed_point}. Since $\taustar_1(N)>1$, there exists a $\tau>1$ such that $\gbar_1(\tau)\leq C_f-Np\cbar(\tau)$. This contradicts the fact that $\gbar_2(2)\leq \gbar_1(1)$. Hence, we conclude that $\taustar_1(N)>1$ implies $\taustar_2(N)>1$.
\end{IEEEproof}
Now, there are three possible cases. 
\paragraph*{Case (a)}  $\taustar_1(N)>1$ and $\taustar_2(N)>1$.
 In this case, $f(\theta_1) =N\cbar(1)q+\gbar_1(2)$ and $f(\theta_2) =N\cbar(1)q+\gbar_2(2)$. 
 from Line~\ref{cond:tau_N>1} of Algorithm~\ref{alg:fixed_point}. 
 Hence, $f(\theta_1)\geq f(\theta_2)$ as $\gbar_1(2)\geq \gbar_2(2)$, and this implies  that $f(\theta_1)\geq f(\theta_2)$ for $\theta_1<\theta_2$.
 \paragraph*{Case (b)} $\taustar_1(N)=\taustar_2(N)=1$. In this case, from Line~\ref{cond:tau_N=1} in Algorithm~\ref{alg:fixed_point},
 \begingroup
 \allowdisplaybreaks
 \begin{align*}
    f(\theta_1) &{=} \sum_{k \leq m_1} B(N,q,k) \Big(k \cbar(1) {+} \gbar_1(2) {-}C_f\Big)  {+} C_f,\\
    f(\theta_2) &{=} \sum_{k \leq m_2} B(N,q,k) (k \cbar(1) {+} \gbar_2(2) {-}C_f\Big) 
     {+} C_f.
\end{align*}
\endgroup
From Algorithm~\ref{alg:fixed_point}, we have that $k \cbar(1) + \gbar_1(2) -C_f \leq 0$ for  $k \leq m_2$ and $k \cbar(1) + \gbar_2(2) -C_f \leq 0$ for $k \leq m_2$. Furthermore, we have already argued that $m_1 \leq m_2$ and $\gbar_1(2) \geq \gbar_2(2)$. Hence, 
$f(\theta_1) \geq f(\theta_2)$.

\paragraph*{Case (c)} $\taustar_1(N)=1,\taustar_2(N)>1$. In this case,
     \begin{align}
         f(\theta_1) =& \sum_{k\leq m_1}B(N,q,k)(k\cbar(1)+\gbar_1(2))\nonumber\\
         &+\sum_{k> m_1}B(N,q,k)C_f,\label{eq:f_theta_1}\\
    f(\theta_2)=&N\cbar(1)q+\gbar_2(2)\nonumber\\
         =&\sum_{k=1}^NB(N,q,k)(N\cbar(1)q+\gbar_2(2)).\label{eq:f_theta_2}
     \end{align}
     We will compare the terms under the summations in~\eqref{eq:f_theta_1} and~\eqref{eq:f_theta_2}. Note that 
     $k\cbar(1)+\gbar_2(2)\leq k\cbar(1)+\gbar_1(2)$ for $k\leq m_1$ and $k\cbar(1)+\gbar_2(2)\leq C_f$ For $k>m_1$. These imply that $f(\theta_1) \geq f(\theta_2)$. 

Finally, we argue that there exists a $\theta$ such that $f(\theta)=\theta$. Let us consider $\theta=\thetabar$ obtained by solving the Bellman's equations~\eqref{eq:bellman_hbar}. Note that, $\thetabar$ satisfies~\eqref{eq:g_tau_fort=_tau_geq_taustar} and~\eqref{eq:taustar_1_in_theta} and hence, also Lines~\eqref{taustar_in_terms_theta} and~\eqref{g_taustar_in_terms_theta} in Algorithm~\ref{alg:fixed_point}. The value of $\taustar(m)$ in Line~\ref{update_taustar} of Algorithm~\ref{alg:fixed_point} follows from the definition of $\taustar(m)$.  The value of $\gbar(\tau)$ in Line~\ref{update_g_tau_minus_1} follows from~\eqref{eqn:taustarm<tau<taustarm1}.
If $\taustar(N)>1$, then from~\eqref{eq:g_2_and_theta_first_case}, 
\[\thetabar=Nq\cbar(1)+\gbar(2)\]
which is same as Line~\ref{cond:tau_N>1} in Algorithm~\ref{alg:fixed_point}. Finally, if $\taustar(N)=1$, then from~\eqref{eq:g_2_and_theta_second_case} ,\begin{equation*}
    \thetabar=\sum_{k<m}B(N,q,k)(k\cbar(1)+\gbar(2)){+}\sum_{k\geq m}B(N,q,k)C_f.
\end{equation*}
which is same as $f(\thetabar)$ in Line~\ref{cond:tau_N=1} in Algorithm~\ref{alg:fixed_point}. Hence $f(\thetabar)=\thetabar$ in either case.
\remove{
\color{black}
\section{Proof of Lemma 4}
\label{Appendix:lemmaAggBacklog}.
\begin{align*}
    \hbarav_{j,k}(\tau):=\sum_{i=0}^{N-j}\binom{N-j}{i}q^i(1-q)^{N-i}h_{k}(\tau,i).
\end{align*}

Fix $(\tau',j')$.
\begin{align*}
    &h_{k+1}(\tau,j)=\nonumber\\
    &\min_{a \in \{0,1\}}\{\cbar(\tau) j(1-a)+\hbarav_{j,{k}}(\tau+1),C_fa+\hbarav_{0,{k}}(1)\}-\nonumber\\
    &\min_{a \in \{0,1\}}\{\cbar(\tau)' j'(1-a)+\hbarav_{j,{k}}(\tau'+1),C_fa+\hbarav_{0,{k}}(1)\}
\end{align*}
Let $h_0(\tau,j)=0, \forall (\tau,j)$. Then, $\hbarav_0(\tau)=0$. 
\begin{align}\label{equation:h1taujj}
    h_{1}(\tau,j)+\theta=
    &\min_{a \in \{0,1\}}\{\cbar(\tau) j(1-a),C_fa\}-\nonumber\\
    &\min_{a \in \{0,1\}}\{\cbar(\tau)' j'(1-a),C_fa\}
\end{align}
Clearly, $h_{1}(\tau,j)$ is non-decreasing in $\tau$. Therefore, $\hbarav_{j,1}(\tau)$ is also non-decreasing in $\tau$. 
Suppose $h_{T}(\tau,j)$ is non-decreasing in $\tau$. Then $\hbarav_{j,T}(\tau)$ is non-decreasing in $\tau$. We need to show that $h_{T+1}(\tau,m)$ is non-decreasing in $\tau$. Since $\hbarav_{j,T}(\tau)$ is non-decreasing in $\tau$, $h_{T+1}(\tau,m)$ is non-decreasing in $\tau$. Then $\hbarav_{j,{T+1}}(\tau)$ is non-decreasing in $\tau$.

Before proving that $\hbarav_j{(\tau)}$ is non-decreasing in $j$, recall the definitions of stochastic ordering from \cite{ross}.

\paragraph*{Definition } Stochastically Dominance: 
We say that the random variable $X$  stochastically dominates the random variable $Y$, written $X \geq_{\text{st}}Y$, if $P(X>a)\geq P(Y>a), \forall a \in \mathbb{R}$. 

We use method of induction and stochastic dominance to argue that $\hbarav_j(\tau)$ is non-decreasing in $\tau$. Let $h_0(\tau,j)=0, \forall (\tau,j)$. From \ref{equation:h1taujj}, $h_1(\tau,j)$ is non-decreasing in $j$. Hence $\hbarav_{j,1}(\tau)$ is non-decreasing in $j$. Suppose $h_T(\tau,j)$ is non-decreasing in $j$. 
$\hbarav_{j,T}(\tau)=\mathbb{E}h_T(\tau,j+i)$, where $i{\sim Binomial}(N-j,q)$. 
Suppose $X_1=j_1+i_1$ and $X_2=j_2+i_2$, such that
 $j_2>j_1$ and  $i_k{\sim Binomial}(N-j_k,q),k\in\{1,2\}$. 

Clearly, $X_2\geq_{\text{st}}X_1$. Since $h_T(\tau,j)$ is non-decreasing in $j$,
\begin{align*}
    h_{T}(\tau,j_2+i_2)\geq_{\text{st}}h_T(\tau,j_1+i_1).
\end{align*}

 From proposition 9.1.2\cite{ross}, $\mathbb{E}h_T(\tau,j_2+i_2)\geq \mathbb{E}h_T(\tau,j_!+i_1)$. Hence $\hbarav_{j,T}(\tau)$ is non-decreasing in $j$. Consequently, $h_{T+1}(\tau,j)$ is non-decreasing in $j$. Hence $\hbarav_{j,T+1}(\tau)$ is non-decreasing in $j$.
 \remove{  
\paragraph*{Definition } 
Stochastic Coupling: If $X \geq_{\text{st}}Y$, then there exist random variables $X^{\ast}$ and $Y^{\ast}$ having the same distribution of $X$ and $Y$ and such that $X^{\ast} \geq Y^{\ast}$, with probability 1.
}
\section{Proof of Theorem}
\label{Appendix:TheoremAggBacklog}
Recall that $\hbarr(\tau,j)+\theta=\min_{a \in \{0,1\}}\{\cbar(\tau) j(1-a)+\hbarav_j(\tau+1),(C_f)a+\hbarav_0(1)\}$. Since $\hbarav_j(\tau)$ is non-decreasing in $\tau$, from Lemma \ref{lemmaAggBacklog}, for a fixed $j$, $\hbarr(\tau,j)$ is non-decreasing in $\tau$. Therefore, $\exists \taustar(j)$ such that,
\begin{align*}
     \forall \tau \geq \taustar(j), \cbar(\tau) j+\hbarav_j(\tau+1)\geq C_f+\hbarav_0(1).
\end{align*}
Therefore,
\begin{align*}
     \taustar(j)=\min\{\tau:\hbarav_j(\tau+1)\geq C_f+\hbarav_0(1)-\cbar(\tau) m\}.
\end{align*}}
\remove{
Now, we prove that $\taustar(j)$ is non-decreasing in $j\in\{1,2,...,N\}$. We prove that by contradiction.
Let us assume that $\taustar(j)<\taustar(j+1)$. This implies that, $\exists \tau: \taustar(j)\leq\tau<\taustar(j+1)$.
Since $\tau\geq\taustar(j)$, 
\begin{align*}
    &\hbarav_j(\tau+1)+\cbar(\tau) j\geq C_f+\hbarav(1)\nonumber\\
    \implies&\hbarav_j(\tau+1)+\cbar(\tau) (j+1)\geq C_f+\hbarav(1)\nonumber\\
\end{align*}
However, since $\tau<\taustar(j+1)$, $\hbarav_{j+1}(\tau+1)+\cbar(\tau) (j+1)< C_f+\hbarav_0(1)$. Therefore our assumption that $\taustar(j)<\taustar(j+1)$ is wrong. Therefore, $\taustar(j)$ is non-decreasing in $j\in\{1,2,...,N\}$.
}
\remove{
Now we will see how to compute the exact value of $\tau$. Under this regime, $Q(d,\tau,2)-Q(d,\tau,1)$ is independent of $d$. Alternatively, we can write the Bellman equation as follows
\begin{align*}
    \label{Alt Bellman}
    &\tilde{h}( \tau)+\tilde{\theta}=\min\{qC_a\taup(1-\alpha)+(1-\alpha)\tilde{h}( \tau+1)+\alpha \tilde{h}( 1),\\
    &q(1-\alpha)C_f+(q+(1-q)\alpha)\tilde{h}( 1)+(1-q)(1-\alpha)\tilde{h}( \tau+1)\}
\end{align*}
It is easy to see that $\tilde{h}( \tau)$ is increasing in $\tau$. 
\paragraph*{For $\tau\geq \taustar$}
\begin{align*}
    &\tilde{h}( \tau)+\tilde{\theta}=qC_{f}+(q+(1-q)\alpha)\tilde{h}( 1)\\
    &+(1-q)(1-\alpha)\tilde{h}( \tau+1)\nonumber\\
    \implies & \tilde{h}( \tau)=qC_{f}+(q+(1-q)\alpha)\tilde{h}( 1)-\tilde{\theta}\\
    &+(1-q)(1-\alpha)\tilde{h}( \tau+1).
\end{align*}
Since the above expression is valid for all $\tau\geq \taustar$, we can substitute $\tilde{h}( \tau+1)$ and get
\begin{align*}
    &\tilde{h}( \tau)=qC_{f}+(q+(1-q)\alpha)\tilde{h}( 1)-\tilde{\theta}\\
    &+(1-q)(1-\alpha)(C_{f}q+(q+(1-q)\alpha)\tilde{h}( 1)-\tilde{\theta}\\
    &+(1-q)(1-\alpha)\tilde{h}( \tau+2))\nonumber\\
    \implies & \tilde{h}( \tau)=(qC_{f}+(q+(1-q)\alpha)\tilde{h}( 1)-\tilde{\theta})+\\
    &(1+(1-q)(1-\alpha))+(1-q)^{2}(1-\alpha)^2\tilde{h}( \tau+2).
\end{align*}
Repeating this procedure $n$ times yields the following 
\begin{align*}
    \tilde{h}( \tau) &=(qC_{f}+(q+(1-q)\alpha)\tilde{h}( 1))(\sum_{k=0}^{n}(1-q)^k(1-\alpha)^k)\nonumber\\
    &-\tilde{\theta}(\sum_{k=0}^{n}(1-q)^k(1-\alpha)^k)\\
    &+(1-q)^{n+1}(1-\alpha)^{n+1}\tilde{h}( \tau+n+1)\nonumber\\
    &=\frac{(q(1-\alpha)C_{f}+(q+(1-q)\alpha)\tilde{h}( 1))(1-((1-q)(1-\alpha))^{n+1})}{1-(1-q)(1-\alpha)}\\
    &-\tilde{\theta} \frac{(1-((1-q)(1-\alpha))^{n+1})}{1-(1-q)(1-\alpha)}\\
    &+(1-q)^{n+1}(1-\alpha)^{n+1}\tilde{h}( \tau+n+1).
\end{align*}

In the limit as $n\to \infty$, we have
\begin{align}
    \tilde{h}( \tau)=\frac{(qC_{f}+(q+(1-q)\alpha)\tilde{h}( 1)-\tilde{\theta})}{1-(1-q)(1-\alpha)}
\end{align}
since $(1-q)^{n+1}\tilde{h}( \tau+n+1)\to 0$ as $n \to \infty$.

\paragraph*{For $\tau < \taustar$}
\begin{align*}
    &\tilde{h}( \tau)+\tilde{\theta}=C_{a}\taupq(1-\alpha)+(1-\alpha)\tilde{h}( \tau+1)+\alpha \tilde{h}( 1)\nonumber\\
    \implies & \tilde{h}( \tau)=C_{a}\taupq(1-\alpha)-\tilde{\theta}+(1-\alpha)\tilde{h}( \tau+1)+\alpha \tilde{h}( 1).
\end{align*}

From the above equation, 
\begin{align*}
     \tilde{h}( 1)=C_{a}pq(1-\alpha)+(1-\alpha)\tilde{h}( 2)-(1-\alpha)\tilde{\theta}.
\end{align*}
We substitute $\tilde{h}( 2)$ in the above equation to yield
\begin{align}
     \tilde{h}( 1)=C_{a}pq(1-\alpha)(1+2(1-\alpha))-(1-\alpha)(1+(1-\alpha))\tilde{\theta}+(1-\alpha)^2\tilde{h}( 3).
\end{align}
Repeating this procedure $\taustar-2$ times yields the following 
\begin{align*}
     &\tilde{h}(1)=C_{a}pq(1-\alpha)(1+2(1-\alpha)+..+(\taustar-1)(1-\alpha)^{\taustar-2})\\
       &-(1-\alpha) (1+(1-\alpha)+...+(1-\alpha)^{\taustar-2})\tilde{\theta}+(1-\alpha)^{\taustar-1}\tilde{h}( \taustar)\nonumber\nonumber\\
       =& \frac{C_{a}pq(1-\alpha)(1-\taustar(1-\alpha)^{(\taustar-1)}+(\taustar-1)(1-\alpha)^{\taustar})}{\alpha^2}\\
        &-\frac{(1-\alpha)(1-(1-\alpha)^{\taustar-1})\tilde{\theta}}{\alpha}+(1-\alpha)^{\taustar-1}\tilde{h}( \taustar) \\ 
\end{align*}
We have $ \tilde{h}( \taustar)=\frac{(C_{f}q+(q+(1-q)\alpha)\tilde{h}( 1)-\tilde{\theta})}{1-(1-q)(1-\alpha)}$ and WLOG $\tilde{h}( 1)=0$.
 Substituting $\tilde{h}( \taustar$) and $\tilde{h}( 1)$ and rearranging the above equation,
\begin{align}
\label{eqn: thetatilda}
     &\tilde{\theta}=K_\alpha\{
     \frac{C_{a}pq(1-\alpha)(1-\taustar(1-\alpha)^{(\taustar-1)}+(\taustar-1)(1-\alpha)^{\taustar})}{\alpha^2}\nonumber\\
     &+\frac{C_{f}q(1-\alpha)^{(\taustar-1)}}{1-(1-q)(1-\alpha)}\}
\end{align}
where $K_\alpha=\frac{\alpha(1-(1-q)(1-\alpha))}{(1-\alpha)(1-(1-q)(1-\alpha)+(1-q)(1-\alpha)^{\taustar}+\alpha(1-\alpha)^{(\taustar-2)}-(1-\alpha)^{(\taustar-1)})}$
Optimal threshold value $\taustar_{opt}$ is the threshold that minimise $\tilde{\theta}$ in equation~\eqref{eqn: thetatilda}.
\begin{align*}
    \taustar_{opt}=\argmin_{\taustar\in \mathbb Z_{+}}\tilde{\theta}.
\end{align*}

\remove{
\begin{align*}
    \label{Alt Bellman}
    &\hbarr(\tau)+\theta=\min\{qC_a\taup+(1-\alpha)\hbarr(\tau+1)+\alpha \hbarr(1),\\
    &qC_f+(q+(1-q)\alpha)\hbarr(1)+(1-q)(1-\alpha)\hbarr(\tau+1)\}
\end{align*}
It is easy to see that $\hbarr(\tau)$ is increasing in $\tau$. 
\paragraph*{For $\tau\geq \taustar$}
\begin{align*}
    &\hbarr(\tau)+\theta=C_{f}q+(q+(1-q)\alpha)\hbarr(1)\\
    &+(1-q)(1-\alpha)\hbarr(\tau+1)\nonumber\\
    \implies & \hbarr(\tau)=C_{f}q+(q+(1-q)\alpha)\hbarr(1)-\theta\\
    &+(1-q)(1-\alpha)\hbarr(\tau+1).
\end{align*}
Since the above expression is valid for all $\tau\geq \taustar$, we can substitute $\hbarr(\tau+1)$ and get
\begin{align*}
    &\hbarr(\tau)=C_{f}q+(q+(1-q)\alpha)\hbarr(1)-\theta\\
    &+(1-q)(1-\alpha)(C_{f}q+(q+(1-q)\alpha)\hbarr(1)-\theta\\
    &+(1-q)(1-\alpha)\hbarr(\tau+2))\nonumber\\
    \implies & \hbarr(\tau)=(C_{f}q+(q+(1-q)\alpha)\hbarr(1)-\theta)+\\
    &(1+(1-q)(1-\alpha))+(1-q)^{2}(1-\alpha)^2\hbarr(\tau+2).
\end{align*}
Repeating this procedure $n$ times yields the following 
\begin{align*}
    \hbarr(\tau) &=(C_{f}q+(q+(1-q)\alpha)\hbarr(1))(\sum_{k=0}^{n}(1-q)^k(1-\alpha)^k)\nonumber\\
    &-\theta(\sum_{k=0}^{n}(1-q)^k(1-\alpha)^k)\\
    &+(1-q)^{n+1}(1-\alpha)^{n+1}\hbarr(\tau+n+1)\nonumber\\
    &=\frac{(C_{f}q+(q+(1-q)\alpha)\hbarr(1))(1-((1-q)(1-\alpha))^{n+1})}{1-(1-q)(1-\alpha)}\\
    &-\theta \frac{(1-((1-q)(1-\alpha))^{n+1})}{1-(1-q)(1-\alpha)}\\
    &+(1-q)^{n+1}(1-\alpha)^{n+1}\hbarr(\tau+n+1).
\end{align*}

In the limit as $n\to \infty$, we have
\begin{align}
    \hbarr(\tau)=\frac{(C_{f}q+(q+(1-q)\alpha)\hbarr(1)-\theta)}{1-(1-q)(1-\alpha)}
\end{align}
since $(1-q)^{n+1}\hbarr(\tau+n+1)\to 0$ as $n \to \infty$.

\paragraph*{For $\tau < \taustar$}
\begin{align*}
    &\hbarr(\tau)+\theta=C_{a}\taupq+(1-\alpha)\hbarr(\tau+1)+\alpha \hbarr(1)\nonumber\\
    \implies & \hbarr(\tau)=C_{a}\taupq-\theta+(1-\alpha)\hbarr(\tau+1)+\alpha \hbarr(1).
\end{align*}

From the above equation, $\theta=\hbarr(1)$. Also,
\begin{align*}
     \hbarr(1)=C_{a}pq+(1-\alpha)\hbarr(2)-(1-\alpha)\theta.
\end{align*}
We substitute $\hbarr(2)$ in the above equation to yield
\begin{align}
     \hbarr(1)=C_{a}pq(1+2(1-\alpha))-(1-\alpha)(1+(1-\alpha))\theta+(1-\alpha)^2\hbarr(3).
\end{align}
Repeating this procedure $\taustar-2$ times yields the following 
\begin{align*}
     &\hbarr(1)=C_{a}pq(1+2(1-\alpha)+..+(\taustar-1)(1-\alpha)^{\taustar-2})\\
     &-(1-\alpha) (1+(1-\alpha)+...+(1-\alpha)^{\taustar-2})\theta+(1-\alpha)^{\taustar+1}\hbarr(\taustar)\nonumber\\
     &=\frac{C_{a}pq(1-(\taustar-1)(1-\alpha)^{\taustar-2}+(\taustar-2)(1-\alpha)^{\taustar-1})}{\alpha^2}\\
     &-\frac{(1-\alpha)(1-(1-\alpha)^{\taustar-1})\theta}{\alpha}+(1-\alpha)^{\taustar-1}\hbarr(\taustar)  
\end{align*}
We have $ \hbarr(\taustar)=\frac{(C_{f}q+(q+(1-q)\alpha)\hbarr(1)-\theta)}{1-(1-q)(1-\alpha)}$ and $\hbarr(1)=\theta$.
Let 
$D_\alpha=1-(1-\alpha)^{\taustar}-(1-q)(1-\alpha)(1-(1-\alpha)^{\taustar}-\alpha(1-\alpha)^{\taustar-2})$.  Substituting $\hbarr(\taustar$) and $\hbarr(1)$ and rearranging the above equation,
\begin{align*}
     &\theta=\frac{\alpha(1-(1-q)(1-\alpha))}{D_\alpha}\{\frac{C_{f}q}{1-(1-q)(1-\alpha)}\\
     &+\frac{C_{a}pq(1-(\taustar-1)(1-\alpha)^{\taustar-2}+(\taustar-2)(1-\alpha)^{\taustar-1})}{\alpha^2}\\\}
\end{align*}
}
}
\section{Proof of Lemma~\ref{lemma:opt_pol_sngle_user}}\label{app:proof_opt_pol_sngl_user}
If $\lambda_i+\alpha_i c_f\leq 0$, then the single stage cost for action~$1$ is negative or zero, hence the optimal action is $1$ irrespective of the content being requested or not. Hence, we focus on the case where $\lambda_i+\alpha_i c_f> 0$. Similar to Theorem~\ref{TheoremAgg}, $\pi^{\ast}(\tau,0)=0\,\,\forall\,\,\tau\geq0$. Also from Theorem~\ref{TheoremAgg}, if $s_i=1$, the optimal policy is threshold type. There exists a $\taustar$ such that for $\tau\geq \taustar$, $\pi^{\ast}(\tau,1)=1$. Let a cycle be defined as the time difference between
two time slots when packets are fetched. Then the expected length of the cycle is $\taustar-1+\frac{1}{q_i}$. The total age cost in $\{1,\dots,\taustar-1\}$ incurred in the cycle is $q_i\sum_{x=1}^{\taustar-1}C_{av,i}(x)$ and the fetching cost is $\alpha_ic_f+\lambda_i$. Hence the optimal average cost from the renewal reward theorem, $\theta_i= \frac{q_i\sum_{x=1}^{\tau-1}C_{av,i}(x)+\alpha_i c_f+\lambda_i}{\tau+\frac{1-q_i}{q_i}}$. Hence, by definition of $\taustar$,\[\taustar={\arg\min}_\tau\frac{q_i\sum_{x=1}^{\tau-1}C_{av,i}(x)+\alpha_i c_f+\lambda_i}{\tau+\frac{1-q_i}{q_i}}\eqqcolon f_i(\alpha_i c_f+\lambda_i).\]
\section{Proof of Lemma~\ref{lemma:WI_indices_mult_user}}
\label{app:lemma_WI}
\begin{enumerate}
\item If $s_i=0$, then if $\lambda> -\alpha_i c_f$, then the optimal action from Lemma~\ref{lemma:opt_pol_sngle_user} $\pi^{\ast}(\tau,0)=0$. The minimum value of $\lambda$ beyond which $\pi^{\ast}(\tau,0)=0$ is $-\alpha_i c_f$. Hence, by the definition of the Whittle index (Definition~\ref{def:WI_index}), $W^i(\tau,0)=-\alpha_i c_f$. 
    If $s_i=1$, the optimal action from Lemma~\ref{lemma:opt_pol_sngle_user}, $\pi^{\ast}(\tau,0)=0$ if $\tau<f(\alpha_i c_f+\lambda_i)\Rightarrow f^{-1}(\tau)-\alpha_i c_f<\lambda\Rightarrow g_i(\tau)-\alpha_i c_f<\lambda$.  
    Again by the definition of Whittle index $W^i(\tau,1)=g_i(\tau)-\alpha_i c_f$. Since $\tau$ is an integer and $\lambda$ can take any real value we define $g_i(\bar{\tau})\coloneqq\min\left\{\lambda_i:  f(\lambda_i)>\bar{\tau}\right\}$ for any $\bar{\tau}\in\mathbb{N}$. This implies that $g(\tau)$ represents the minimum  value of $\lambda$ for which it is not optimal to fetch the content.
    \item 

By definition,
\[
W^i(\tau,1)
= \min\{\lambda : f_i(\lambda+\alpha_i c_f)>\tau\}.
\]
Let $\mu=\lambda+\alpha_i c_f$, so that $\lambda=\mu-\alpha_i c_f$. Then the condition 
$f_i(\lambda+\alpha_i c_f)>\tau$ becomes $f_i(\mu)>\tau$. Hence,
\[
\{\lambda : f_i(\lambda+\alpha_i c_f)>\tau\}
= \{\mu-\alpha_i c_f : f_i(\mu)>\tau\}.
\]
Taking the minimum on both sides gives
\begin{align*}
 W^i(\tau,1)
&= \min_{\mu}\{\mu-\alpha_i c_f : f_i(\mu)>\tau\}\\
&= \bigg(\min_{\mu}\{\mu : f_i(\mu)>\tau\}\bigg) - \alpha_i c_f.  
\end{align*}

By the definition of $\bar{W}_i(\tau,1)$, we have
\[
\bar{W}^i(\tau,1) = \min\{\lambda : f_i(\lambda)>\tau\} = \min_{\mu}\{\mu : f_i(\mu)>\tau\}.
\]
Combining the two expressions yields
\[
\bar{W}^i(\tau,1)=W^i(\tau,1)+\alpha_i c_f.
\]
Hence $\sum_{i:s_i=1}W^i(\tau,1)+\sum_{i:s_i=1}W^i(\tau,0)=\sum_{i=1}^N\bar{W}^i(\tau,1)-c_f$ is independent of $\alpha$ since $\bar{W}_i(\cdot)$ is independent of $\alpha$.
\end{enumerate}
\end{document}